%% file: main.tex
\documentclass[a4paper,11pt]{article}
\pdfoutput=1 

\usepackage{jheppub} 

\usepackage[T1]{fontenc} 

\usepackage{float}
\usepackage{slashed}
\usepackage{xcolor}
\usepackage{xspace}
\usepackage{tablefootnote}
\usepackage{multirow}
\usepackage{booktabs}
\usepackage{array}
\usepackage{lineno}
\usepackage{graphicx,color}
\usepackage{subcaption}
\usepackage{arydshln}
\usepackage{pifont}
\usepackage{dsfont}

\usepackage[normalem]{ulem}

\definecolor{ctcolorblack}{gray}{0}
\definecolor{ctcolorgray}{gray}{.5}
\definecolor{ctcolorgraylight}{gray}{.8}
\definecolor{ctcolorgraylighter}{gray}{.95}

\usepackage{listings}			
\title{Automated event generation for S-wave quarkonium and leptonium production in NRQCD and NRQED}

\author[a]{Alice Colpani Serri,}
\emailAdd{alice.colpani\_serri.dokt@pw.edu.pl}
\author[b,c]{Chris A. Flett,}
\emailAdd{christopher.flett@ijclab.in2p3.fr}
\author[b]{Jean-Philippe Lansberg,}
\emailAdd{Jean-Philippe.Lansberg@in2p3.fr}
\author[c]{Olivier Mattelaer,}
\emailAdd{olivier.mattelaer@uclouvain.be}
\author[d]{Hua-Sheng Shao,}
\emailAdd{huasheng.shao@lpthe.jussieu.fr}
\author[d]{and Lukas Simon}
\emailAdd{lsimon@lpthe.jussieu.fr}

\affiliation[a]{Faculty of Physics, Warsaw University of Technology, plac Politechniki 1, 00-661, Warszawa, Poland}
\affiliation[b]{Universit\'e Paris-Saclay, CNRS, IJCLab, 91405 Orsay, France}
\affiliation[c]{Centre for Cosmology, Particle Physics and Phenomenology (CP3), Universit\'e Catholique de Louvain, Chemin du Cyclotron, Louvain-la-Neuve,
B-1348, Belgium}
\affiliation[d]{Laboratoire de Physique Th\'eorique et Hautes Energies (LPTHE), UMR 7589, Sorbonne Universit\'e et CNRS, 4 place Jussieu, 75252 Paris Cedex 05, France}

\def\remove#1#2{#1\hspace{-#2truecm}\backslash}

\newcommand{\gammaupc}{\texttt{gamma-UPC}}
\newcommand{\helaconia}{\texttt{HELAC-Onia}}
\newcommand{\madgraph}{\texttt{MadGraph5\_aMC@NLO}}
\newcommand{\mgshort}{\texttt{MG5\_aMC}}
\newcommand{\madgraphfour}{\texttt{MadGraph/MadEvent}~v4}
\newcommand{\helacphegas}{\texttt{HELAC-PHEGAS}}
\newcommand{\madonia}{\texttt{MadOnia}}
\newcommand\prompt{{\tt MG5\_aMC>}}
\newcommand{\pythia}{\texttt{Pythia}}
\newcommand{\pythiaeight}{\texttt{Pythia}~v8.315}
\newcommand{\herwig}{\texttt{Herwig}}
\newcommand{\ufo}{\textsc{UFO}}

\newcommand\ident{{\cal I}}
\newcommand\Ione{\ident_1}
\newcommand\Itwo{\ident_2}
\usepackage{textcomp}
\newcommand{\mgtilde}{\raisebox{0.5ex}{\texttildelow}}

\abstract{We present an extension of the \madgraph\ framework that enables the automated calculation of leading-order cross sections for S-wave quarkonium and leptonium production within the non-relativistic QCD (NRQCD) and non-relativistic QED (NRQED) factorisation formalisms. The framework has been validated against a variety of benchmark processes, demonstrating robustness and flexibility for phenomenological studies. A key advantage of this implementation is its seamless integration with existing \madgraph\ features, allowing computations not only within the Standard Model but also in a wide range of Beyond the Standard Model or Effective Field Theory scenarios via a modified Universal Feynman Output (\ufo) interface. Furthermore, the framework maintains compatibility with standard Monte Carlo event generators for parton showering and hadronisation. Through numerous examples, we highlight that theoretical studies of quarkonium processes require careful consideration: the impact of subleading contributions is often difficult to predict using simple counting arguments based solely on the hierarchy of couplings and velocity-scaling rules.
}

\begin{document} 
\maketitle
\flushbottom

\input{01_introduction}

\input{02_methodology}

\input{03_program}

\input{04_setup}

\input{05_benchmarks}

\input{06_quarkonium}

\input{07_leptonium}

\input{08_conclusions}

\input{acknowledgements}

\bibliographystyle{JHEP} 
\bibliography{references}

\end{document}

%% file: 01_introduction.tex
\section{Introduction}

Heavy quark-antiquark and lepton-antilepton composite systems, known as quarkonia and leptonia respectively, are among the simplest bound states in the subatomic world. Due to its multi-scale nature, quarkonium production serves as an essential tool for exploring both perturbative and non-perturbative aspects of quantum chromodynamics (QCD). 
Motivated by recent advances in both the theoretical descriptions and experimental measurements of quarkonium~\cite{Brambilla:2010cs,Lansberg:2019adr,Chapon:2020heu} and leptonium~\cite{dEnterria:2023yao,Blumer:2024fvc}, it is appealing to improve event generators for Monte Carlo (MC) simulations of quarkonium and leptonium processes in full generality. This work presents the first step toward developing a general-purpose automated program for quarkonium and leptonium production within the widely-used \madgraph\ (\mgshort\ hereafter) framework~\cite{Alwall:2014hca,Frederix:2018nkq}, enabling the automatic generation of matrix elements and event samples for a broad range of processes and collider environments.

Quarkonium production is a branch of heavy-quark physics focused on studying bound states of charm and bottom quarks.
Such bound states, the strong-interaction counterparts of the electromagnetically bound positronium $(e^+e^-)$ atom, are among the simplest hadronic systems from a theoretical perspective. Notably, the discovery of the $J/\psi$ particle, the first observed charmonium state, sparked the `November Revolution' in particle physics more than 50 years ago. Despite its apparent simplicity, the quarkonium production mechanism remains to be fully understood~\cite{Brambilla:2010cs,Lansberg:2019adr,Chapon:2020heu}. In particular, it remains challenging to reconcile theoretical predictions with experimental measurements across a wide range of observables simultaneously. Nonetheless, quarkonia have far-reaching applications, serving as one of the most effective probes of gluon dynamics in the proton. For example, the upcoming Electron-Ion Collider~(EIC) is anticipated to provide smoking-gun signals of gluon saturation~\cite{Boer:2024ylx}.

Quarkonium production is commonly described within the framework of non-relativistic QCD (NRQCD)~\cite{Bodwin:1994jh}, the most widely adopted approach for quarkonium studies. Within this framework, the short-distance cross section for producing a heavy-quark pair in a quantum state $n = {}^{2S+1}L^{[C]}_{J}$ can be computed in perturbative QCD (pQCD).
Here, the quantum numbers of quarkonium intermediate Fock states are characterised by the relative orbital angular momentum $L = 0,1,\dots$ of the constituent heavy quarks (S-wave, P-wave, etc.), the quarkonium spin $S=0,1$ (singlet/triplet), the total angular momentum $J$, and the colour $C = 1,8$ (singlet/octet).
The probability that a heavy-quark pair in a given quantum state $n$ hadronises into a physical quarkonium state is described by a long-distance matrix element (LDME), whose value is expected to follow the velocity-scaling rules of NRQCD. In this way, the entire quarkonium spectrum, consisting of hierarchical Fock states, can be constructed in a manner consistent with quantum-field theory.

Let us briefly review the existing public theoretical tools for calculating cross sections and/or simulating MC events for quarkonium processes:
\begin{itemize}
\item\textbf{General-purpose MC event generators}: Particle-level events involving quarkonia can be generated in four principal ways within general-purpose MC event generators, such as \herwig~\cite{Corcella:2000bw,Corcella:2002jc,Bahr:2008pv,Bellm:2013hwb,Bellm:2015jjp}, \pythia~\cite{Sjostrand:2000wi,Sjostrand:2006za,Sjostrand:2007gs,Sjostrand:2014zea,Bierlich:2022pfr}, and \texttt{Sherpa}~\cite{Gleisberg:2008ta,Sherpa:2019gpd,Sherpa:2024mfk}. In these tools, quarkonia can be produced (i) from decays of other particles, (ii) during the hadronisation phase from a heavy-quark pair, (iii) within parton showers, or (iv) in hard interactions modelled by matrix elements. Recently, quarkonium-specific parton showers have been introduced in \pythia~v8.3~\cite{Cooke:2023ldt} and \herwig~v7.4~\cite{Masouminia:2025kec}. Matrix elements for several dedicated NRQCD-based quarkonium-production processes have also been implemented in \pythia. However, these two functionalities -- matrix elements and parton showers -- are not yet fully compatible.
\item \textbf{Process-independent parton-level event generators}: More self-contained MC tools for generating leading-order (LO) parton-level quarkonium events in arbitrary processes have been developed through several dedicated efforts, such as \madonia~\cite{Artoisenet:2007qm}, based on \madgraphfour~\cite{Alwall:2007st} and \helaconia~\cite{Shao:2012iz,Shao:2015vga}, which is based on \helacphegas~\cite{Kanaki:2000ey,Papadopoulos:2000tt,Kanaki:2000ms,Papadopoulos:2006mh,Cafarella:2007pc}. Although each comes with certain limitations, these tools are built upon the NRQCD formalism and offer process-independent implementations. For instance, \madonia\ enables automated event generation for processes involving a single S-wave or P-wave quarkonium in the final state, while \helaconia\ can also handle processes with multiple S-wave and P-wave quarkonia. The former, however, has not been ported to the current state-of-the-art \mgshort\ framework. Particle-level event samples can be readily obtained by interfacing parton-level events with general-purpose MC generators through the Les Houches Event File (LHEF) format~\cite{Alwall:2006yp}.
\item\textbf{Process-specific event generators}: Several process-specific implementations of quarkonium production have been developed in dedicated event generators, including \texttt{SuperChic}~\cite{Harland-Lang:2020veo}, \texttt{STARlight}~\cite{Klein:2016yzr}, \texttt{eSTARlight}~\cite{Lomnitz:2018juf}, \texttt{EPOS4}~\cite{Werner:2023zvo,Zhao:2023ucp,Zhao:2024ecc,Zhao:2025wtx}, and \texttt{BCVEGPY}~\cite{Chang:2003cq,Chang:2005hq,Chang:2006xka}. \texttt{SuperChic}, \texttt{STARlight}, and \texttt{eSTARlight} focus on exclusive processes in hadronic collisions. In particular, \texttt{SuperChic} models central exclusive quarkonium production with intact forward protons or ions, whereas \texttt{STARlight} simulates photon-induced reactions in ultraperipheral collisions, and \texttt{eSTARlight} focuses on photo- and electro-production in electron-ion collisions. In contrast, \texttt{EPOS4} can be used to simulate inclusive quarkonium production in heavy-ion collisions, accounting for the effects of the quark–gluon plasma and the soft-QCD environment. Instead of employing the NRQCD formalism, \texttt{EPOS4} adopts the quarkonium Wigner density matrix formalism~\cite{Villar:2022sbv}. Meanwhile, \texttt{BCVEGPY} is a dedicated parton-level event generator for $B_c$-meson hadroproduction, which can be interfaced with \pythia.
\item\textbf{Process-specific cross-section calculators}: There are also cross-section calculators developed for specific quarkonium-production processes. One such tool is \texttt{FDCHQHP}~\cite{Wan:2014vka}, which was generated using the semi-automated private code \texttt{FDC}~\cite{Wang:2004du}. \texttt{FDCHQHP} can compute differential cross sections as well as quarkonium polarisation observables for single inclusive S- and P-wave charmonium and bottomonium hadroproduction processes at next-to-leading order (NLO) in the strong coupling~$\alpha_s$. For non-prompt $J/\psi$ or $\psi(2S)$ inclusive production at hadron colliders, the differential cross sections can be computed with \texttt{FONLL}~\cite{Cacciari:1998it,Cacciari:2001td,Cacciari:2012ny,Cacciari:2015fta} at NLO plus next-to-leading logarithmic (NLL) accuracy in QCD.
\item \textbf{Toolkits for semi-automatic symbolic calculations}: There are also public software packages for semi-automatic symbolic or analytic calculations of matrix elements for quarkonium processes within non-relativistic effective field theories (EFTs). Two representative examples are \texttt{FeynOnium}~\cite{Brambilla:2020fla} and \texttt{AmpRed}~\cite{Chen:2024xwt}.
\end{itemize}

\sloppy

Analogous to quarkonium, leptons of opposite electric charge ($\ell^\pm = e^\pm, \mu^\pm, \tau^\pm$) can form bound states, collectively known as \emph{leptonia}, through their quantum electrodynamics (QED) interaction. Among the six possible leptonium systems -- $(e^+e^-)$, $(\mu^\pm e^\mp)$, $(\mu^+\mu^-)$, $(\tau^\pm e^\mp)$, $(\tau^\pm\mu^\mp)$, and $(\tau^+\tau^-)$ -- only the first two, positronium $(e^+e^-)$~\cite{PhysRev.82.455} and muonium $(\mu^\pm e^\mp)$~\cite{Hughes:1960zz}, have been experimentally observed to date. Consequently, leptonium production has received comparatively little attention in high-energy physics, apart from a few theoretical explorations. Because leptonia consist solely of leptons, these systems are free from the complexities of strong interactions that affect hadronic atoms, allowing for extremely precise theoretical predictions. Comparing such predictions with high-precision spectroscopic and lifetime measurements provides sensitive probes of higher-order QED effects, possible contributions from new physics, and fundamental constants such as the fine-structure constant $\alpha$ and lepton mass $m_\ell$. The lightest system, positronium, has been extensively studied for precision tests of QED~\cite{Karshenboim:2005iy} and in searches for violations of the discrete spacetime CPT symmetries~\cite{Bernreuther:1988tt,Yamazaki:2009hp}, where C, P, and T denote charge conjugation, parity, and time reversal, respectively. Moreover, muonium–antimuonium conversion searches test charged-lepton flavour violation~\cite{Feinberg:1961zza,Abela:1996dm,Cvetic:2005gx,Fukuyama:2021iyw}, possibly providing windows into physics beyond the Standard Model (BSM). Finally, observation of true tauonium (often called ditauonium) $(\tau^+\tau^-)$ would help refine our understanding of tau-lepton properties, in particular its mass~\cite{dEnterria:2023yao,Fu:2023uzr}. In contrast to quarkonium, publicly available theoretical tools for studying leptonium processes remain very limited.

The main motivation of this work is to develop and ultimately provide the community with an all-encompassing tool for bound-state production studies in NRQCD and non-relativistic QED (NRQED)~\cite{Caswell:1985ui}. The goal is to deliver reliable, efficient, and fast automated state-of-the-art computations, with NLO accuracy based on an extended Frixione–Kunszt–Signer (FKS) infrared-divergence subtraction formalism~\cite{Frixione:1995ms,Frixione:1997np,AH:2024ueu} planned for near-term implementation. This work lays the foundation for that endeavour and falls within the aforementioned category of process-independent parton-level event generators, designed to simulate processes involving one or more quarkonia and/or leptonia. As mentioned, our implementation is embedded within the well-established \mgshort\ framework, which is widely used in the high-energy physics community, thereby extending its capabilities to include S-wave quarkonium and leptonium production. To the best of our knowledge, no other publicly available tool currently provides similar functionality for leptonium processes.

The structure of this paper is as follows. Section~\ref{sec:theory} presents the theoretical formalism of quarkonium and leptonium production in the collinear factorisation framework, while section~\ref{sec:program} details its implementation in \mgshort. Section~\ref{sec:setup} introduces the general parameter setup employed throughout this work. We validate our implementation in section~\ref{sec:benchmark} before showcasing the capabilities of our tool for both quarkonium and leptonium production, including illustrative examples of phenomenological relevance, in sections~\ref{sec:quarkonium} and~\ref{sec:leptonium}, respectively. Finally, conclusions are drawn in section~\ref{sec:conclusions}.

%% file: 02_methodology.tex
\section{Theoretical framework}\label{sec:theory}

In this section, we introduce the formalisms for describing quarkonium- and leptonium-production processes within NRQCD and NRQED. For completeness, we present the general structure of the factorised cross sections for both S-wave and P-wave configurations and describe in detail the covariant projection method used to isolate individual Fock states~\cite{Kuhn:1979bb,Guberina:1980dc,Berger:1980ni,Petrelli:1997ge,Maltoni:1997pt}. The current implementation in our framework supports only the colour and spin projectors required for S-wave production. The incorporation of orbital and total angular momentum projectors, which is necessary for P-wave states, is left for future development.

In the QCD collinear factorisation framework, the differential cross section for the inclusive production of a non-relativistic bound state $\mathcal{B}$ in hadron-hadron collisions,
\begin{equation}
    \mathrm{N}_1 + \mathrm{N}_2 \rightarrow \mathcal B + X\,,    
\end{equation}
can be expressed as
\begin{equation}\label{eq:xsec}
\begin{aligned}
    {\rm d}\sigma( \mathrm{N}_1 + \mathrm{N}_2 \rightarrow \mathcal B + X) &= \sum_{\Ione,\Itwo,n} \int {\rm d}x_1 {\rm d}x_2 f_{\Ione/\mathrm{N}_1}(x_1) f_{\Itwo/\mathrm{N}_2}(x_2) \\
    &\quad\times{\rm d}\hat{\sigma}(\Ione\Itwo \rightarrow (C_1 C_2)[n] + X_p) \langle \mathcal O_{n}^{\mathcal B} \rangle\,,
\end{aligned}
\end{equation}
where ${\rm d}\hat{\sigma}$ denotes the partonic (or short-distance) cross section for producing the heavy constituents $C_1$ and $C_2$ in a quantum state $n$, which subsequently evolve into the physical bound state $\mathcal B$ with probability characterised by the LDME $\langle \mathcal O_n^{\mathcal B} \rangle$. For quarkonium production, the constituents are a heavy quark and antiquark, $(C_1 C_2) = (Q \bar{Q}^\prime)$, and $\mathcal B = \mathcal Q$, while for leptonium production, the constituents are a pair of massive, oppositely-charged leptons, $(C_1 C_2) = (\ell^- \ell^{\prime +})$, and $\mathcal B = \mathcal L$.  In Eq.~\eqref{eq:xsec}, $X$ represents additional associated radiation and beam remnants, $X_p$ denotes parton-level radiation, and $f_{\Ione/\mathrm{N}1}$ ($f_{\Itwo/\mathrm{N}_2}$) is the parton distribution function (PDF) describing the probability of finding a parton $\Ione$ ($\Itwo$) carrying a longitudinal momentum fraction $x_1$ ($x_2$) of its parent hadron momentum. At face value, this expression involves an infinite sum over intermediate Fock states $n$; however, the NRQCD and NRQED velocity-scaling rules in powers of $v^2$~\cite{Caswell:1985ui,Bodwin:1994jh} impose a hierarchy (cf.\@\xspace table \ref{tab:scaling} for charmonium and bottomonium) that determines the phenomenological relevance of the corresponding Fock states. 

The short-distance cross section ${\rm d}\hat{\sigma}(\Ione\Itwo \rightarrow (C_1 C_2)[n] + X_p)$ can be obtained from the production amplitude of $C_1$ and $C_2$ by projecting it onto the specific quantum state $n$, following section 2 of ref.~\cite{AH:2024ueu}. As mentioned previously, the quantum state $n$ is specified by its $S$, $L$, $J$ and $C$ quantum numbers. For leptonium ($\mathcal B = \mathcal L$), only the colour singlet configuration ($C=1$) is possible.

With the same notation as in refs.~\cite{Frederix:2009yq,AH:2024ueu}, let us consider a generic $2\to n$ partonic process
\begin{equation}
r = (\mathcal{I}_1,\dots,\mathcal{I}_{n+2})
\end{equation}
describing the production of $C_1$ and $C_2$,
\begin{equation}
    \mathcal{I}_1(k_1) \mathcal{I}_2(k_2) \to \mathcal{I}_3(k_3)\mathcal{I}_4(k_4)\dots\mathcal{I}_{n+2}(k_{n+2})\,,
\end{equation}
where the $j$-th parton has identity $\mathcal{I}_j$, and $k_j$ denotes its four-momentum. Without loss of generality, we assign $\mathcal{I}_3 = C_1$ and $\mathcal{I}_4 = C_2$. The amplitude can then be written as
\begin{equation}
    \mathcal{A}^{(n,0)}(r) = \bar{u}_{\lambda_{C_1}}(k_3)\, \Gamma^{(n,0)}(r)\,  v_{\lambda_{C_2}}(k_4)\,,
\end{equation}
where $\bar{u}$ and $v$ denote the outgoing Dirac spinors of $C_1$ and $C_2$ with helicities $\lambda_{C_1}$ and $\lambda_{C_2}$, respectively. The amputated amplitude $\Gamma^{(n,0)}(r)$ encodes the Dirac-algebra structure.

For $(C_1C_2)=(Q\bar{Q}^\prime)$, we first apply the colour projection. Let $c_3$ and $c_4$ denote the colour indices of $C_1$ and $C_2$. The colour-projected amplitude is 
\begin{equation}\label{eq:proj_c}
    \mathcal{A}^{(n,0)}_{\left\{[C]\right\}}(r) = \sum_{c_3,c_4} ~\mathbb{P}_C ~\mathcal{A}^{(n,0)}(r)\,,
\end{equation}
where the operator $\mathbb{P}_{C=1} = \delta_{c_3 c_4}/\sqrt{N_c}$~\footnote{$N_c=3$ in QCD.} projects the heavy-quark pair onto a colour-singlet $(C=1)$ Fock state, and $\mathbb{P}_{C=8} = \sqrt{2}\,t_{c_4 c_3}^{c_{34}}$ projects onto a colour-octet $(C=8)$ one, with $t_{c_4 c_3}^{c_{34}}$ the Gell-Mann matrix element. For leptonium, $\mathbb{P}_{C=1} = \delta_{c_3 c_4}$ with $c_3 = c_4 = 0$, which leaves the amplitude $\mathcal{A}^{(n,0)}(r)$ unchanged. We keep the colour projection for leptonium here to maintain a unified formalism for both quarkonium and leptonium production.

The spin-projected amplitude can be expressed as
\begin{equation}\label{eq:proj_s}
    \mathcal{A}^{(n,0)}_{\left\{[C] ,S \right\}}(r) = \sum_{\lambda_{C_1}, \lambda_{C_2}} ~\mathbb{P}_S ~\mathcal{A}^{(n,0)}_{\left\{[C]\right\}}(r)\,,
\end{equation}
where the spin projector is given by
\begin{equation}
    \mathbb{P}_S = \frac{\bar{v}_{\lambda_{C_2}}(k_4)\Gamma_{\!S}\,u_{\lambda_{C_1}}(k_3)}{2\sqrt{2m_{C_1}m_{C_2}}}\,,
\end{equation} 
with $\Gamma_{\!S=0}=\gamma_5$ for the production of a spin-singlet $(C_1C_2)$ and $\Gamma_{\!S=1}=\slashed{\varepsilon}^{*}_{\lambda_s}(K)$ for a spin-triplet $(C_1C_2)$. Here, $\varepsilon^{*}_{\lambda_s}(K)$ denotes the polarisation vector of the $(C_1C_2)$ system with total four-momentum $K = k_3 + k_4$ and spin-related helicity $\lambda_s = \pm1,0$. The constituent momenta are parameterised as
\begin{align}
k_3^{\mu} &= \frac{m_{C_1}}{m_{C_1} + m_{C_2}} K^{\mu} + q^{\mu}\,, \\
k_4^{\mu} &= \frac{m_{C_2}}{m_{C_1} + m_{C_2}} K^{\mu} - q^{\mu}\,,
\end{align}
where $m_{C_1}$ ($m_{C_2}$) denotes the mass of $C_1$ ($C_2$), and $q^{\mu}$ is the relative momentum of the constituents.

To project onto a Fock state with a given orbital angular momentum $L=0,1$ (corresponding to S-wave and P-wave configurations, respectively), one differentiates the colour- and spin-projected amplitude above $L$ times with respect to the relative momentum $q^{\mu}$. The projection is then completed by taking the limit $q \rightarrow 0$. For the cases $L = 0$ and $L = 1$, the orbital-angular-momentum–projected amplitude takes the form
\begin{equation}\label{eq:proj_l}
    \mathcal{A}^{(n,0)}_{\left\{[C] ,S ,L \right\}}(r) = \left[\mathbb{P}_L ~\mathcal{A}^{(n,0)}_{\left\{[C] ,S \right\}}(r)\right]_{q=0} =  \bigg[ \bigg( \varepsilon_{\lambda_l}^{\mu,*}(K) \frac{{\rm d}}{{\rm d}q^\mu}\bigg)^L \mathcal{A}^{(n,0)}_{\left\{[C] ,S \right\}}(r)\bigg]_{q=0}\,,
\end{equation}
where $\varepsilon_{\lambda_l}^{\mu,*}(K)$ denotes the polarisation vector associated with the $L=1$ orbital-angular-momentum state, and $\lambda_l = \pm 1, 0$ labels its polarisation components.

Finally, when $L\neq0$ and $S\neq 0$, the amplitude must be projected onto a Fock state with a given total angular momentum $J = |L - S|, \dots, L + S$,
\begin{equation}\label{eq:proj_j}
    \mathcal{A}^{(n,0)}_{\left\{[C] ,S ,L, J \right\}}(r) = \sum_{\lambda_s, \lambda_l} \mathbb{P}_J ~\mathcal{A}^{(n,0)}_{\left\{[C] ,S ,L \right\}}(r)\,,
\end{equation}
where $\mathbb{P}_J = \langle J, \lambda_j | L, \lambda_l; S, \lambda_s \rangle$ is the Clebsch–Gordan coefficient, and $\lambda_j = -J, -J+1, \dots, J-1, J$. If either $L = 0$ or $S = 0$, the value of $J$ is uniquely determined.

After the projection procedure is carried out, the constituents $\mathcal{I}_3=C_1$ and $\mathcal{I}_4=C_2$ are combined into a new effective particle $\mathcal{I}_{3\oplus4}=(C_1C_2)[n]$ in the quantum state $n={}^{2S+1}L_J^{[C]}$, with four-momentum $K^\mu$ and invariant mass $m_\mathcal{B}=m_{C_1}+m_{C_2}$. The resulting process, 
\begin{equation}
\dot{r}=r^{3\oplus4,\remove{4}{0.145}}=(\mathcal{I}_1,\mathcal{I}_2,\mathcal{I}_{3\oplus4},\remove{\mathcal{I}}{0.200}_4,\dots,\mathcal{I}_{n+2})\,,
\end{equation}
is described by the amplitude
\begin{equation}\label{eq:amp_quarkonium}
    {\mathds A}^{(n-1,0)}(\dot{r}) = \mathcal{A}^{(n,0)}_{\left\{[C] ,S ,L, J \right\}}(r)\,. 
\end{equation}
The corresponding differential partonic cross section can be written as
\begin{equation}\label{eq:xsec_quarkonia}
{\rm d}\hat{\sigma}=\frac{1}{\mathcal{N}(\dot{r})}\frac{1}{(2J+1)N_{[C]}}\frac{m_{C_1}+m_{C_2}}{2m_{C_1}m_{C_2}}\!\left(\frac{1}{2\hat{s}} \frac{1}{\omega(\mathcal{I}_1)\omega(\mathcal{I}_2)} \right)\! \sum_{\substack{\rm colour\\ \rm spin}} |  {\mathds A}^{(n-1,0)}(\dot{r})|^2\,{\rm d}\phi_{n-1}(\dot{r})\,,
\end{equation}
where $\mathcal{N}(\dot{r})$ is the final-state symmetry factor at the level of the process $\dot{r}$. The colour factors are given by $N_{[C=1]} = 2N_c$ and $N_{[C=8]} = N_c^2-1$ for quarkonium, and $N_{[C=1]} = 1$ for leptonium. The Lorentz-invariant flux factor is given by $1/(2\hat{s})$ with $\hat{s}=(k_1+k_2)^2$, where the masses of $\Ione$ and $\Itwo$ have been neglected.\footnote{In the code implementation of \mgshort, however, these mass effects are included whenever relevant.} The factor $\omega(\mathcal{I})$ accounts for the averaging over colour and spin degrees of freedom for particle $\mathcal{I}$. The phase-space element ${\rm d}\phi_{n-1}(\dot{r})$ represents the $(n-1)$-body phase-space measure. We emphasise that both the symmetry factor and the phase-space integration are evaluated after the constituents $C_1$ and $C_2$ have been projected into the Fock state, and hence the relevant phase space is that of the $(n-1)$-body final state.

In contrast to quarkonia, the LDMEs of leptonia can be reliably calculated by solving the Schr\"odinger equation with the Coulomb potential. These calculations can be systematically improved by including higher-order radiative and relativistic corrections. For leptonium production, the LO LDME of the $N$-th radial excitation of an S-wave leptonium is given by
\begin{equation}
    \langle \mathcal O_{^{2S+1}\mathrm{S}_{S}^{[1]}}^{{\mathcal L}(N\mathrm{S})} \rangle = (2S+1)\dfrac{\alpha^3}{\pi N^3}\left(\dfrac{m_{\ell^-}m_{\ell^{\prime +}}}{m_{\ell^-}+m_{\ell^{\prime +}}}\right)^{\!3}\,.\label{eq:leptoniumLDME}
\end{equation}
Higher-order radiative and relativistic corrections to the leptonium LDMEs can be found, for instance, in ref.~\cite{dEnterria:2022alo}. 

For processes producing multiple quarkonia, leptonia, or combinations thereof, the projection procedure described above can be applied iteratively.

%% file: 03_program.tex
\section{Implementation}\label{sec:program}

\subsection{Interface and generation syntax}\label{sec:interface}

\begin{table}[t!]
\centering
\renewcommand*{\arraystretch}{1.3}
\begin{tabular}[t]{ccccc}
\toprule
\textbf{power counting}& $\eta_Q$ & $J/\psi,\Upsilon$ & $h_Q$ & $\chi_{QJ}$ \\
\midrule
$v^3$ & $^1S_0^{[1]}$ & $^3S_1^{[1]}$ & --- & --- \\
$v^5$ & --- & --- & $^1P_1^{[1]},{}^1S_0^{[8]}$ & $^3P_J^{[1]},{}^3S_1^{[8]}$ \\
$v^7$ & $^1S_0^{[8]},{}^3S_1^{[8]},{}^1P_1^{[8]}$ & $^1S_0^{[8]},{}^3S_1^{[8]},{}^3P_J^{[8]}$ & --- & --- \\
\bottomrule
\end{tabular}
\caption{Hierarchy of Fock-state contributions to different quarkonium states, organised according to their leading powers in the NRQCD velocity-scaling rules~\cite{Bodwin:1994jh}.}
\label{tab:scaling}
\end{table}

Within the NRQCD framework, a physical quarkonium state is described as a superposition of different Fock states. According to the velocity-scaling rules~\cite{Bodwin:1994jh}, a physical bound state can be systematically expanded in powers of the relative velocity
\begin{equation}
v^{\mu} = q^{\mu}/(2\mu_\mathcal{Q})\,,
\end{equation}
where $q^\mu$ denotes the relative momentum of the two constituents and $\mu_\mathcal{Q}=m_Qm_{\bar{Q}'}/(m_Q+m_{\bar{Q}^\prime})$ is their reduced mass. For charmonium states such as $J/\psi$, a typical value is $v^2=-v^\mu v_\mu\sim0.3$. The velocity scaling arises from the scaling behaviour of both the Fock-state amplitudes and their associated LDMEs. This hierarchy governs the relative importance of each term in the expansion, allowing one to truncate the series at a given order in $v$ to achieve the desired precision in cross-section predictions. The corresponding power counting for the most relevant charmonia and bottomonia is summarised in table~\ref{tab:scaling}. Adopting these scaling relations in combination with Eq.~\eqref{eq:xsec}, the partonic cross section for $J/\psi$ production becomes
\begin{equation}
\begin{aligned}
    &\quad{\rm d}\hat{\sigma}(\Ione\Itwo \rightarrow  J/\psi + X) \\
    &= {\rm d}\hat{\sigma}(\Ione\Itwo \rightarrow (c \bar{c}) \left[^3S_1^{[1]}\right] + X) \langle {\cal O}^{J/\psi}_{^3S_1^{[1]}}\rangle + {\rm d}\hat{\sigma}(\Ione\Itwo \rightarrow (c \bar{c}) \left[^3P_J^{[8]}\right]  + X) \langle {\cal O}^{J/\psi}_{^3P_J^{[8]}}\rangle \\
    &\quad+ {\rm d}\hat{\sigma}(\Ione\Itwo \rightarrow (c \bar{c}) \left[^1S_0^{[8]}\right] + X) \langle {\cal O}^{J/\psi}_{^1S_0^{[8]}}\rangle + {\rm d}\hat{\sigma}(\Ione\Itwo \rightarrow (c \bar{c}) \left[^3S_1^{[8]}\right]   + X) \langle {\cal O}^{J/\psi}_{^3S_1^{[8]}}\rangle+\ldots\,,
\end{aligned}
\end{equation}
where the ellipses ``$\ldots$'' represent the omitted relativistic corrections and Fock-state contributions scaling at higher orders in $v^2$.

To implement a physical bound state as an expansion in terms of different Fock states, we adopt a scheme in which each Fock state is treated similarly to an elementary \textit{particle}, while physical bound states are handled analogously to the \textit{multiparticles} class in \mgshort. Generating processes involving bound states therefore requires loading a Universal Feynman Output (\ufo)  model~\cite{Degrande:2011ua,Darme:2023jdn}, which includes a new class, \texttt{Boundstate}, introduced in analogy to the \texttt{Particle} class for elementary particles. Along with this article, we provide a new \ufo\ model of the SM, dubbed \texttt{sm\_onia}. This model contains a dedicated file, \texttt{boundstates.py}, which defines the Fock-state content of various bound states. These can be used in \mgshort\ to generate processes in the same way as ordinary elementary particles, either by specifying the Fock-state \textit{name} or its \textit{pdg\_code}.~\footnote{Currently, only S-wave Fock states are implemented.} The naming convention is structured as follows: the \textit{name} begins with the label of the physical bound state ({e.g.,\@\xspace} \texttt{Jpsi} for the $J/\psi$ meson), followed by parentheses containing the Fock-state information. The first entry inside the parentheses denotes the radial excitation or principal quantum number $N$ ({e.g.,\@\xspace} \texttt{1} for the ground state, \texttt{2} for the first excited state). Following a vertical bar (\texttt{|}) are the quantum numbers of the Fock state: spin multiplicity $2S+1$ ({e.g.,\@\xspace} \texttt{1} for spin singlet, \texttt{3} for spin triplet), orbital angular momentum $L$ ({e.g.,\@\xspace} \texttt{S}), total angular momentum $J$ ({e.g.,\@\xspace} \texttt{1}), colour representation $C$ ({e.g.,\@\xspace} \texttt{1} for colour-singlet states).~\footnote{For the \textit{name} of leptonium states, we omit the colour index, as their colour configuration is trivial.} The numbering scheme for Fock state PDG identity codes follows, where possible, the convention established by \pythia~\cite{Bierlich:2022pfr}, ensuring compatibility with standard event-generation workflows. Colour-singlet states of quarkonia are assigned the same PDG identity codes as their corresponding physical mesons, while colour-octet states are given non-standard PDG codes, following the structure $99n_qn_sn_rn_Ln_J$, where: $n_q$ denotes the quark flavour (\texttt{4} for charmonia, \texttt{5} for bottomonia, \texttt{7} for $B_c$ mesons), $n_s$ identifies the type of colour-octet (\texttt{0} for $^3S_1^{[8]}$, \texttt{1} for $^1S_0^{[8]}$, \texttt{2} for $^3P_J^{[8]}$)~\footnote{We could not find any PDG identity code for $^1P_1^{[8]}$ in \pythia.}, and the remaining digits $n_r, n_L, n_J$ encode additional quantum numbers in line with the MC conventions~\cite{ParticleDataGroup:2024cfk} (cf. section 45 therein). For leptonia, we instead introduce new eight-digit PDG identity codes, $\pm990n_{\ell_1}n_{\ell_2}n_rn_Ln_J$, where $n_{\ell_{1}}=\max\{n_{\ell^-},n_{\ell^{\prime +}}\}$ and $n_{\ell_2}=\min\{n_{\ell^-},n_{\ell^{\prime +}}\}$. Here, $n_{\ell^-}$ ($n_{\ell^{\prime+}}$) denotes the last digit of the constituent lepton (antilepton) PDG code (e.g.,\@\xspace \texttt{1} for $e^\pm$, \texttt{3} for $\mu^\pm$, and \texttt{5} for $\tau^\pm$). The positive sign is chosen if $n_{\ell^-}\geq n_{\ell^{\prime +}}$, and the negative sign otherwise. The remaining digits, $n_r$, $n_L$, and $n_J$, follow the general MC convention guidelines~\cite{ParticleDataGroup:2024cfk}. Examples of the Fock-state configurations $^{3}S_1^{[1]}$ for the $J/\psi$ meson and $^1S_0$ for the positronium atom are provided in listing~\ref{lst:boundstates.py}. A complete list of available Fock states can be displayed on the \mgshort\ interface using

\vskip 0.25truecm
\noindent
~~\prompt\ {\tt ~display~fockstates}
\vskip 0.25truecm

\noindent
in analogy to the \texttt{display~particles} command for elementary particles.

\begin{lstlisting}[caption={Definitions of the $J/\psi$ meson Fock state $^{3}S_1^{[1]}$ and positronium atom (denoted $\mathit{Ps}$) Fock state $^1S_0$ in \texttt{boundstates.py} of the new \ufo\ model \texttt{sm\_onia}.},captionpos=b,language=Python,label={lst:boundstates.py},float=tp,floatplacement=tbp]
# J/psi meson
jpsi_13s11 = Boundstate(pdg_code = 443,
                        name = 'Jpsi(1|3S11)',
                        particles = ['c','c~'],
                        principal = 1,
                        spin = 3,
                        orbital = 0,
                        J = 1,
                        color = 1,
                        charge = 0,
                        texname = 'jpsi13S11')

# positronium atom
ps_11s0 = Boundstate(pdg_code = 99011001,
                     name = 'Ps(1|1S0)',
                     particles = ['e-','e+'],
                     principal = 1,
                     spin = 1,
                     orbital = 0,
                     J = 0,
                     color = 1,
                     charge = 0,
                     texname = 'Ps11S0')
\end{lstlisting}

As an illustration, to generate a process containing a quarkonium meson in a specific Fock state, such as
\begin{equation}
pp\to J/\psi\!\left[^3S_1^{[1]}\right]+j+X\,,    
\end{equation}
the corresponding generation syntax in \mgshort\ is:
\vskip 0.25truecm
\noindent
~~\prompt\ {\tt ~import~model~sm\_onia-c\_mass}

\noindent
~~\prompt\ {\tt ~generate~p p > Jpsi(1|3S11) j}

\noindent
~~\prompt\ {\tt ~output;~launch}
\vskip 0.25truecm
\noindent
The first command loads the \ufo\ model \texttt{sm\_onia}, while imposing the restriction that the charm quark is treated as massive, a necessary condition to form the $J/\psi$ meson within the NRQCD formalism. In the second command, the generation stage, the desired Fock state is specified by its \textit{name} as defined in listing~\ref{lst:boundstates.py}. After generating the process and creating the output, the numerical values of the LDMEs can be specified in the input card \texttt{onia\_card.dat}, following the interface instructions.~\footnote{It is also possible to modify the values through the interface with the command \texttt{set onia\_card ldme pdg\_code value}. This command can also be used in \mgshort\ input scripts. Note that this adaptability of the LDMEs is an improvement over the deprecated \texttt{MadOnia}, where the LDMEs were hardcoded internally and not adjustable at runtime.}
The input card follows the SUSY Les Houches Accord format~\cite{Skands_2004,Allanach_2009}, with LDME values provided in the block labeled \texttt{Block ldme}. Within this block, each LDME is associated with its corresponding Fock state by specifying the Fock state \textit{pdg\_code}, as defined in \texttt{boundstates.py}, at the beginning of a new line followed by the numerical value, separated by at least one space, as shown in listing~\ref{lst:onia_card}. The default values assigned to the LDMEs are summarised in table~\ref{tab:ldme}.~\footnote{We emphasise that new bound states can be easily added in \texttt{boundstates.py}. In such cases, the newly added Fock states are automatically assigned a default LDME value of unity.} For S-wave colour-singlet states, the LDMEs are computed through their relation to the quarkonium radial wave function at the origin, $R_\mathcal{Q}(0)$, via
\begin{equation}\label{eq:ldme_cs}
    \langle \mathcal O_{^{2S+1}S_S^{[1]}}^{\mathcal Q} \rangle = (2S+1)N_{[C=1]}\frac{\left|R_\mathcal{Q}(0)\right|^2}{4\pi}\,,
\end{equation}
with wave-function values taken from refs.~\cite{Eichten:1995ch,Eichten:2019hbb}. Colour-octet LDMEs are taken from fits to experimental data: ref.~\cite{Han:2014jya} for charmonium mesons and ref.~\cite{Han:2014kxa} for bottomonium mesons. Since no experimental measurements are currently sensitive to $B_c$ and $B_c^\ast$ colour-octet LDMEs, the octet LDMEs of the pseudoscalar state are assigned with a default value that is a factor of 100 smaller than the corresponding colour-singlet value, while the octet LDMEs of the vector $B_c^\ast$ are determined using the heavy-quark spin-symmetry theorem~\cite{Bodwin:1994jh}. On the other hand, the LDMEs of leptonia, which are not adjustable in \texttt{onia\_card.dat}, are fixed according to Eq.~\eqref{eq:leptoniumLDME}.
\begin{lstlisting}[caption={Definition of $J/\psi$ LDMEs (in units of GeV$^3$) in the input file \texttt{onia\_card.dat}},captionpos=b,language=Python,label={lst:onia_card},float=tp,floatplacement=tbp]
#*********************************************************************
# Long distance matrix elements (LDME)                               *
#*********************************************************************
Block ldme
   443      1.160000000000000   # LDME for Jpsi(1|3S11)
   9940003  0.009029230000000   # LDME for Jpsi(1|3S18)
   9941003  0.014600000000000   # LDME for Jpsi(1|1S08)
\end{lstlisting}

\begin{lstlisting}[caption={Definition of the physical $J/\psi$ meson as a collection of contributing Fock states in \texttt{boundstates\_default.txt}.},captionpos=b,language=Python,label={lst:boundstates_default.txt},float=tp,floatplacement=tbp]
# Syntax: label = Fock states (separated by spaces)
Jpsi = Jpsi(1|3S11) Jpsi(1|1S08) Jpsi(1|3S18)
\end{lstlisting}

\begin{table}[ht!]
\centering
\renewcommand*{\arraystretch}{1.4}
\begin{tabular}[t]{cc|cc}
\toprule
${\mathcal Q}[n]$ & $\langle \mathcal O_{n}^{\mathcal Q} \rangle\ \left[\mathrm{GeV}^{3}\right]$ & ${\mathcal Q}[n]$ & $\langle \mathcal O_{n}^{\mathcal Q} \rangle\ \left[\mathrm{GeV}^{3}\right]$ \\
\midrule
$\eta_c\!\left[^1S_0^{[1]}\right]$ & $0.386666666666667$ & $J/\psi\!\left[^3S_1^{[1]}\right]$ & $1.16$ \\
$\eta_c\!\left[^3S_1^{[8]}\right]$ & $0.0146$ & $J/\psi\!\left[^1S_0^{[8]}\right]$ & $0.0146$ \\
$\eta_c\!\left[^1S_0^{[8]}\right]$ & $0.003009743333333$ & $J/\psi\!\left[^3S_1^{[8]}\right]$ & $0.00902923$ \\
$\eta_c(2S)\!\left[^1S_0^{[1]}\right]$ & $0.253333333333333$ & $\psi(2S)\!\left[^3S_1^{[1]}\right]$ & $0.76$ \\
$\eta_c(2S)\!\left[^3S_1^{[8]}\right]$ & $0.02$ & $\psi(2S)\!\left[^1S_0^{[8]}\right]$ & $0.02$ \\
$\eta_c(2S)\!\left[^1S_0^{[8]}\right]$ & $0.0004$ & $\psi(2S)\!\left[^3S_1^{[8]}\right]$ & $0.0012$ \\ \midrule
$\eta_b\!\left[^1S_0^{[1]}\right]$ & $3.093333333333333$ & $\Upsilon\!\left[^3S_1^{[1]}\right]$ & $9.28$ \\
$\eta_b\!\left[^3S_1^{[8]}\right]$ & $0.000170128$ & $\Upsilon\!\left[^1S_0^{[8]}\right]$ & $0.000170128$ \\
$\eta_b\!\left[^1S_0^{[8]}\right]$ & $0.0099142$ & $\Upsilon\!\left[^3S_1^{[8]}\right]$ & $0.0297426$ \\
$\eta_b(2S)\!\left[^1S_0^{[1]}\right]$ & $1.543333333333333$ & $\Upsilon(2S)\!\left[^3S_1^{[1]}\right]$ & $4.63$ \\
$\eta_b(2S)\!\left[^3S_1^{[8]}\right]$ & $0.0612263$ & $\Upsilon(2S)\!\left[^1S_0^{[8]}\right]$ & $0.0612263$ \\
$\eta_b(2S)\!\left[^1S_0^{[8]}\right]$ & $0.003197393333333$ & $\Upsilon(2S)\!\left[^3S_1^{[8]}\right]$ & $0.00959218$\\
$\eta_b(3S)\!\left[^1S_0^{[1]}\right]$ & $1.18$ & $\Upsilon(3S)\!\left[^3S_1^{[1]}\right]$ & $3.54$ \\
$\eta_b(3S)\!\left[^3S_1^{[8]}\right]$ & $0.0272909$ & $\Upsilon(3S)\!\left[^1S_0^{[8]}\right]$ & $0.0272909$ \\
$\eta_b(3S)\!\left[^1S_0^{[8]}\right]$ & $0.002567153333333$ & $\Upsilon(3S)\!\left[^3S_1^{[8]}\right]$ & $0.00770146$\\ \midrule
$B_c^\pm\!\left[^1S_0^{[1]}\right]$ & $0.736$ & $B_c^{\ast \pm}\!\left[^3S_1^{[1]}\right]$ & $2.208$ \\
$B_c^\pm\!\left[^3S_1^{[8]}\right]$ & $0.00736$ & $B_c^{\ast \pm}\!\left[^1S_0^{[8]}\right]$ & $0.00736$ \\
$B_c^\pm\!\left[^1S_0^{[8]}\right]$ & $0.00736$ & $B_c^{\ast \pm}\!\left[^3S_1^{[8]}\right]$ & $0.02208$ \\
$B_c^\pm(2S)\!\left[^1S_0^{[1]}\right]$ & $0.469348$ & $B_c^{\ast \pm}(2S)\!\left[^3S_1^{[1]}\right]$ & $1.40804$ \\
$B_c^\pm(2S)\!\left[^3S_1^{[8]}\right]$ & $0.00469348$ & $B_c^{\ast \pm}(2S)\!\left[^1S_0^{[8]}\right]$ & $0.00469348$ \\
$B_c^\pm(2S)\!\left[^1S_0^{[8]}\right]$ & $0.00469348$ & $B_c^{\ast \pm}(2S)\!\left[^3S_1^{[8]}\right]$ & $0.0140804$ \\
\bottomrule
\end{tabular}
\caption{Default LDME values in \mgshort. Colour-singlet values are computed according to Eq.~\eqref{eq:ldme_cs} using wave functions from ref.~\cite{Eichten:1995ch}, while colour-octet numbers for charmonium and bottomonium mesons are taken from refs.~\cite{Han:2014jya,Han:2014kxa}.}
\label{tab:ldme}
\end{table}

Following the concept of \textit{multiparticles} in \mgshort, such as the proton (\texttt{p}) and the jet (\texttt{j}), which represent collections of light (anti)quarks and the gluon ({e.g.,\@\xspace} \texttt{p = g u c d s u\mgtilde\ c\mgtilde\ d\mgtilde\ s\mgtilde} in the four-flavour number scheme), physical bound states are treated as collections of Fock states, referred to as \textit{boundstates} in \mgshort. Each bound state is defined along with all its Fock states in the file \texttt{boundstates\_default.txt}, located in the \texttt{input} directory. For instance, the $J/\psi$ meson (\texttt{Jpsi}) can be defined as \texttt{Jpsi = Jpsi(1|3S11) Jpsi(1|1S08) Jpsi(1|3S18)}, as illustrated in listing~\ref{lst:boundstates_default.txt}. An overview of all available \textit{boundstates} and their constituents can be displayed in the interface using the command

\vskip 0.25truecm
\noindent
~~\prompt\ {\tt ~display~boundstates}
\vskip 0.25truecm

\noindent
in analogy to the command \texttt{display~multiparticles}.

Since processes are generated at the level of Fock states, a physical bound state corresponds to the sum over all its related Fock-state subprocesses. For example, the command to generate $J/\psi$ production in association with a jet:

\vskip 0.23truecm
\noindent
~~\prompt\ {\tt ~import~model~sm\_onia-c\_mass}

\noindent
~~\prompt\ {\tt ~generate~p p > Jpsi j}

\noindent
~~\prompt\ {\tt ~output;~launch}
\vskip 0.23truecm
\noindent

\noindent
is equivalent to generating each of its Fock states separately:

\vskip 0.23truecm
\noindent
~~\prompt\ {\tt ~import~model~sm\_onia-c\_mass}

\noindent
~~\prompt\ {\tt ~generate~p p > Jpsi(1|3S11) j}

\noindent
~~\prompt\ {\tt ~add~process~p p > Jpsi(1|1S08) j}

\noindent
~~\prompt\ {\tt ~add~process~p p > Jpsi(1|3S18) j}

\noindent
~~\prompt\ {\tt ~output;~launch}
\vskip 0.23truecm
\noindent

\noindent
We emphasise that, aside from computational limitations, \mgshort\ at LO can handle processes with an arbitrary number of final-state S-wave quarkonia and leptonia, along with any associated elementary particles.\vspace{-0.3\baselineskip}

\subsection{Running modes}\vspace{-0.3\baselineskip}
Processes involving non-relativistic bound states can be generated in two different running modes within \mgshort, depending on the intended application. The current implementation supports both the \textit{leading-order mode} and the \textit{standalone mode}.

\paragraph{Leading-order mode:} The \textit{leading-order mode} is the default working mode, designed for parton-level event simulations at LO in perturbation theory. In this mode, \mgshort\ automatically handles the generation of Feynman diagrams, the construction of matrix elements, and of the phase-space integration. The results are total or fiducial cross sections and unweighted parton-level events, which can be exported in the LHEF format for further processing, such as showering and hadronisation with external tools. The generation syntax rules for obtaining an output in this mode are described in the last subsection, section~\ref{sec:interface}.

\paragraph{Standalone mode:} The \textit{standalone mode}, in contrast, generates only the code for evaluating the matrix elements of a given process, without embedding them into a full event-generation framework. This mode is especially useful when the amplitudes are to be integrated into external frameworks, or for testing and validation purposes. Only the \texttt{Fortran}-based output is supported in standalone mode for amplitudes with bound states so far. The \texttt{C++} version of the standalone matrix element mode, which is available for purely elementary-particle processes, is not yet compatible with bound-state configurations.
To obtain the matrix element code in standalone form, the usual command

\vskip 0.23truecm
\noindent
~~\prompt\ {\tt ~output;~launch}
\vskip 0.23truecm

\noindent
should be replaced with:

\vskip 0.23truecm
\noindent
~~\prompt\ {\tt ~output~standalone;~launch}
\vskip 0.23truecm

\noindent
This will generate a lightweight, portable \texttt{Fortran} code that evaluates helicity amplitudes for fixed kinematic configurations but does not produce events or cross sections by itself.~\footnote{Specifically, the \textit{standalone mode} returns the spin- and colour-summed squared amplitudes, including initial-state averages and final-state symmetry factors. Moreover, the squared amplitude is already multiplied by the LDME and the related normalisation factors. More details can be found in section~\ref{sec:benchmark}.}

\subsection{Technical details of the implementation}
To provide further insight into the practical implementation within the \mgshort\ framework, we outline here a number of technical aspects relevant to the treatment of bound-state processes.

\paragraph{Colour projection:} The application of the colour projector, as defined in Eq.~\eqref{eq:proj_c}, is handled at the \texttt{Python} level within \mgshort\ during diagram generation. The projector is applied when constructing the Feynman rules, enforcing the required colour structure for the specified Fock state. The corresponding normalisation factor, however, is evaluated within the generated \texttt{Fortran} code to ensure consistent treatment alongside the matrix-element evaluation.

\paragraph{Spin projection:} Spin projectors, in accordance with Eq.~\eqref{eq:proj_s}, are implemented via a dedicated subroutine generated by the \texttt{ALOHA}~\cite{deAquino:2011ub} module. This subroutine relies on the \texttt{HELAS}~\cite{Hagiwara:1990dw,Murayama:1992gi} library, which enables the numerical evaluation of helicity amplitudes. The spin projector is applied on-the-fly for each phase-space point and contributes directly to the construction of the matrix elements in the \texttt{Fortran} code.

\paragraph{Long-distance matrix elements:} In the implementation, the LDMEs are directly incorporated into the calculation of the squared amplitudes rather than being treated as external factors. Technically, the \texttt{Fortran} subroutine \texttt{SMATRIX} returns, up to a global factor, the product $\left|{\mathds A}^{(n-1,0)}(\dot{r})\right|^2 \langle {\cal O}^\mathcal{B}_{n}\rangle$. This design choice implies that the partonic cross section $\mathrm{d}\hat{\sigma}$ is \textit{not} first computed and subsequently multiplied by the global LDME factor, as suggested by Eq.~\eqref{eq:xsec}.

\paragraph{Phase-space integration:} Phase-space integration is currently performed using a basic sampling strategy, in contrast to the single-diagram enhanced (SDE)~\cite{Maltoni:2002qb} multi-channel approach~\cite{Kleiss:1994qy} and the helicity-recycling method~\cite{Mattelaer:2021xdr} employed for processes involving only elementary particles. We rely on a pseudo multi-channel strategy in \mgshort, where the \texttt{VEGAS}~\cite{Lepage:1977sw,lepage1980vegas} integrator is parallelised over multiple CPU threads to improve sampling performance. In this setup, the channels are still divided according to the SDE prescription, but the phase space is always parametrised using $s$-channel variables~\cite{Byckling:1971vca}, without further optimisation for the dominant topologies in each channel. While this approach may limit numerical efficiency for arbitrarily complex processes, it is sufficient for most phenomenologically relevant cases. The multi-channel feature is currently disabled because of additional constraints on the phase space at the elementary-particle level: the momenta of the constituents forming a bound state are not independent. Consequently, the number of available phase-space degrees of freedom is smaller than the number of internal propagators, making it non-trivial to identify an optimal parametrisation for each channel and to perform the sampling according to the propagator structure in processes involving bound states.

%% file: 04_setup.tex
\section{Computational setup}\label{sec:setup}

Before validating the implementation and presenting the first results obtained with the extended \mgshort\ framework, we begin by outlining the baseline setup used throughout the illustrative calculations discussed in the following sections. To avoid repetition, we define here a global set of default input parameters. Any deviations from these defaults, such as the application of fiducial cuts to regulate infrared singularities when relevant, are specified on a case-by-case basis whenever they occur.

Depending on the final state under consideration, we employ either a four-flavour number scheme (4FS) or a three-flavour number scheme (3FS), corresponding to the inclusion of four or three massless quark flavours, respectively. The 4FS is used exclusively for bottomonium production, whereas the 3FS is applied to processes involving charmonia, charmed $B_c$ mesons, and leptonia. Consequently, for processes involving mesons that contain only massive bottom quarks, we use the SM \ufo\ model extended to handle bound states, which can be imported with the command

\vskip 0.25truecm
\noindent
~~\prompt\ {\tt ~import~model~sm\_onia}

\vskip 0.25truecm
\noindent
which automatically adopts the 4FS. If, instead, at least one charm quark is a constituent of a quarkonium, the 3FS version of the model must be used. This can be achieved by appending the \texttt{c\_mass} restriction:

\vskip 0.25truecm
\noindent
~~\prompt\ {\tt ~import~model~sm\_onia-c\_mass}

\vskip 0.25truecm
\noindent
This setup assigns a non-zero mass to the charm quark. Moreover, in this model, the charm quark is automatically removed from the definitions of the \textit{multiparticles} \texttt{p} and \texttt{j}, which specify the partonic content of the proton and a jet, respectively. For leptonium production, we employ the SM in the 3FS with leptons treated as massive. This configuration can be loaded via

\vskip 0.25truecm
\noindent
~~\prompt\ {\tt ~import~model~sm\_onia-lepton\_masses}

\noindent
~~\prompt\ {\tt ~define~p~=~g u d s u\mgtilde\ d\mgtilde\ s\mgtilde}

\noindent
~~\prompt\ {\tt ~define~j~=~g u d s u\mgtilde\ d\mgtilde\ s\mgtilde}

\vskip 0.25truecm
\noindent
where it is mandatory to redefine the \textit{multiparticles} explicitly in order to emulate a 3FS.~\footnote{In the \texttt{lepton\_masses} restriction card, the charm quark is taken to be massless. In our leptonium-production computations, only processes that do not receive contributions from diagrams containing charm quarks are considered. Therefore, the described adjustment ensures a consistent 3FS configuration.}

The overall parameter choices follow the default settings of the \ufo\ model \texttt{sm\_onia} and its optional restrictions, including the default LDMEs listed in table~\ref{tab:ldme}. These settings are consistent with those of the default \texttt{sm} model in \mgshort. The numerical values for the masses of the charm quark ($m_c$), bottom quark ($m_b$), top quark ($m_t$), electron ($m_e$), muon ($m_\mu$), tau lepton ($m_\tau$), $Z$ boson ($m_Z$), and Higgs boson ($m_H$) are summarised in table~\ref{tab:setup}. As mentioned previously, $m_c$ is assigned a non-zero value only when using the 3FS. In the electroweak sector, we take as free parameters the $Z$-boson mass $m_Z$, the fine-structure constant $\alpha_{G_\mu}$ in the $G_\mu$ scheme, and the Fermi constant $G_F$ extracted from muon decay. These values are also listed in table~\ref{tab:setup}. According to this choice of input parameters, the $W$-boson mass is not treated as an independent quantity and is determined from the given inputs to be\vspace{-0.1\baselineskip}
\begin{equation}
    m_W = 80.419\,\mathrm{GeV}\,.
\end{equation}
The CKM matrix is taken to be the identity. In the default model settings, all particle widths are set to zero except those for the top quark and the $W$, $Z$, and Higgs bosons, whose widths are 
\begin{equation}
\begin{aligned}
\Gamma_t&=1.4915\,\mathrm{GeV}\,,\quad &\Gamma_W&=2.441404\,\mathrm{GeV}\,,\\ \Gamma_Z&=2.0476\,\mathrm{GeV}\,,\quad\text{and}\quad &\Gamma_H&=6.3823393\,\mathrm{MeV}\,,
\end{aligned}
\end{equation}
respectively.

\begin{table}[t!]
\centering
\renewcommand*{\arraystretch}{1.4}
\begin{tabular}[t]{cc|cc}
\toprule
\textbf{parameter}& value & \textbf{parameter}& value \\
\midrule
$m_c$ & $1.55\,\mathrm{GeV}$ \textsuperscript{\ref{fnt:c_mass_restriction}} & $m_Z$ & $91.188\,\mathrm{GeV}$ \\
$m_b$ & $4.7\,\mathrm{GeV}$ & $m_H$ & $125\,\mathrm{GeV}$ \\
$m_t$  & $173\,\mathrm{GeV}$ & $\alpha_{G_\mu}^{-1}$ & $132.507$ \\
$m_e$ & $511\,\mathrm{keV}$ \textsuperscript{\ref{fnt:lepton_masses_restriction}} & $G_F$ & $1.166390\cdot10^{-5}\,\mathrm{GeV}^{-2}$ \\
$m_\mu$ & $105.66\,\mathrm{MeV}$ \textsuperscript{\ref{fnt:lepton_masses_restriction}} & & \\
$m_\tau$ & $1.777\,\mathrm{GeV}$ \textsuperscript{\ref{fnt:lepton_masses_restriction}} & & \\
\bottomrule
\end{tabular}
\caption{Summary of the global SM parameter settings employed in all analyses presented in this article. The charm-quark mass, $m_c$, takes a non-zero value only in the 3FS when the \texttt{c\_mass} restriction is enabled, whereas the $u$, $d$, and $s$ quarks are kept massless. Similarly, non-zero lepton masses are included only upon loading the \texttt{lepton\_masses} restriction.}
\label{tab:setup}
\end{table}
\addtocounter{footnote}{1}
\footnotetext{Takes a non-zero value only under the \texttt{c\_mass} restriction.\label{fnt:c_mass_restriction}}
\addtocounter{footnote}{1}
\footnotetext{Takes a non-zero value only under the \texttt{lepton\_masses} restriction.\label{fnt:lepton_masses_restriction}}

For all processes, the central renormalisation and factorisation scales are set equal to
\begin{equation}
    \mu_R=\mu_F=H_T/2=\frac{1}{2}\sum_{i}{\sqrt{k_{i,T}^2+m_i^2}}\,,\label{eq:HTovertwo}
\end{equation}
with the sum running over all final-state particles, where $k_{i,T}$ denotes the transverse momentum of the $i$-th particle.~\footnote{Advanced users may specify alternative dynamical scale choices by providing dedicated \texttt{Fortran} functions in an external file, which can be linked to \mgshort\ through the \texttt{run\_card.dat}. Further details are available in the online documentation at \href{https://answers.launchpad.net/mg5amcnlo/+faq/3325}{https://answers.launchpad.net/mg5amcnlo/+faq/3325}.}

In sections~\ref{sec:quarkonium} and \ref{sec:leptonium}, we investigate the production of quarkonia and leptonia, respectively, at both electron-positron ($e^+e^-$) and proton-proton ($pp$) colliders. For the former, we consider asymmetric beams with an electron beam energy of $E^{e^-}_{\mathrm{beam}}=7\,\mathrm{GeV}$ and a positron beam energy of $E^{e^+}_{\mathrm{beam}}=4\,\mathrm{GeV}$, resulting in a total centre-of-mass (c.m.) energy of $\sqrt{s}=10.58\,\mathrm{GeV}$, corresponding to the setup used at SuperKEKB~\cite{Ohnishi:2013fma} operating at the $\Upsilon(4S)$ resonance. Hadronic collisions are studied at the LHC~\cite{Evans:2008zzb} Run 2 energy, $\sqrt{s}=13\,\mathrm{TeV}$, using the \texttt{PDF4LHC21\_40}~\cite{PDF4LHCWorkingGroup:2022cjn} NNLO PDF set in all calculations.

%% file: 05_benchmarks.tex
\section{Benchmark processes and validation}\label{sec:benchmark}

In this section, we benchmark our LO implementation by comparing its numerical results for various observables across multiple processes in $e^+ e^-$ and $pp$ collisions with the corresponding results from \helaconia~\cite{Shao:2012iz,Shao:2015vga}. The comparisons include processes with single and multiple quarkonium final states, as well as quarkonium production in association with SM elementary particles. Our aim is to validate our calculations and assess their consistency with other established tools. This benchmarking provides an important cross check, ensuring the reliability of our results for the applications discussed in the following two sections.

In the first instance, we performed a cross check of our implementation using \mgshort\ in its \textit{standalone mode}. Such a comparison is carried out at the level of the square of the helicity amplitude averaged (summed) over the initial-state (final-state) particle colours/helicities for a randomly generated physical phase-space point by \texttt{Rambo}~\cite{Kleiss1986RAMBO}. In summary, this mode returns the quantity
\begin{equation}\label{eq:me_square}
    \left|\mathcal{M}\right|^2=\frac{1}{\mathcal{N}(\dot{r})}\frac{1}{(2J+1)N_{[C]}}\frac{m_{C_1}+m_{C_2}}{2m_{C_1}m_{C_2}}\!\left(\frac{1}{\omega(\mathcal{I}_1)\omega(\mathcal{I}_2)} \right)\! \sum_{\substack{\rm colour\\ \rm spin}}\left|{\mathds A}^{(n-1,0)}(\dot{r})\right|^2\langle {\cal O}^\mathcal{B}_{n}\rangle\,,
\end{equation}
where $C_1$ and $C_2$ denote the constituents of the bound state $\mathcal{B}$, which may be either a quarkonium or a leptonium. The amplitude ${\mathds A}^{(n-1,0)}(\dot{r})$ is computed according to Eq.~\eqref{eq:amp_quarkonium}, depending on the nature of the bound state.

In table~\ref{tab:me}, as a showcase, we present selected benchmark results for various partonic processes involving different quarkonium Fock states. These processes span a range of final-state multiplicities and include associations with SM elementary particles from across all sectors. In this way, we maximise the scope of our benchmarking, making it as inclusive and exhaustive as possible. For all comparisons, we adopt the input parameters defined in section~\ref{sec:setup} and fix the partonic c.m. energy to $\sqrt{\hat{s}} = 1\,\mathrm{TeV}$, even for processes with $e^+e^-$ initial states. To quantify the agreement between the two tools, we define the relative deviation as
\begin{equation}
    \Delta_\mathrm{rel.}=\left|\dfrac{\left|\mathcal{M}\right|^2_{\mgshort}-\left|\mathcal{M}\right|^2_\helaconia}{\left|\mathcal{M}\right|^2_\helaconia}\right|\,,
    \label{relrat}
\end{equation}
where $\left|\mathcal{M}\right|_\mgshort^2$ and $\left|\mathcal{M}\right|_\helaconia^2$ denote the squared amplitudes, as defined in Eq.~\eqref{eq:me_square}, evaluated with \mgshort\ and \helaconia, respectively. As expected, we find excellent agreement between \mgshort\ and \helaconia, with differences only at the level of floating-point precision. This outcome provides strong validation of our projector implementation and ensures the fidelity of our LO matrix-element calculations. The complete list of 42 processes that we have cross-checked with these two tools at the matrix-element level can be found in the supplementary material.

\begin{center}
\begin{minipage}{\textwidth}
\begin{table}[H]
\centering
\renewcommand*{\arraystretch}{1.2}
\begin{tabular}[h!]{lcc}
\toprule
\multirow{2}{*}{\textbf{process}}& \madgraph & \multirow{2}{*}{$\Delta_\mathrm{rel.}$}  \\
 & \helaconia  \\
\midrule
\multirow{2}{*}{$gg \rightarrow J/\psi\! \left[^3S_1^{[8]}\right]+g$} &   $4.132917971335\cdot 10^{-3}\phantom{\,\mathrm{GeV}^{-2}}$ & \multirow{2}{*}{$7.8 \cdot 10^{-11}$}  \\
& $4.132917971659\cdot 10^{-3}\phantom{\,\mathrm{GeV}^{-2}}$  \\ \hdashline
\multirow{2}{*}{$u \bar{u} \rightarrow \eta_c\!\left[^1S_0^{[1]}\right] + c \bar{c}$} & $1.381481341714290\cdot 10^{-12}$  GeV$^{-2}$ & \multirow{2}{*}{$1.3 \cdot 10^{-15}$} \\
& $1.381481341714289\cdot 10^{-12}$   GeV$^{-2}$ \\ \hdashline
\multirow{2}{*}{$g g \rightarrow B_c^{+}\left[^1S_0^{[1]}\right] + b \bar{c}$ } & $3.781983565900759 \cdot 10^{-12}$ GeV$^{-2}$ & \multirow{2}{*}{$1.1 \cdot 10^{-15}$} \\
& $ 3.781983565900755\cdot 10^{-12}$  GeV$^{-2}$ \\ \hdashline
\multirow{2}{*}{$u \bar{u} \rightarrow B_c^+\!\left[^3S_1^{[8]}\right] + b  \bar{c}$} & $6.87392368906369\cdot 10^{-14}$ GeV$^{-2}$ & \multirow{2}{*}{$3.1 \cdot 10^{-13}$} \\
& $6.87392368906585\cdot 10^{-14}$  GeV$^{-2}$  \\ \hdashline
\multirow{2}{*}{$gg \rightarrow \eta_c\!\left[^1S_0^{[1]}\right] + g g g$ } & $5.3063891680528 \cdot 10^{-11}$ GeV$^{-4}$ & \multirow{2}{*}{$1.1 \cdot 10^{-14}$} \\
& $5.3063891680527 \cdot 10^{-11}$ GeV$^{-4}$ \\ \hdashline
\multirow{2}{*}{$g g \rightarrow \Upsilon\!\left[^3S_1^{[1]}\right] + \Upsilon\!\left[^3S_1^{[8]}\right] + H$ } & $4.753109487454\cdot 10^{-21}$ GeV$^{-2}$ & \multirow{2}{*}{$1.5 \cdot 10^{-7}$} \\
& $4.753108764824\cdot 10^{-21}$  GeV$^{-2}$ \\ \hdashline
\multirow{2}{*}{$gg \rightarrow J/\psi\!\left[^3S_1^{[8]}\right]+J/\psi\! \left[^3S_1^{[8]}\right]+g$} &  $7.885077765242 \cdot 10^{-11}$ GeV$^{-2}$ & \multirow{2}{*}{$1.8 \cdot 10^{-10}$}  \\
&$7.885077766678 \cdot 10^{-11}$  GeV$^{-2}$ \\ \hdashline
\multirow{2}{*}{$gg \rightarrow J/\psi\! \left[^3S_1^{[1]}\right] + \Upsilon\! \left[^3S_1^{[1]}\right] + g$ \footnotemark} &$1.156767080922 \cdot 10^{-15}$ GeV$^{-2}$ & \multirow{2}{*}{$2.9 \cdot 10^{-7}$} \\
&$1.156766744184 \cdot 10^{-15}$ GeV$^{-2}$  \\ \hdashline
\multirow{2}{*}{$gg \rightarrow J/\psi\! \left[^3S_1^{[1]}\right] + \Upsilon\! \left[^3S_1^{[1]}\right] + J/\psi\! \left[^3S_1^{[8]}\right]\!\!\!\!\!\!$} & $2.510898856056 \cdot 10^{-27}$ GeV$^{-2}$ & \multirow{2}{*}{$3.9 \cdot 10^{-9}$} \\
& $2.510898865795 \cdot 10^{-27}$ GeV$^{-2}$  \\ \hdashline
\multirow{2}{*}{$g g\rightarrow J/\psi\! \left[^1S_0^{[8]}\right] + \gamma Z$} &  $1.419801780011\cdot 10^{-16}$ GeV$^{-2}$ & \multirow{2}{*}{$4.9 \cdot 10^{-7}$}  \\
& $1.419801072960 \cdot 10^{-16}$  GeV$^{-2}$ \\ \hdashline
\multirow{2}{*}{$e^+ e^- \rightarrow J/\psi\left[^1S_0^{[8]}\right] + c \bar{c}$ } & $1.350294584265\cdot 10^{-15}$ GeV$^{-2}$ & \multirow{2}{*}{$1.5 \cdot 10^{-7}$} \\
& $1.350294377299 \cdot 10^{-15}$  GeV$^{-2}$ \\ \hdashline
\multirow{2}{*}{$e^+ e^- \rightarrow J/\psi\left[^3S_1^{[1]}\right] + g g$ } & $1.843805235534  \cdot 10^{-14}$ GeV$^{-2}$ & \multirow{2}{*}{$1.4 \cdot 10^{-7}$} \\
& $ 1.843804974567\cdot 10^{-14}$  GeV$^{-2}$ \\ \hdashline
\multirow{2}{*}{$\gamma g \rightarrow \eta_c\!\left[^1S_0^{[1]}\right] + g g g$ } & $2.168626920441142  \cdot 10^{-14}$ GeV$^{-4}$ & \multirow{2}{*}{$1.6 \cdot 10^{-14}$} \\
& $2.168626920441177 \cdot 10^{-14}$  GeV$^{-4}$ \\ \bottomrule
\end{tabular}
\caption{Benchmark squared amplitudes $ \left|\mathcal{M}\right|^2$ (as defined in Eq.~\eqref{eq:me_square}) in the \textit{standalone mode} for selected partonic processes relevant for various collider types, evaluated at random physical phase-space points. We compare \mgshort\ (top) calculations against \helaconia\ (bottom). The last column of the table shows the relative deviation, as defined in Eq.~\eqref{relrat}.}
\label{tab:me}
\end{table}
\vspace{-1.5\baselineskip}
\end{minipage}\end{center}\footnotetext{To ensure that the leading contribution, $\mathcal{O}(\alpha_s^3 \alpha^2)$, is included, this process was generated using the command \texttt{generate g g > Jpsi(1|3S11) Upsilon(1|3S11) g QCD=99 QED=99}, where the options \texttt{QCD=99 QED=99} enable all relevant diagrams to be considered.\label{fnt:standalone}}

\begin{table}[hbt!]
\centering
\renewcommand*{\arraystretch}{1.2}
\begin{tabular}[t!]{lcc}
\toprule
\multirow{2}{*}{\textbf{process}}& \madgraph & \multirow{2}{*}{\hspace{10mm}\textbf{pull}\hspace{10mm}} \\
 & \helaconia & \\
\midrule
\multirow{2}{*}{$pp\to u\bar{u} \to J/\psi\!\left[^3S_1^{[8]}\right]$} & $78.757(3)\,\mathrm{nb}$ & \multirow{2}{*}{$0.89$} \\
& $78.754(1)\,\mathrm{nb}$ & \\ \hdashline
\multirow{2}{*}{$pp\to u\bar{u} \to J/\psi\!\left[^3S_1^{[8]}\right]+H$} & $42.055(2)\,\mathrm{yb}$ & \multirow{2}{*}{$0.28$} \\
& $42.056(3)\,\mathrm{yb}$ & \\ \hdashline
\multirow{2}{*}{$pp\to gg \to J/\psi\!\left[^3S_1^{[8]}\right]+H$} & $1.8530(7)\,\mathrm{ab}$ & \multirow{2}{*}{$0.71$} \\
& $1.8523(7)\,\mathrm{ab}$ & \\ \hdashline
\multirow{2}{*}{$pp\to gg \to J/\psi\!\left[^3S_1^{[8]}\right]+HH$} & $15.927(3)\,\mathrm{yb}$ & \multirow{2}{*}{$0.15$} \\
& $15.93(2)\,\mathrm{yb}$ & \\ \hdashline
\multirow{2}{*}{$pp\to gg \to J/\psi\!\left[^3S_1^{[8]}\right]+HHH$} & $1.9802(5)\,\mathrm{rb}$ & \multirow{2}{*}{$3.27$} \\
& $1.967(4)\,\mathrm{rb}$ & \\ \hdashline
\multirow{2}{*}{$pp\to gg \to J/\psi\!\left[^3S_1^{[8]}\right]+g$} & $8.9215(7)\,$µ$\mathrm{b}$ & \multirow{2}{*}{$2.60$} \\
& $8.927(2)\,$µ$\mathrm{b}$ & \\ \hdashline
\multirow{2}{*}{$pp\to gg \to J/\psi\!\left[^3S_1^{[1]}\right]+J/\psi\!\left[^3S_1^{[1]}\right]$} & $8.921(2)\,\mathrm{nb}$ & \multirow{2}{*}{$1.12$} \\
& $8.916(4)\,\mathrm{nb}$ & \\ \hdashline
\multirow{2}{*}{$pp\to gg \to J/\psi\!\left[^3S_1^{[8]}\right]+J/\psi\!\left[^3S_1^{[8]}\right]$} & $86.240(7)\,\mathrm{pb}$ & \multirow{2}{*}{$1.42$} \\
& $86.27(2)\,\mathrm{pb}$ & \\ \hdashline
\multirow{2}{*}{$pp\to gg \to \eta_c\!\left[^1S_0^{[8]}\right]+\eta_b\!\left[^1S_0^{[8]}\right]$} & $195.984(9)\,\mathrm{fb}$ & \multirow{2}{*}{$0.24$} \\
&$195.987(9)\,\mathrm{fb}$ & \\ \hdashline
\multirow{2}{*}{$pp\to u\bar{u} \to \eta_c\!\left[^1S_0^{[1]}\right]+J/\psi\!\left[^3S_1^{[8]}\right]$} & $152.79(1)\,\mathrm{fb}$ & \multirow{2}{*}{$0.99$} \\
&$152.73(6)\,\mathrm{fb}$ & \\ \hdashline
\multirow{2}{*}{$pp\to u\bar{u} \to \eta_c\!\left[^1S_0^{[1]}\right]+\Upsilon\!\left[^3S_1^{[1]}\right]$} & $212.90(2)\,\mathrm{zb}$ & \multirow{2}{*}{$0.00$} \\
& $212.9(1)\,\mathrm{zb}$ & \\ \hdashline
\multirow{2}{*}{$pp\to u\bar{u} \to B_c^{+}\!\left[^1S_0^{[1]}\right]+B_c^{\ast -}\!\left[^3S_1^{[1]}\right]$} & $2.7920(5)\,\mathrm{pb}$ & \multirow{2}{*}{$0.58$} \\
& $2.7925(7)\,\mathrm{pb}$ & \\ \hdashline
\multirow{2}{*}{$e^+e^- \to J/\psi\!\left[^3S_1^{[1]}\right]+Z$} & $1.61586(9)\,\mathrm{fb}$ & \multirow{2}{*}{$0.17$} \\
& $1.61584(8)\,\mathrm{fb}$ & \\
\bottomrule
\end{tabular}
\caption{Benchmark cross sections for selected $pp$ ($\sqrt{s}=13\,\mathrm{TeV}$) and $e^+e^-$ ($\sqrt{s}=1\,\mathrm{TeV}$) collision processes. We compare \mgshort\ (top) calculations against \helaconia\ (bottom), using the setup introduced in section~\ref{sec:setup}. Numerical MC integration uncertainties on the last printed digit are given in brackets. The last column of the table shows the pull as defined in Eq.~\eqref{eq:pull}.}
\label{tab:xsec}
\end{table}

To test the \textit{leading-order mode}, we compute total cross sections with \mgshort\ and compare the results against those obtained with \helaconia. A summary of this comparison is shown in table~\ref{tab:xsec}. The selected processes provide a technically robust validation of our implementation, rather than addressing specific phenomenological questions. For the $pp$ collision processes, we employ the setup described in section~\ref{sec:setup}. We compute the cross section for single-quarkonium production increasing phase-space complexity by including additional SM particles in the final state. This includes processes ranging from simple $2\to 1$ up to more complex $2\to 4$ scatterings. In addition, we evaluate quarkonium pair-production cross sections for various combinations of charmonia and bottomonia in different Fock state configurations. To quantify the agreement between the two tools, we compute the pull, defined as
\begin{equation}\label{eq:pull}
    \textbf{pull}=\left|\dfrac{\sigma_\mgshort-\sigma_\helaconia}{\sqrt{\Delta_{\mgshort}^2+\Delta_{\helaconia}^2}}\right|\,,
\end{equation}
where $\sigma_{\mgshort}$ and $\sigma_{\helaconia}$ denote the cross sections evaluated with \mgshort\ and \helaconia, respectively, and $\Delta_{\mgshort}$, $\Delta_{\helaconia}$ are the corresponding MC uncertainties. Overall, we observe excellent agreement between \mgshort\ and \helaconia, with pull values consistent with statistical expectations. This serves as a validation of our LO implementation.

In addition to the $pp$ collision processes, we also perform a cross-check for an $e^+e^-$ collision process, $e^+e^- \to J/\psi\!\left[{}^3S_1^{[1]}\right]+Z$. Here, we adopt the general setup from section~\ref{sec:setup}, but increase the collision energy to $\sqrt{s} = 1\,\mathrm{TeV}$ with symmetric beam energies, $E_\mathrm{beam}^{e^+} = E_\mathrm{beam}^{e^-} = 500\,\mathrm{GeV}$, to allow for on-shell production of the $Z$ boson. As in the hadronic case, we find excellent agreement between the results from \mgshort\ and \helaconia\ within the quoted MC uncertainties.

Finally, we emphasise that the benchmarks presented in tables~\ref{tab:me} and \ref{tab:xsec} represent only a subset of the more extensive tests we have performed across different processes, for both fully-inclusive and fiducial cross sections. The additional results are not shown, as they do not provide further insight beyond what is already illustrated. Nevertheless, these tests, like the examples reported, serve to validate our code and consistently confirm the correctness of our implementation.

%% file: 06_quarkonium.tex
\section{Quarkonium production}\label{sec:quarkonium}

To highlight the broad applicability of our implementation across different collision systems, we present cross sections for
quarkonium production in proton-proton ($pp$) collisions  (section~\ref{sec:pp_collider}), electron-proton~($e^-p$) collisions  (section~\ref{sec:ep_collider}), and electron-positron~($e^+ e^-$) collisions  (section~\ref{sec:lepton_collider}). 

In $pp$ collisions at the LHC with a c.m. energy of $\sqrt{s}=13$ TeV, we discuss single-quarkonium production as a simple final state, and its production in association with elementary particles. We extend our analysis to double- and triple-quarkonium production to illustrate the capabilities of our code. These studies are complemented by investigations of deep-inelastic scattering (DIS) and photoproduction studies in $e^- p$ collisions at HERA and the upcoming EIC, as well as quarkonium production in association with a (di-)jet and a (di-)photon in $e^+ e^-$ collisions at $B$-factories, such as Belle and BaBar, and at LEP experiments. Finally, we perform dedicated studies within Higgs Effective Field Theory~(HEFT)
and include parton showering with \pythia, demonstrating how our new implementation seamlessly integrates with existing \mgshort\ features.

We emphasise that the broad range of processes considered in what follows is intended to demonstrate the versatility of our implementation; the numerical values presented should not be directly compared with experimental measurements, where available. As previously stated, our current implementation is restricted to S-wave production at LO and does not, for instance, account for feeddown effects where they occur. Nevertheless, we provide a sufficiently detailed exposition of the results to allow the reader to reproduce all numerical values shown using \mgshort\ with the same parameter inputs. In addition, we make our analyses scripts available as auxiliary files in the supplemental material accompanying the arXiv submission.

\subsection{Quarkonium production in proton-proton collisions}
\label{sec:pp_collider}

\subsubsection{Inclusive single charmonium and bottomonium production}

As a first application of the new NRQCD feature of \mgshort, we compute the cross sections for $2\to 1$ scattering processes involving a single charmonium or bottomonium state in $pp$ collisions. Although these cross sections have been known for decades -- and even their NLO QCD corrections were computed analytically~\footnote{See refs.~\cite{Brodsky:2009cf,Feng:2015cba,Lansberg:2020ejc} for recent NLO phenomenological studies of such cross sections.} almost 30 years ago~\cite{Petrelli:1997ge} -- the user-friendly interface of \mgshort\ now makes such computations readily accessible to everyone,~\footnote{\mgshort\ can also be used at \url{http://nloaccess.in2p3.fr}.} providing a valuable resource for the high-energy physics community on both theoretical and experimental fronts. 

Extensive experimental efforts have been devoted to the study of inclusive heavy quarkonium production -- charmonium, bottomonium, and $B_c$ -- in high-energy collisions at the LHC and other hadron-hadron colliders such as RHIC and the Tevatron. Interested readers may consult the reviews in refs.~\cite{Kramer:2001hh,QuarkoniumWorkingGroup:2004kpm,Lansberg:2006dh,Brambilla:2010cs,ConesadelValle:2011fw} which address Tevatron (and HERA) results, as well as more recent ones~\cite{Andronic:2015wma,Lansberg:2019adr,Tang:2020ame} discussing advances driven by RHIC and LHC data.

\begin{table}[t]
\centering
\renewcommand*{\arraystretch}{1.4}
\begin{tabular}[t]{lc|lc}
\toprule
\textbf{process}& $\sigma$ & \textbf{process}& $\sigma$  \\
\midrule
$pp \to \eta_c+X$ & $ 2.9366(5)\,$µ$\mathrm{b}$ & $pp \to \eta_b+X$ & $5.4935(7)\,$µ$\mathrm{b}$  \\
$pp \to J/\psi+X$ & $536.14(6)\,\mathrm{nb}$ & $pp \to \Upsilon+X$ & $6.0655(4)\,\mathrm{nb}$  \\
\bottomrule
\end{tabular}
\caption{Total LO ($\mathcal{O}(\alpha_s^2)$) cross sections for single charmonium and bottomonium production in $pp$ collisions at $\sqrt{s}=13\,\mathrm{TeV}$, based on the setup described in section~\ref{sec:setup}. The numbers in parentheses indicate the numerical uncertainties from MC phase-space integration. These results are provided for illustration only: they do not include feeddown effects or experimental cuts, and are specific to the chosen LDME values (cf.\@\xspace table \ref{tab:ldme}).}
\label{tab:pp2Q}
\end{table}

We consider four different processes to illustrate the usage of the code: charmonium production as a spin-singlet $\eta_c$ or spin-triplet $J/\psi$, and bottomonium production as a spin-singlet $\eta_b$ or spin-triplet $\Upsilon$. We compute the fully-inclusive LO cross sections for these states using the setup described in section~\ref{sec:setup}, with results summarised in table~\ref{tab:pp2Q}. It is important to note that we always include all relevant colour-singlet and colour-octet S-wave contributions unless stated otherwise. For example, $\eta_c$ production includes the colour-singlet state ${}^1S_0^{[1]}$ as well as the two colour-octet states ${}^3S_1^{[8]}$ and ${}^1S_0^{[8]}$. Hence, we have
\begin{equation}\label{eq:etac}
\begin{aligned}
    \sigma(pp\to\eta_c+X)&=\sigma(pp\to\eta_c\!\left[{}^1S_0^{[1]}\right]+X)+\sigma(pp\to\eta_c\!\left[{}^3S_1^{[8]}\right]+X)\\
    &\quad+\sigma(pp\to\eta_c\!\left[{}^1S_0^{[8]}\right]+X)
\end{aligned}
\end{equation}
and, analogously, for $J/\psi$ production,
\begin{equation}
\begin{aligned}
	\sigma(pp\to J/\psi+X)&=\sigma(pp\to J/\psi\!\left[{}^3S_1^{[1]}\right]+X)+\sigma(pp\to J/\psi\!\left[{}^1S_0^{[8]}\right]+X)\\
    &\quad+\sigma(pp\to J/\psi\!\left[{}^3S_1^{[8]}\right]+X)\,.
\end{aligned}
\end{equation}

As a side note, due to the modular architecture of \mgshort, individual Fock-state contributions, along with their decomposition into initial-state partonic channels, are listed separately in the generated \texttt{HTML} output. For instance, even when using the \textit{boundstates} syntax:
\vskip 0.25truecm
\noindent
~~\prompt\ {\tt ~import~model~sm\_onia-c\_mass}

\noindent
~~\prompt\ {\tt ~generate~p p > etac}

\noindent
~~\prompt\ {\tt ~output;~launch}
\vskip 0.25truecm

\noindent
the \texttt{HTML} output lists the six subprocesses~\footnote{The notation with $q\bar{q}$ initial state implicitly covers both possibilities: a quark from beam~1 and an antiquark from beam~2, as well as the reversed configuration with the antiquark from beam~1 and the quark from beam~2.}
\begin{equation}
    \begin{aligned}
    &pp\to gg \to \eta_c\!\left[{}^1S_0^{[1]}\right]\,,\qquad &pp\to gg &\to \eta_c\!\left[{}^3S_1^{[8]}\right]\,,\qquad
    &pp\to gg \to \eta_c\!\left[{}^1S_0^{[8]}\right]\,,\\
    &pp\to q\bar{q} \to \eta_c\!\left[{}^1S_0^{[1]}\right]\,,\qquad &pp\to q\bar{q} &\to \eta_c\!\left[{}^3S_1^{[8]}\right]\,,\quad\text{and}\quad
    &pp\to q\bar{q} \to \eta_c\!\left[{}^1S_0^{[8]}\right]\,.
    \end{aligned}
\end{equation}
For $\eta_c$ production, the \texttt{HTML} output shows the following six cross sections~\footnote{According to the \texttt{HTML} output file, the cross section for the process $gg \to \eta_c\!\left[{}^3S_1^{[8]}\right]$ (cf.\@\xspace Eq.~\eqref{eq:gg_etac3s18}) is $4.0(1)\cdot 10^{-27}\,\mathrm{pb}$. This extremely small value can be interpreted as numerically zero, consistent with the Landau-Yang theorem~\cite{Landau:1948kw,Yang:1950rg,Cacciari:2015ela}, which forbids this process at LO.}:
{\allowdisplaybreaks
\begin{align}
        \sigma(pp\to gg \to \eta_c\!\left[{}^1S_0^{[1]}\right])&= 2.2995(5)\,\text{µ}\mathrm{b}\,,\\
        \sigma(pp\to gg \to \eta_c\!\left[{}^3S_1^{[8]}\right])&=0\,,\label{eq:gg_etac3s18}\\
        \sigma(pp\to gg \to \eta_c\!\left[{}^1S_0^{[8]}\right])&=33.43(6)\,\mathrm{nb}\,,\\
        \sigma(pp\to q\bar{q} \to \eta_c\!\left[{}^1S_0^{[1]}\right])&=0\,,\\
        \sigma(pp\to q\bar{q} \to \eta_c\!\left[{}^3S_1^{[8]}\right])&=603.7(2)\,\mathrm{nb}\,,\\
        \sigma(pp\to q\bar{q} \to \eta_c\!\left[{}^1S_0^{[8]}\right])&=0\,,
\end{align}
}
which sum, according to Eq.~\eqref{eq:etac}, to
\begin{equation}
    \sigma(pp \to \eta_c+X)=2.9366(5)\,\text{µ}\mathrm{b}\,,
\end{equation}
as listed in table~\ref{tab:pp2Q}.

To understand qualitatively the numbers reported in table~\ref{tab:pp2Q}, it is useful to recall that the LDME values used in this paper (cf.\@\xspace table \ref{tab:ldme}) follow the hierarchy expected from velocity-scaling rules: the colour-octet LDMEs are generally one or two orders of magnitude smaller than their colour-singlet counterparts. Unless there are particular reasons, such as kinematic or dynamical enhancements~\cite{Artoisenet:2008fc,Lansberg:2008gk,Lansberg:2009db,Lansberg:2013qka,Lansberg:2014swa,Shao:2018adj} or suppressions due to quantum-number conservation, this hierarchy would naturally govern the ordering of each partonic cross section. However, as we shall see later in this section, there are many counter-examples that violate this na\"ive expectation. A first example, shown in table~\ref{tab:pp2Q}, is that the total cross sections of spin-singlet states are much larger than those of spin-triplet states. This is because the $2\to 1$ QCD partonic processes for $(Q\bar{Q})\!\left[^3S_1^{[1]}\right]$ are forbidden due to either colour or C-number conservation. However, the electroweak channel $q\bar{q}\to \gamma^*/Z^{(*)}\to (Q\bar{Q})\!\left[^3S_1^{[1]}\right]$ can be non-zero, which we do not include here. \footnote{An apparently counterintuitive observation regarding the cross sections in table~\ref{tab:pp2Q} is that the $\eta_c$ cross section is smaller than the $\eta_b$ one. This occurs because these cross sections are very sensitive to the low-scale and low-$x$ regions of the PDFs~\cite{Lansberg:2020ejc}, where constraints are essentially absent. We have verified that using an alternative PDF set than \texttt{PDF4LHC21\_40} can yield a completely different $\eta_c$ total cross section, which may even exceed the $\eta_b$ cross section.}

\subsubsection{Inclusive $B_c$-meson production}

Unlike hidden-flavour charmonium and bottomonium production, the inclusive production of the $B_c$ meson at the LHC is dominantly associated with an open bottom and charm quark pair. We take the process $pp\to B_c^++b\bar{c}+X$ as an example. It can be generated via the following syntax:
\vskip 0.25truecm
\noindent
~~\prompt\ {\tt ~import~model~sm\_onia-c\_mass}

\noindent
~~\prompt\ {\tt ~generate~p p > Bc+ b c\mgtilde}

\noindent
~~\prompt\ {\tt ~output;~launch}
\vskip 0.25truecm

\noindent
The total cross section at $\sqrt{s}=13$ TeV is 

\begin{equation}
    \sigma(pp\to B_c^++b\bar{c}+X) = 14.77(6)\,\mathrm{nb}\,.
\end{equation}
Because $B_c^+$ carries electric charge, it has no well-defined C parity. All six partonic channels are non-zero. The total cross section is dominated by the subprocess $gg\to B_c^+\!\left[^1S_{0}^{[1]}\right]+b\bar{c}$, due to the gluon luminosity and the value of the colour-singlet LDME.

\subsubsection{Associated production with an electroweak boson or a jet\label{sec:quarkoniumplusboson}}

We now turn to quarkonium production in $pp$ collisions in association with other SM elementary particles. In particular, we consider charmonium and bottomonium final states produced alongside a $W^{\pm}$ or $Z$ boson (table~\ref{tab:3}), and a photon or a jet (table~\ref{tab:3_}).  These processes provide complementary insight into the quarkonium-production mechanism~\cite{Gong:2012ah,Li:2014ava,Lansberg:2009db,Lansberg:2013wva,Shao:2018adj,Li:2019anc,Butenschoen:2022wld,Brambilla:2024iqg}, and are also broadly relevant for rare-decay studies~\cite{Bodwin:2013gca,ATLAS:2015vss,ATLAS:2018xfc,CMS:2022fsq}, probing linearly polarised gluons in the proton~\cite{denDunnen:2014kjo,Lansberg:2017tlc}, and BSM searches~\cite{Grifols:1987iq,Robinett:1991us,Clarke:2013aya,Gonzalez-Alonso:2014rla}, as well as for investigations of the double-parton-scattering~(DPS) mechanism~\cite{Lansberg:2016rcx,Lansberg:2016muq,Lansberg:2017chq}.~\footnote{In this work, only the single-parton-scattering~(SPS) contribution is computed.} For a recent review, we refer the interested reader to ref.~\cite{Lansberg:2019adr} and to measurements at the LHC~\cite{ATLAS:2014yjd,ATLAS:2014ofp,ATLAS:2019jzd,CMS:2025xlt}.

\begin{table}[t!]
\centering
\renewcommand*{\arraystretch}{1.4}
\begin{tabular}[t]{lc|lc}
\toprule
\textbf{process}& $\sigma$ & \textbf{process}& $\sigma$  \\
\midrule
$pp \to \eta_c+W^++X$ & $3.125(1)\,\mathrm{pb}$ & $pp \to \eta_b+W^++X$ & $588.1(7)\,\mathrm{ab}$  \\
$pp \to \eta_c+W^-+X$ & $2.097(1)\,\mathrm{pb}$ & $pp \to \eta_b+W^-+X$ & $422.0(2)\,\mathrm{ab}$  \\
$pp \to \eta_c+Z+X$ & $2.2183(7)\,\mathrm{pb}$ & $pp \to \eta_b+Z+X$ & $131.89(3)\,\mathrm{fb}$ \\
$pp \to J/\psi+W^++X$ & $1.9328(6)\,\mathrm{pb}$ & $pp \to \Upsilon+W^++X$ & $102.81(4)\,\mathrm{fb}$ \\
$pp \to J/\psi+W^-+X$ & $1.2968(4)\,\mathrm{pb}$ & $pp \to \Upsilon+W^-+X$ & $73.77(3)\,\mathrm{fb}$ \\
$pp \to J/\psi+Z+X$ & $1.3425(4)\,\mathrm{pb}$ & $pp \to \Upsilon+Z+X$ & $340.7(1)\,\mathrm{fb}$ \\
\bottomrule
\end{tabular}
\caption{Total cross sections for single charmonium and bottomonium production at $\mathcal{O}(\alpha_s^2\alpha)$ in $pp$ collisions at $13\,\mathrm{TeV}$ in association with a $W^{\pm}$ or $Z$ boson, based on the setup described in section~\ref{sec:setup}. The numbers in parentheses represent the numerical uncertainties from the MC phase-space integration. These results are provided for illustration only: they do not include feeddown effects or experimental cuts, and are specific to the chosen LDME values (cf.\@\xspace table \ref{tab:ldme}).}
\label{tab:3}
\end{table}

For charmonium or bottomonium production in association with a $W^\pm$ boson, only the colour-octet partonic channel $q\bar{q}^{\,\prime} \to (Q\bar{Q})\!\left[^3S_1^{[8]}\right]+W^\pm$ contributes at LO (i.e., $\mathcal{O}(\alpha_s^2\alpha)$).~\footnote{As discussed in ref.~\cite{Lansberg:2013wva} for the $J/\psi$ case, important contributions are also expected from $q\bar{q}^{\,\prime} \to (Q\bar{Q})\!\left[^3S_1^{[1]}\right]+W^\pm$ at $\mathcal{O}(\alpha^3)$ via an off-shell photon and $q g \to (Q\bar{Q})\!\left[^3S_1^{[1]}\right]+W^\pm+Q$ at $\mathcal{O}(\alpha_s^3\alpha)$ which can also be computed with our extension.} In contrast, for production with a $Z$ boson, five partonic channels contribute, except for the quark-induced colour-singlet ones, $q\bar{q}\to (Q\bar{Q})\!\left[^3S_1^{[1]},{}^1S_0^{[1]}\right]+Z$, which vanish. Even in the $Z$-boson case, however, the charmonium cross sections are dominated by the $q\bar{q} \to (c\bar{c})\!\left[^3S_1^{[8]}\right]+Z$ channel, while for bottomonium the leading contributions arise from $gg\to (b\bar{b})\!\left[^1S_0^{[1]},{}^3S_1^{[1]}\right]+Z$. This difference can be attributed to two main reasons. First, the colour-octet LDMEs are relatively less suppressed with respect to the colour-singlet ones in the charmonium case. Second, the colour-octet channels receive an additional enhancement from the propagators in the $g^*\!\to (Q\bar{Q})\left[^3S_1^{[8]}\right]$ transitions. This propagator enhancement is stronger for charmonium than for bottomonium because of the smaller heavy-quark mass. These considerations imply that the spin-singlet to spin-triplet cross-section ratios should approximately follow the corresponding LDME ratios for the dominant channels.

So far, no experimental studies of inclusive quarkonium–photon production have been reported, apart from searches for exclusive radiative decays of the $Z$ or $H$ bosons, such as from $H \rightarrow J/\psi + \gamma$ and $\Upsilon + \gamma$. These Higgs rare decay channels offer sensitivity to the Higgs–quark Yukawa couplings to $c$ and $b$ quarks. Ongoing inclusive studies at ATLAS~\cite{Tee:2021mfd}, focusing on the back-to-back configuration, aim to probe the gluon transverse momentum distribution (TMD) inside the proton~\cite{denDunnen:2014kjo}.

When considering a final-state jet in addition to the quarkonium, or studying quarkonium plus photon production, we impose cuts of $p_{T,j}>10\,\mathrm{GeV}$ and $|\eta_j|<5$ on the jet, and $p_{T,\gamma}>2\,\mathrm{GeV}$ and $|\eta_\gamma|<2.5$ on the photon. These cuts ensure sufficient separation from the colliding beams, suppress soft and collinear configurations, and mimic typical experimental isolation criteria.

\begin{table}[t!]
\centering
\renewcommand*{\arraystretch}{1.4}
\begin{tabular}[t]{lc|lc}
\toprule
\textbf{process}& $\sigma$ & \textbf{process}& $\sigma$  \\
\midrule
$pp \to \eta_c+j+X$ & $805.4(4)\,\mathrm{nb}$ & $pp \to \eta_b+j+X$ & $315.4(2)\,\mathrm{nb}$  \\
$pp \to J/\psi+j+X$ & $329.8(2)\,\mathrm{nb}$ & $pp \to \Upsilon+j+X$ & $19.85(1)\,\mathrm{nb}$ \\
$pp \to \eta_c+\gamma+X$ & $789.3(4)\,\mathrm{pb}$ & $pp \to \eta_b+\gamma+X$ & $2.257(1)\,\mathrm{pb}$  \\
$pp \to J/\psi+\gamma+X$ & $19.13(1)\,\mathrm{nb}$ & $pp \to \Upsilon+\gamma+X$ & $ 897.4(5)\,\mathrm{pb}$ \\
\bottomrule
\end{tabular}
\caption{Fiducial cross sections for single charmonium and bottomonium production in $pp$ collisions at $13\,\mathrm{TeV}$ in association with a photon ($\mathcal{O}(\alpha_s^2\alpha)$) or a jet ($\mathcal{O}(\alpha_s^3)$), based on the setup described in section~\ref{sec:setup} and with cuts defined in the main text. The numbers in parentheses represent the numerical uncertainties from the MC phase-space integration. These results are given for illustration only: they do not include feeddown effects or experimental cuts, and are specific to the chosen LDME values (cf.\@\xspace table~\ref{tab:ldme}).}
\label{tab:3_}
\end{table}

In jet-associated production processes, the leading channel is $gg\! \to\! (Q\bar{Q})\!\left[^3S_1^{[8]}\right]+g$ in the $\eta_c$, $J/\psi$, and $\Upsilon$ cases, since they share the leading-$p_T$ behaviour arising from gluon fragmentation, $g^* \to (Q\bar{Q})\!\left[^3S_1^{[8]}\right]$. In the $\eta_c$ case, the colour-singlet channel $gg \to \eta_c\!\left[^1S_0^{[1]}\right]+g$ also gives rise to significant contributions. For $\eta_b+j$ production, due to the very small LDME value of $\eta_b\!\left[^3S_1^{[8]}\right]$ (cf.\@\xspace table~\ref{tab:ldme}), the leading contribution comes from the colour-singlet channel $gg \to \eta_b\!\left[^1S_0^{[1]}\right]+g$. Such a conclusion is valid only under the specific conditions considered here (e.g.,\@\xspace LO, $p_T$ cut, LDME values, etc.).

For photon-associated production processes, due to C-parity conservation, colour-singlet channels for pseudoscalar states are forbidden. In contrast, for spin-triplet states, the gluon-induced colour-singlet process $gg\to (Q\bar{Q})\!\left[^3S_1^{[1]}\right]+\gamma$ is dominant. This qualitatively explains the observed cross-section hierarchy between spin-singlet and spin-triplet states in table~\ref{tab:3_}.

\subsubsection[$J/\psi$ production in association with an open charm-anticharm pair]{\boldmath $J/\psi$ production in association with an open charm-anticharm pair}

The associated production of a heavy quarkonium and an open charm-anticharm pair, where the latter hadronises into a light charm hadron such as a $D^0$, $\bar{D}^0$, $D^\pm$, $D_s^\pm$, or $\Lambda_c^\pm$, is of significant phenomenological interest for a variety of reasons. These processes serve as valuable probes for understanding the underlying mechanisms~\cite{Berezhnoy:1998aa,Baranov:2006dh,Artoisenet:2007xi,Li:2011yc,Shao:2018adj,Shao:2020kgj} of heavy-quarkonium production, as they can constitute an important source of inclusive charmonium production~\cite{Qiao:2003pu,Lansberg:2010vq,Belyaev:2017lbo,Feng:2025btt}, for studying the DPS mechanism~\cite{Shao:2020kgj}, and for exploring~\cite{Brodsky:2009cf} the intrinsic charm content of the proton~\cite{Brodsky:1980pb}. Moreover, they provide an opportunity to study potential factorisation-breaking effects arising from colour transfer between the quarkonium and the open-charm hadron near the mass threshold~\cite{Nayak:2007mb,Nayak:2007zb}. In the context of heavy-ion collisions, such processes can also be used to probe the impact-parameter-dependent cold nuclear-matter modifications of nuclear PDFs through DPS processes~\cite{Shao:2020acd}.

The LHCb experiment has observed the associated production of $J/\psi+D$~\cite{LHCb:2012aiv} and $\Upsilon+D$~\cite{LHCb:2015wvu}, with results indicating that SPS contributions alone are insufficient to describe the data, thus requiring the inclusion of DPS effects. Motivated by this strong phenomenological relevance, numerous theoretical studies have been dedicated to these processes~\cite{Berezhnoy:2015jga,Likhoded:2015fdr,Karpishkov:2019vyt,Shao:2020kgj,Chernyshev:2023qea}.

Even though a detailed phenomenological study is beyond the scope of this work, we aim to illustrate the potential of the \mgshort\ implementation for such applications. To this end, we consider the production of a $J/\psi$ meson in association with an open charm-anticharm pair in a $pp$ collision at $\sqrt{s}=13$ TeV, i.e.,\@\xspace the process $pp\to J/\psi+c\bar{c}+X$. Using the baseline setup outlined in section~\ref{sec:setup}, we obtain the total cross section
\begin{equation}
    \sigma(pp\to J/\psi+c\bar{c}+X) = 585^{+987}_{-401}\,\mathrm{nb}\,,
\end{equation}
with contributions from individual $J/\psi$ Fock states given by 
\begin{equation}
    \begin{aligned}
        \sigma(pp\to J/\psi\!\left[{}^3S_1^{[1]}\right]+c\bar{c}+X) &= 429^{+735}_{-306}\,\mathrm{nb}\,,\\
        \sigma(pp\to J/\psi\!\left[{}^1S_0^{[8]}\right]+c\bar{c}+X) &= 77.5^{+137.9}_{-51.3}\,\mathrm{nb}\,,\\
        \sigma(pp\to J/\psi\!\left[{}^3S_1^{[8]}\right]+c\bar{c}+X) &= 78.4^{+114.4}_{-43.9}\,\mathrm{nb}\,.
    \end{aligned}
\end{equation}
The quoted uncertainties reflect 7-point scale variations with $\mu_{R/F}\in\{\mu_0/2,\mu_0,2\mu_0\}$, enforcing $1/2\le\mu_{R}/\mu_{F}\le2$, around the central scale $\mu_0=H_T/2$ as defined in Eq.~\eqref{eq:HTovertwo}. 

The SPS cross section of this process is dominated by the colour-singlet contributions, while the colour-octet ones are subdominant but not entirely negligible. This observation is consistent with the findings of ref.~\cite{Shao:2020kgj} (see table III therein). However, as pointed out in ref.~\cite{Shao:2020kgj}, the cross section reported here may significantly underestimate the true $J/\psi+D$ SPS contribution, since the calculation is performed strictly in the 3FS. To obtain a reliable prediction of the SPS contribution for this process, an appropriate matching or combination of the three- and four-flavour schemes~(3FS and 4FS) is essential.

\begin{figure}[!hbt]
    \begin{subfigure}[b]{0.5\textwidth}
        \includegraphics[width=\textwidth]{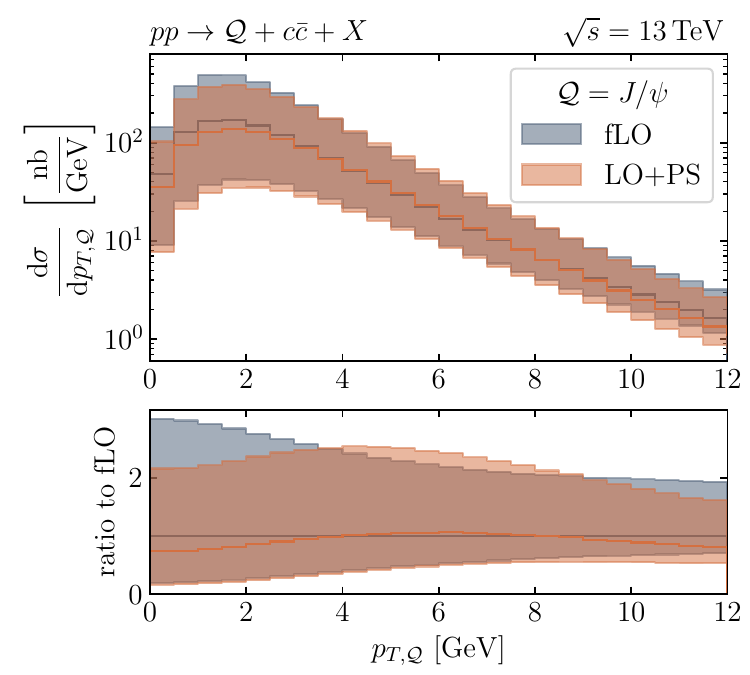} 
        \caption{}
        \label{fig:ps_a}
    \end{subfigure}
    \begin{subfigure}[b]{0.5\textwidth}
        \includegraphics[width=\textwidth]{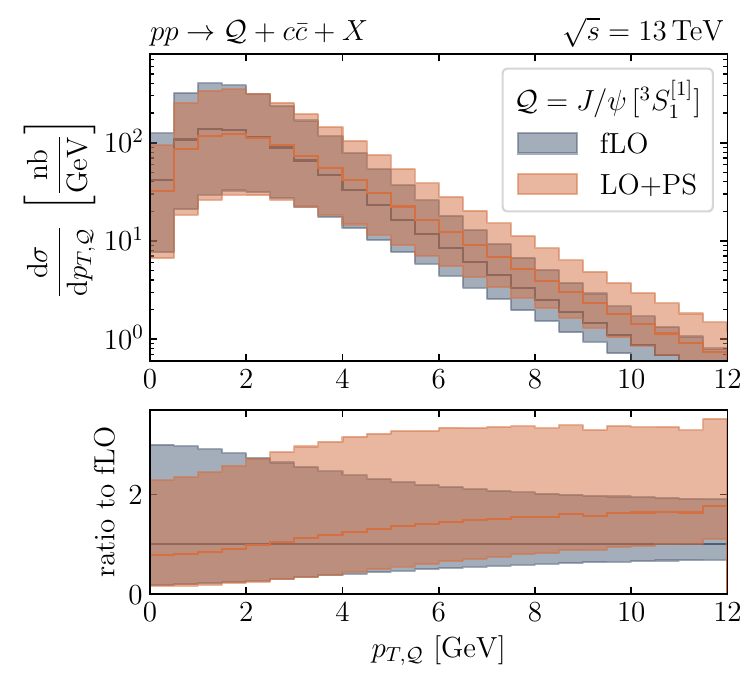}
        \caption{}
        \label{fig:ps_b}
    \end{subfigure}\\
    \begin{subfigure}[c]{0.5\textwidth}
        \includegraphics[width=\textwidth]{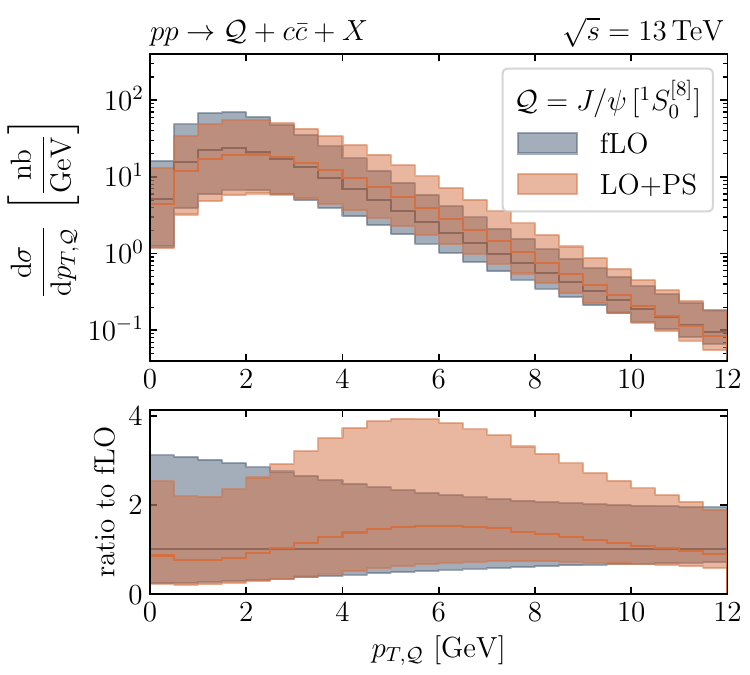}
        \caption{}
        \label{fig:ps_c}
    \end{subfigure}
    \begin{subfigure}[d]{0.5\textwidth}
        \includegraphics[width=\textwidth]{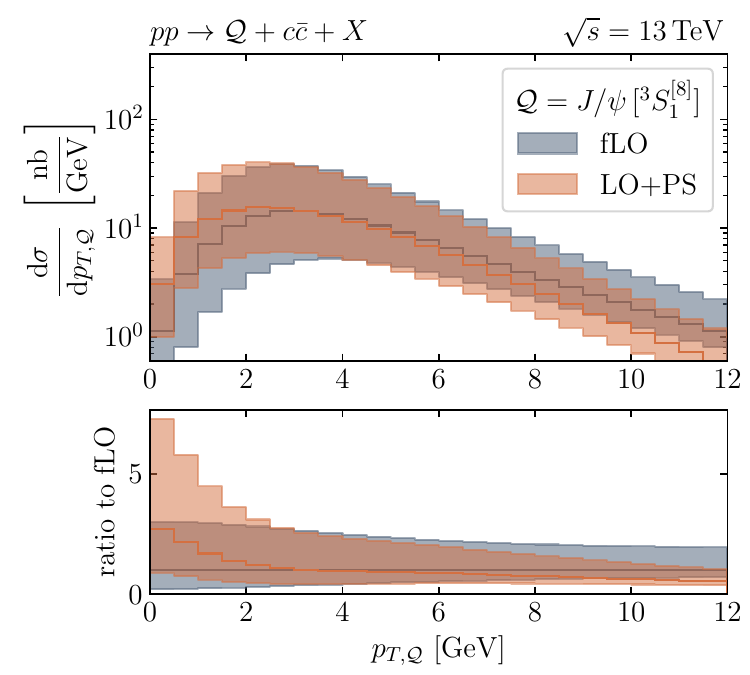}
        \caption{}
        \label{fig:ps_d}
    \end{subfigure}
    \caption{Transverse momentum ($p_{T,\mathcal{Q}}$) distributions for $J/\psi$ production in association with $c\bar{c}$ in $pp$ collisions at $\sqrt{s}=13$ TeV. Panels (b-d) correspond to different $J/\psi$ Fock states, while panel (a) shows the total contribution obtained by combining them. The blue curve represents the fixed-order LO~(fLO) computation from \mgshort, and the orange curve shows the \mgshort\ LO events combined with PS from \pythiaeight~(LO+PS). Both include their respective 7-point scale-variation uncertainty bands. The lower panels display the ratio to the fLO contribution in each case.}
	\label{fig:ps}
\end{figure}\noindent

To further illustrate the potential of \mgshort\ for this process beyond fixed-order accuracy, we employ its interface to \pythiaeight~\cite{Bierlich:2022pfr} to perform parton-shower (PS) simulations on the LHEF events generated by \mgshort.~\footnote{\pythia\ also handles the hadronisation of the open charm and anticharm quarks, thereby enabling studies of exclusive final states such as $J/\psi + D$. However, for the purposes of this work, we consider only the fully inclusive $J/\psi + c\bar{c}$ final state, focusing on the modifications of the quarkonium kinematic distributions.} The default \pythia\ settings are used, with initial-state radiation (ISR), final-state radiation (FSR), multiparton interactions (MPI), and hadronisation enabled. The $J/\psi$ is treated as a stable particle.

In figure~\ref{fig:ps}, we compare the $J/\psi$ transverse momentum spectra from the fixed-order LO (fLO) calculation and the LO events matched to PS (LO+PS). The $p_{T,J/\psi}$ spectrum including all three Fock states is shown in figure~\ref{fig:ps_a}, while figures~\ref{fig:ps_b},~\ref{fig:ps_c}, and~\ref{fig:ps_d} display the spectra for the individual $J/\psi\!\left[^3S_1^{[1]}\right]$, $J/\psi\!\left[^1S_0^{[8]}\right]$, and $J/\psi\!\left[^3S_1^{[8]}\right]$ Fock states, respectively. We observe that PS effects broaden the $p_{T,J/\psi}$ distributions in the $J/\psi\!\left[^3S_1^{[1]}\right]$ and $J/\psi\!\left[^1S_0^{[8]}\right]$ channels, while they narrow the transverse-momentum distribution in the $J/\psi\!\left[^3S_1^{[8]}\right]$ channel. These opposite shifts from the two spin-triplet Fock states partially compensate each other in the total (figure~\ref{fig:ps_a}), resulting in a moderate overall distortion of the $p_{T,J/\psi}$ spectrum. Unlike the processes $pp\to J/\psi\!\left[^3S_1^{[8]},\, ^1S_0^{[8]},\, ^3P_J^{[8]}\right]+j+X$, the cross sections for these processes at small $p_{T,J/\psi}$ are finite. In the former case, one would need to impose cuts, and the effect of showering in presence of such cuts would be more significant~\cite{Lansberg:2024zap}.

For the sake of completeness, we note that the same LDMEs for the colour-octet channels are used in both the fLO and LO+PS analyses in this paper. In extractions of colour-octet LDMEs from the $p_T$ spectra of inclusive quarkonium hadroproduction processes, several works (e.g., ref.~\cite{Sanchis-Lozano:1999syz}) have pointed out that the LDMEs obtained from fLO and LO+PS analyses can differ significantly. This difference can be understood because the inclusion of PS effects in channels such as $pp\to (Q\bar{Q})\!\left[^1S_0^{[8]}\right]+j+X$ modifies the $p_T$ behaviour, in a manner somewhat analogous to what is observed when comparing NLO to LO results~\cite{Shao:2018adj}. The LDMEs used in this paper (see table~\ref{tab:ldme}) are obtained from fixed-order NLO fits to the $p_T$ distributions in quarkonium hadroproduction. Other NLO-based extractions also exist in the literature, though the list is too long to be reproduced here. The interested reader is guided to ref.~\cite{Lansberg:2019adr}.

\subsubsection{Inclusive quarkonium-pair production}

In this section, we discuss the production of final states containing two quarkonia. We consider quarkonium-pair production involving charmonium, bottomonium, and $B_c^{\pm}$ mesons.

Among all quarkonium-pair production processes, double $J/\psi$ production is the most studied final state. The process was first observed by the NA3 collaboration at the CERN-SPS in pion–nucleus and proton–nucleus interactions~\cite{NA3:1982qlq,NA3:1985rmd} in the 1980s. It was subsequently studied at the Tevatron and the LHC~\cite{LHCb:2011kri,D0:2014vql,CMS:2014cmt,ATLAS:2016ydt,LHCb:2016wuo,LHCb:2023ybt} at different energies and in different kinematic regions. These measurements provide valuable constraints on the production mechanisms (see, e.g., refs.~\cite{Lansberg:2013qka,Sun:2014gca,Lansberg:2020rft,Lansberg:2019fgm,He:2019qqr,Sun:2023exa,Baranov:2015cle,Lansberg:2015lva}), on gluon TMDs~\cite{Lansberg:2017dzg,Scarpa:2019fol}, which remain largely unconstrained, and on DPS~\cite{Kom:2011bd,Lansberg:2014swa,Borschensky:2016nkv,Prokhorov:2020owf}. Double $J/\psi$ production also constitutes an irreducible background for fully-charmed tetraquark searches~\cite{LHCb:2020bwg,ATLAS:2023bft,CMS:2023owd} and rare decays such as $H\to J/\psi + J/\psi$ and $Z\to J/\psi + J/\psi$~\cite{CMS:2022fsq}. To date, the only measurements of $\Upsilon$-pair production were performed by CMS~{\cite{CMS:2016liw,CMS:2020qwa}, while LHCb recently reported the first study of $J/\psi+\psi(2S)$ production~\cite{LHCb:2023wsl}. 

The associated production of $J/\psi + \eta_c$ can provide insight into the impact of QCD radiative corrections on the $p_T$ spectrum~\cite{Lansberg:2013qka}. However, it has not yet been observed at hadron colliders. The same applies to $J/\psi + \chi_c$ ~\cite{Likhoded:2016zmk} and to the bottomonium states.

In contrast, $J/\psi + \Upsilon$ production in hadron-hadron collisions has now been measured at both the Tevatron by D0~\cite{D0:2015dyx} and at the LHC by LHCb~\cite{LHCb:2023qgu}. The main motivation for studying charmonium plus bottomonium production is to probe colour-octet quarkonium production mechanisms~\cite{Ko:2010xy,Lansberg:2015lva,Lansberg:2020rft}, DPS~\cite{Likhoded:2015zna,Shao:2016wor}, and potential new exotic hadrons such as fully-heavy tetraquarks ($b\bar{b}c\bar{c}$). Unlike $J/\psi$-pair or $\Upsilon$-pair production, the SPS colour-singlet process $pp\to J/\psi\!\left[^3S_1^{[1]}\right]+\Upsilon\!\left[^3S_1^{[1]}\right]+X$ vanishes in QCD at LO ($\mathcal{O}(\alpha_s^4$)) due to quantum-number conservation. In fact, the lowest-order contribution in QCD is loop-induced at $\mathcal{O}(\alpha_s^6)$~\cite{Shao:2016wor}. This suppression of the SPS colour-singlet channel makes DPS contributions more prominent and increases the relative importance of colour-octet mechanisms in the SPS channel.

\begin{table}[t!]
\centering
\renewcommand*{\arraystretch}{1.4}
\begin{tabular}[t]{lc|lc}
\toprule
\textbf{process}& $\sigma$ & \textbf{process}& $\sigma$  \\
\midrule
$pp \to \eta_c+\eta_c+X$ & $ 35.81(1)\,\mathrm{nb}$ & $pp \to \eta_b+\eta_b+X$ & $ 75.64(3)\,\mathrm{pb}$ \\
$pp \to \eta_c+J/\psi+X$ & $ 7.233(3)\,\mathrm{nb}$ & $pp \to \eta_b+\Upsilon+X$ & $ 1.9244(6)\,\mathrm{pb}$ \\
$pp \to J/\psi+J/\psi+X$ & $ 10.756(3)\,\mathrm{nb}$ & $pp \to \Upsilon+\Upsilon+X$ & $ 44.63(1)\,\mathrm{pb}$ \\
\bottomrule
\end{tabular}
\caption{Total cross sections for charmonium- and bottomonium-pair production at $\mathcal{O}(\alpha_s^4)$ in $pp$ collisions at $\sqrt{s}=13\,\mathrm{TeV}$, based on the setup described in section~\ref{sec:setup}. The numbers in parentheses indicate the numerical uncertainties from the MC phase-space integration. The results are provided for illustration only: they do not include feeddown or DPS effects, nor any experimental cuts and they depend on the chosen LDME values (cf.\@\xspace table~\ref{tab:ldme}).}
\label{tab:4}
\end{table}

Table~\ref{tab:4} shows our results for the production of charmonium or bottomonium pairs, while table~\ref{tab:5} presents the production of mixed charmonium–bottomonium pairs as well as $B_c^{\pm}$ pairs. All cross sections are computed at $\mathcal{O}(\alpha_s^4)$. For charmonium or bottomonium pairs, the cross sections are dominated by colour-singlet configurations when both quarkonia are of the same type, e.g.,\@\xspace in the production of two $\eta_c$ ($\eta_b$) or two $J/\psi$ ($\Upsilon$) mesons. By contrast, at LO in QCD, processes involving one pseudoscalar and one vector quarkonium of the same flavour, such as $\eta_c+J/\psi$ ($\eta_b+\Upsilon$), are typically dominated by configurations involving at least one colour-octet state. This is because, due to C parity, the pure colour-singlet partonic channels $gg\to (Q\bar{Q})\!\left[^1S_0^{[1]}\right]+(Q\bar{Q})\!\left[^3S_1^{[1]}\right]$ and $q\bar{q}\to (Q\bar{Q})\!\left[^1S_0^{[1]}\right]+(Q\bar{Q})\!\left[^3S_1^{[1]}\right]$ are forbidden in QCD. This also explains why the cross sections for  pseudoscalar-vector final states are smaller than those of the other two configurations in table~\ref{tab:4}. 

When charmonium and bottomonium are produced together, only the simultaneous production of two pseudoscalars, $\eta_c+\eta_b$, receives contributions from the pure colour-singlet channel at $\mathcal{O}(\alpha_s^4)$. The remaining charmonium-bottomonium final states shown on the left of table~\ref{tab:5} have vanishing colour-singlet contributions at LO in QCD for both gluon-gluon and quark-antiquark initiated channels, and therefore must involve at least one colour-octet state. This explains the observed pattern of cross-section values in table~\ref{tab:5}. 

In contrast, for $B_c$-meson pair production, all colour-singlet partonic channels contribute and dominate the total cross sections. The cross sections of $pp\to B_c^{\pm}+B_c^{\ast \mp}+X$ are significantly smaller than those of the other two $B_c$-pair processes in the right column of table \ref{tab:5}, as they are suppressed by a factor $\left(\frac{m_b-m_c}{m_b+m_c}\right)^2\approx 0.25$~\cite{Chen:2024dkx}. The observed cross-section pattern is consistent with the findings of refs.~\cite{Li:2009ug,Chen:2024dkx}.

We again emphasise the broad physics potential of quarkonium-pair production studies for probing the interplay between SPS and DPS interactions. However, the numerical values presented in this work account only for the SPS contribution at LO, using generic cuts and LDME values, for illustration purposes.

We conclude the discussion on pair production by noting that an advantage of our implementation over \helaconia\ is that it assigns different LDMEs to the colour-octet states on the fly, making \mgshort\ more intuitive to use. For instance, this simplifies computations such as $\eta_c\!\left[{}^3S_1^{[8]}\right]+J/\psi\!\left[{}^3S_1^{[8]}\right]$, since no rescaling of one of the LDMEs is required anymore. Moreover, \mgshort\ automatically generates all possible Fock-state combinations directly at the level of the physical process, including the long-distance contributions, when using a command such as \texttt{generate p p > etac Jpsi}. In contrast, \helaconia\ performs this task at the level of the short-distance coefficients, meaning that the user must combine them manually to obtain the physical results.

\begin{table}[t!]
\centering
\renewcommand*{\arraystretch}{1.4}
\begin{tabular}[t]{lc|lc}
\toprule
\textbf{process}& $\sigma$ & \textbf{process}& $\sigma$  \\
\midrule
$pp \to \eta_c+\eta_b+X$ & $ 2.7427(9)\,\mathrm{nb}$ & $pp \to B_c^+ + B_c^-+X$ & $472.5(1)\,\mathrm{pb}$ \\
$pp \to \eta_c+\Upsilon+X$ & $ 23.431(7)\,\mathrm{pb}$ & $pp \to B_c^+ + B_c^{\ast -}+X$ & $89.23(2)\,\mathrm{pb}$ \\
$pp \to J/\psi+\eta_b+X$ & $ 381.4(1)\,\mathrm{pb}$ & $pp \to B_c^- + B_c^{\ast +}+X$ & $89.23(2)\,\mathrm{pb}$\\
$pp \to J/\psi+\Upsilon+X$ & $ 10.503(3)\,\mathrm{pb}$ & $pp \to B_c^{\ast +} + B_c^{\ast -}+X$ & $940.9(2)\,\mathrm{pb}$ \\
\bottomrule
\end{tabular}
\caption{Total cross sections for mixed charmonium–bottomonium pair production (left) and $B_c$-pair production (right) at $\mathcal{O}(\alpha_s^4)$ in $pp$ collisions at $\sqrt{s}=13\,\mathrm{TeV}$, based on the setup described in section~\ref{sec:setup}. The numbers in parentheses indicate the numerical uncertainties from the MC phase-space integration. The results are provided for illustration only: they do not include feeddown or DPS effects, nor any experimental cuts, and they depend on the chosen LDME values (cf.\@\xspace table \ref{tab:ldme}).}
\label{tab:5}
\end{table}

\subsubsection[Inclusive triple $J/\psi$ production]{\boldmath Inclusive triple $J/\psi$ production}

To showcase the capabilities of our implementation, we consider triple $J/\psi$ production in $pp$ collisions. This serves as an illustrative example of a non-trivial three-quarkonium final state, proposed in refs.~\cite{Shao:2019qob,dEnterria:2016ids,dEnterria:2017yhd} as a probe of triple parton scattering~(TPS). Recently measured by CMS~\cite{CMS:2021qsn}, this marks the first experimental measurement of TPS, providing new insight into multi-parton scattering dynamics.
The LO ($\mathcal{O}(\alpha_s^6)$) total cross section in QCD at $\sqrt{s}=13\,\mathrm{TeV}$ is
\begin{equation}
    \sigma(pp\to J/\psi+J/\psi+J/\psi+X) = 1.038(3)\,\mathrm{pb}\,.\label{eq:threeJpsixs}
\end{equation}
Breaking it down, the cross section receives contributions from ten different Fock-state combinations:
{\allowdisplaybreaks
\begin{align}
    \sigma(pp\to J/\psi\!\left[{}^3S_1^{[1]}\right]+J/\psi\!\left[{}^3S_1^{[1]}\right]+J/\psi\!\left[{}^3S_1^{[1]}\right]+X) &= 54.57(3)\,\mathrm{fb}\,,\\
    \sigma(pp\to J/\psi\!\left[{}^3S_1^{[1]}\right]+J/\psi\!\left[{}^3S_1^{[1]}\right]+J/\psi\!\left[{}^1S_0^{[8]}\right]+X) &= 423(3)\,\mathrm{fb}\,,\\
    \sigma(pp\to J/\psi\!\left[{}^3S_1^{[1]}\right]+J/\psi\!\left[{}^3S_1^{[1]}\right]+J/\psi\!\left[{}^3S_1^{[8]}\right]+X) &= 533(2)\,\mathrm{fb}\,,\\
    \sigma(pp\to J/\psi\!\left[{}^3S_1^{[1]}\right]+J/\psi\!\left[{}^1S_0^{[8]}\right]+J/\psi\!\left[{}^1S_0^{[8]}\right]+X) &= 2.977(7)\,\mathrm{fb}\,,\\
    \sigma(pp\to J/\psi\!\left[{}^3S_1^{[1]}\right]+J/\psi\!\left[{}^1S_0^{[8]}\right]+J/\psi\!\left[{}^3S_1^{[8]}\right]+X) &= 13.07(3)\,\mathrm{fb}\,,\\
    \sigma(pp\to J/\psi\!\left[{}^3S_1^{[1]}\right]+J/\psi\!\left[{}^3S_1^{[8]}\right]+J/\psi\!\left[{}^3S_1^{[8]}\right]+X) &= 4.253(9)\,\mathrm{fb}\,,\\
    \sigma(pp\to J/\psi\!\left[{}^1S_0^{[8]}\right]+J/\psi\!\left[{}^1S_0^{[8]}\right]+J/\psi\!\left[{}^1S_0^{[8]}\right]+X) &= 0.759(7)\,\mathrm{fb}\,,\\
    \sigma(pp\to J/\psi\!\left[{}^1S_0^{[8]}\right]+J/\psi\!\left[{}^1S_0^{[8]}\right]+J/\psi\!\left[{}^3S_1^{[8]}\right]+X) &= 1.964(4)\,\mathrm{fb}\,,\\
    \sigma(pp\to J/\psi\!\left[{}^1S_0^{[8]}\right]+J/\psi\!\left[{}^3S_1^{[8]}\right]+J/\psi\!\left[{}^3S_1^{[8]}\right]+X) &= 2.728(3)\,\mathrm{fb}\,,\\
    \sigma(pp\to J/\psi\!\left[{}^3S_1^{[8]}\right]+J/\psi\!\left[{}^3S_1^{[8]}\right]+J/\psi\!\left[{}^3S_1^{[8]}\right]+X) &= 2.101(2)\,\mathrm{fb}\,.
\end{align}}\noindent
\begin{figure}
    \centering
    \includegraphics[width=0.5\linewidth]{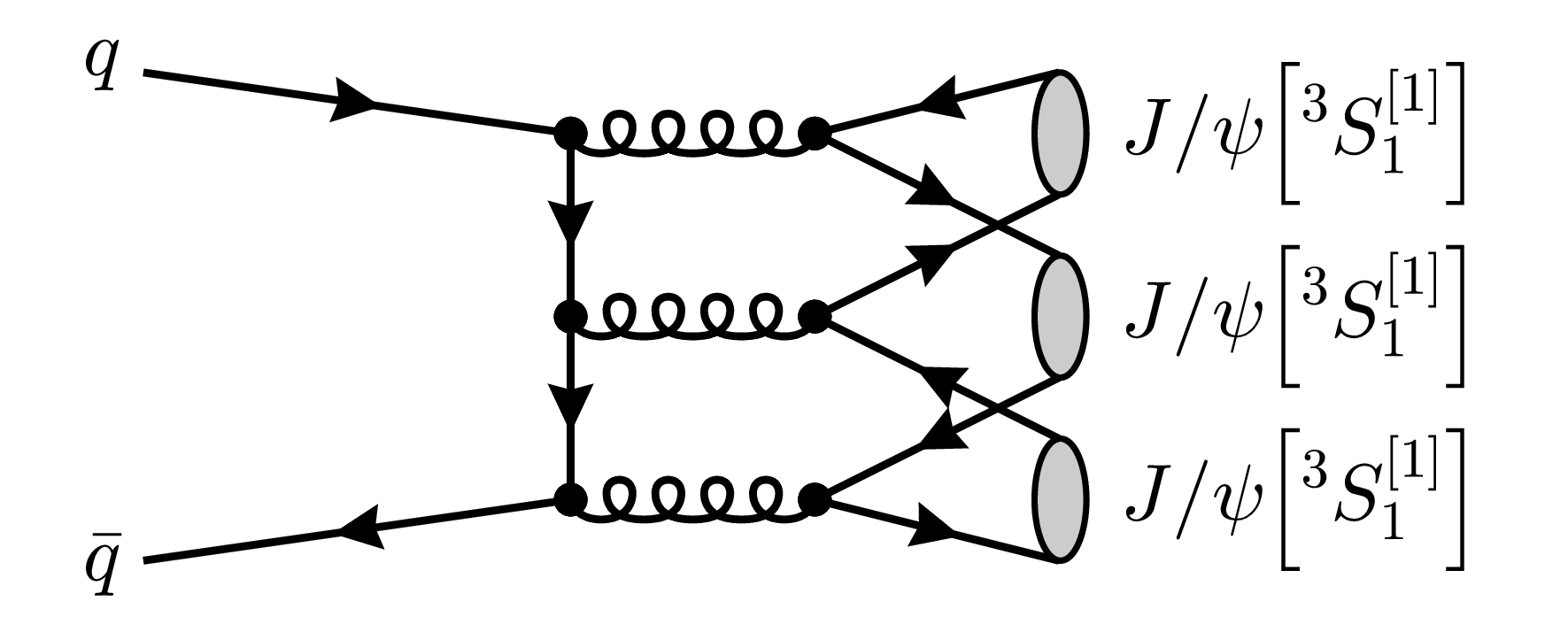}
    \caption{Example Feynman diagram contributing to triple colour-singlet production. All Feynman diagrams in this article were created with \texttt{FeynGame}~\cite{feyngame,feyngame2,feyngame3}.}
    \label{fig:diag_3jpsi}
\end{figure}
For the triple colour-singlet channel, the gluon-induced process $gg\to J/\psi\!\left[{}^3S_1^{[1]}\right]+J/\psi\!\left[{}^3S_1^{[1]}\right]+J/\psi\!\left[{}^3S_1^{[1]}\right]$ is forbidden by C-parity conservation, leaving only the quark–antiquark-initiated process possible at LO (cf.\@\xspace figure~\ref{fig:diag_3jpsi}). The latter is, however, suppressed by the small quark–antiquark luminosity at the LHC. As a result, the NLO QCD corrections, where the $2\to4$ gluon-induced process $gg\to J/\psi\!\left[{}^3S_1^{[1]}\right]+J/\psi\!\left[{}^3S_1^{[1]}\right]+J/\psi\!\left[{}^3S_1^{[1]}\right]+g$ becomes allowed, can be large, as shown in ref.~\cite{Shao:2019qob}. In that study, this gluon-induced SPS contribution at $\mathcal{O}(\alpha_s^7)$ was found to be around $2\,\mathrm{pb}$ at $\sqrt{s}=13\,\mathrm{TeV}$ (cf.\@\xspace table I therein), including feeddown contributions from $\psi(2S)$. This value is of the same order as Eq.~\eqref{eq:threeJpsixs}, highlighting the important role of these radiative corrections for a reliable SPS cross-section prediction. Nevertheless, the SPS cross section remains several orders of magnitude smaller than the DPS and TPS contributions.

\subsubsection{Heavy vector-quarkonium production in association with a Higgs boson}

\begin{figure}[t]
    \center
    \begin{subfigure}[b]{0.7\textwidth}
        \includegraphics[width=0.49\textwidth]{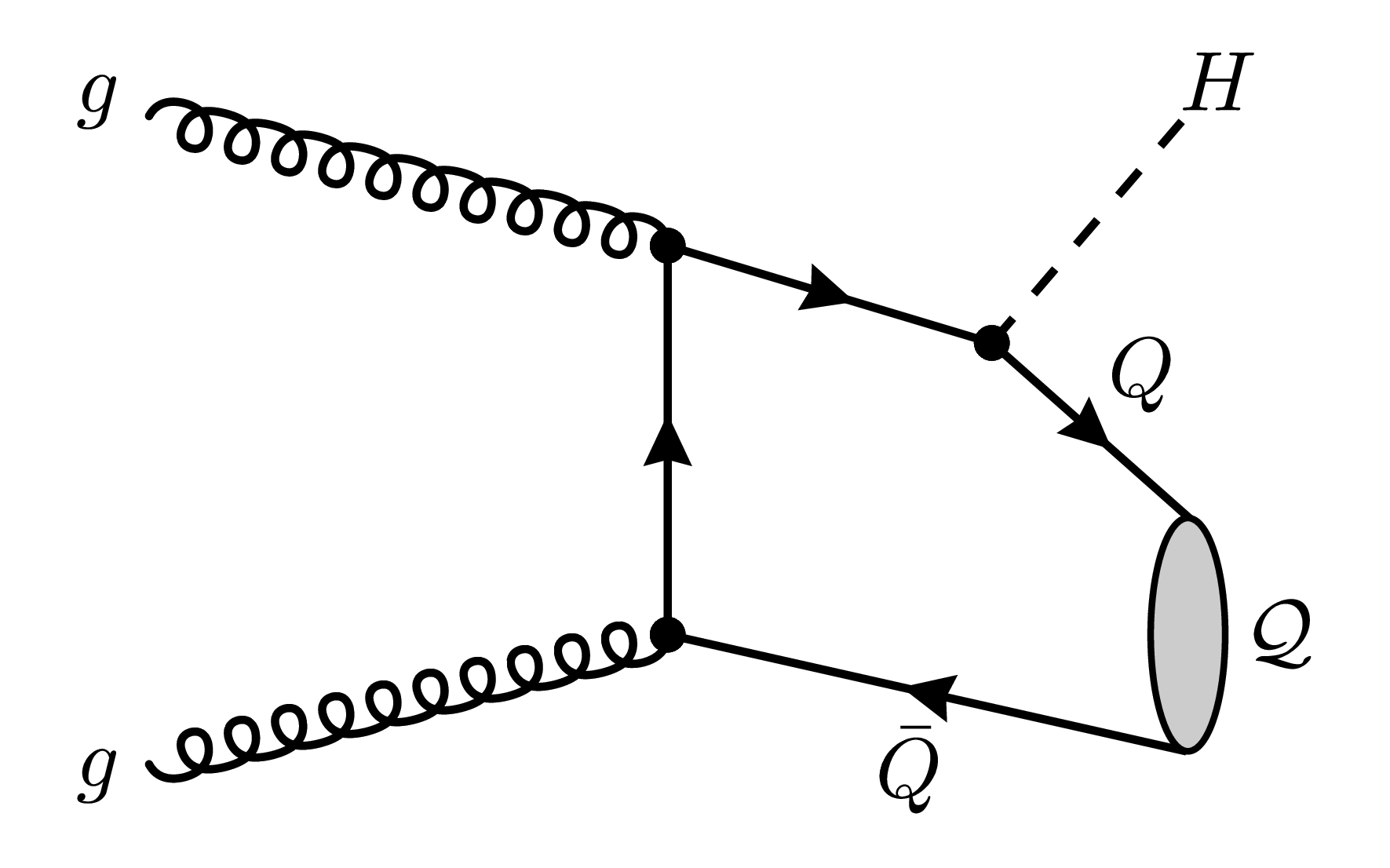}
        \includegraphics[width=0.49\textwidth]{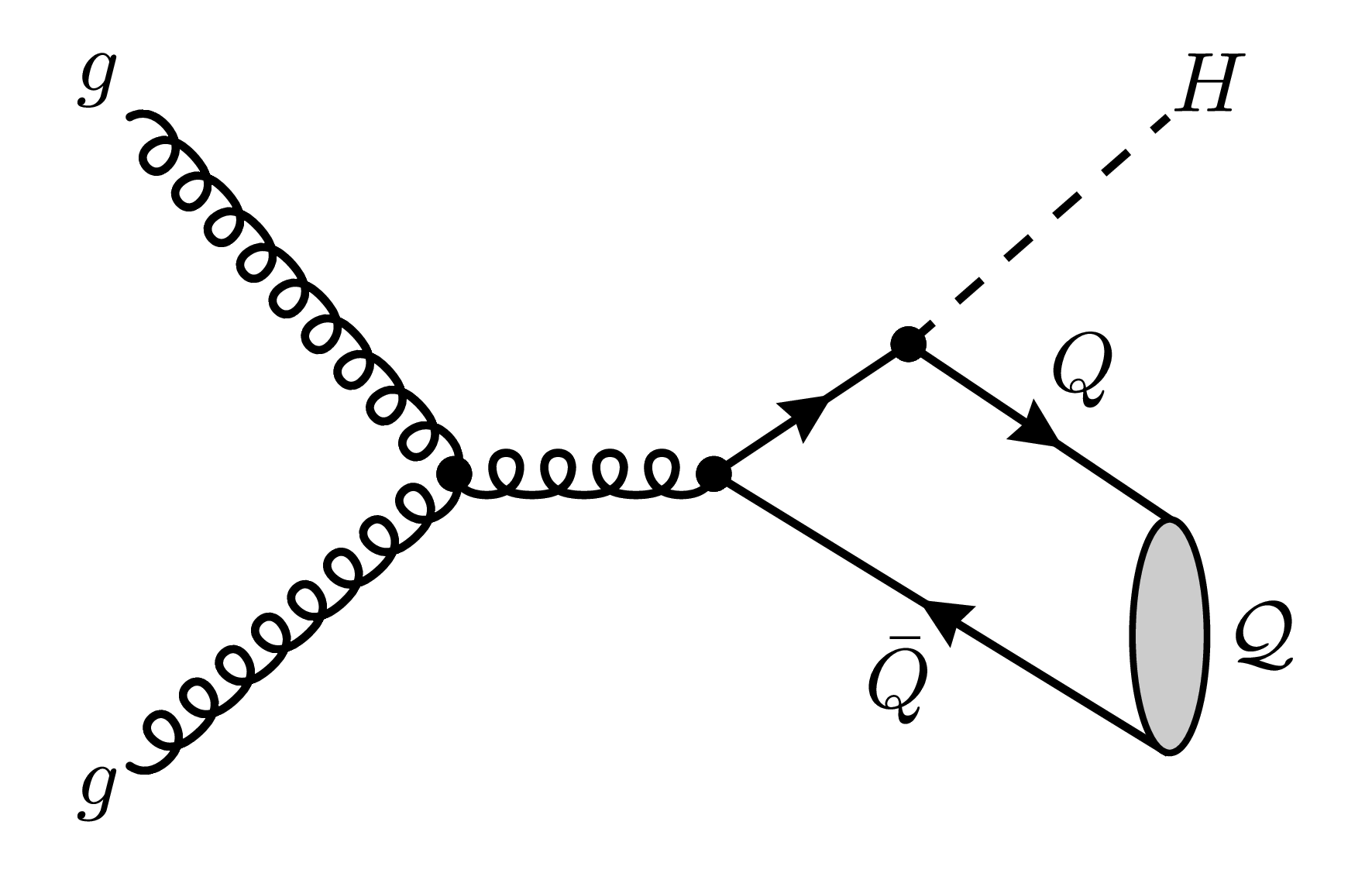}
        \caption{}
        \label{fig:diag_heft_a}
    \end{subfigure}\\
    \begin{subfigure}[b]{0.7\textwidth}
        \includegraphics[width=0.49\textwidth]{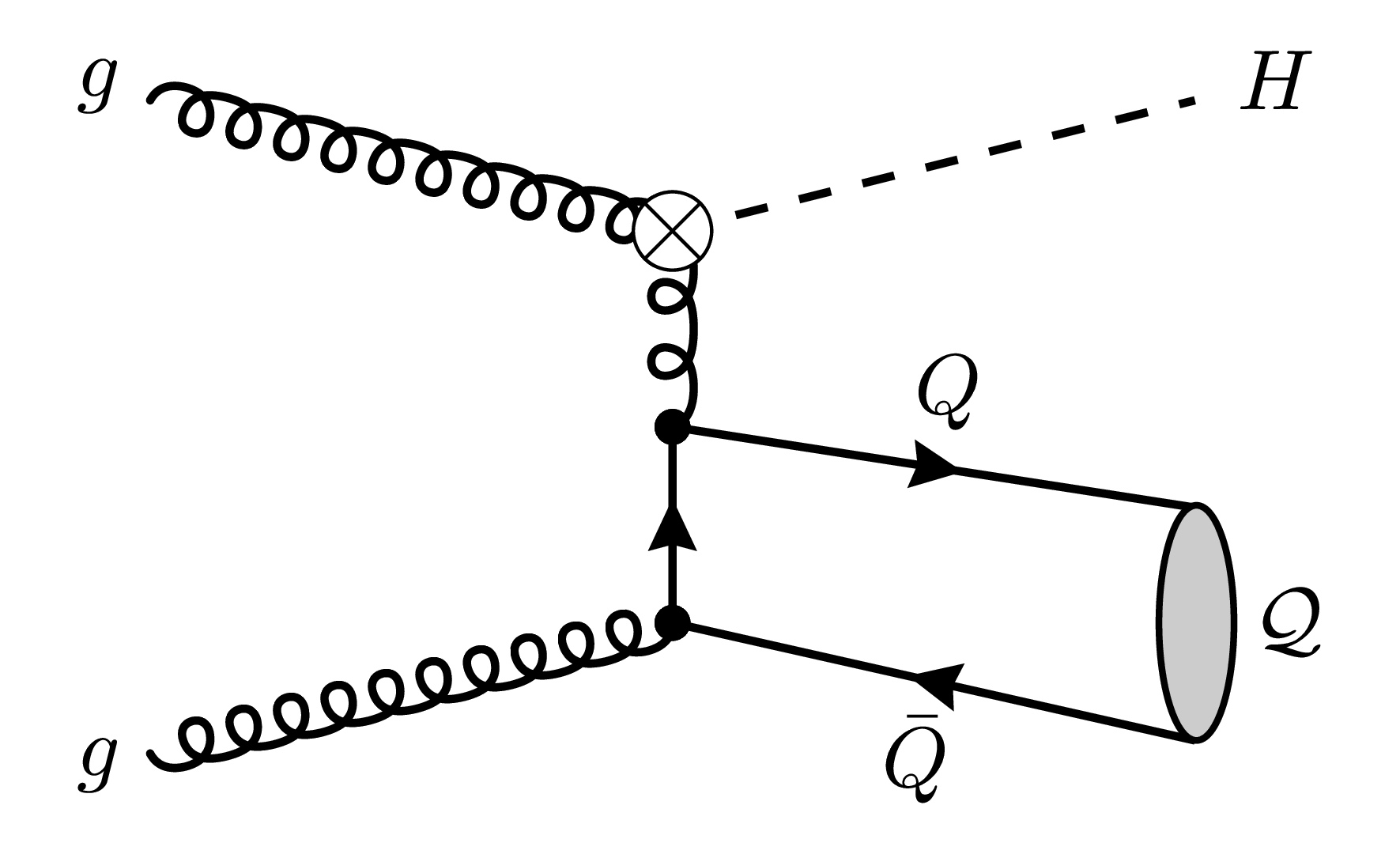}
        \includegraphics[width=0.49\textwidth]{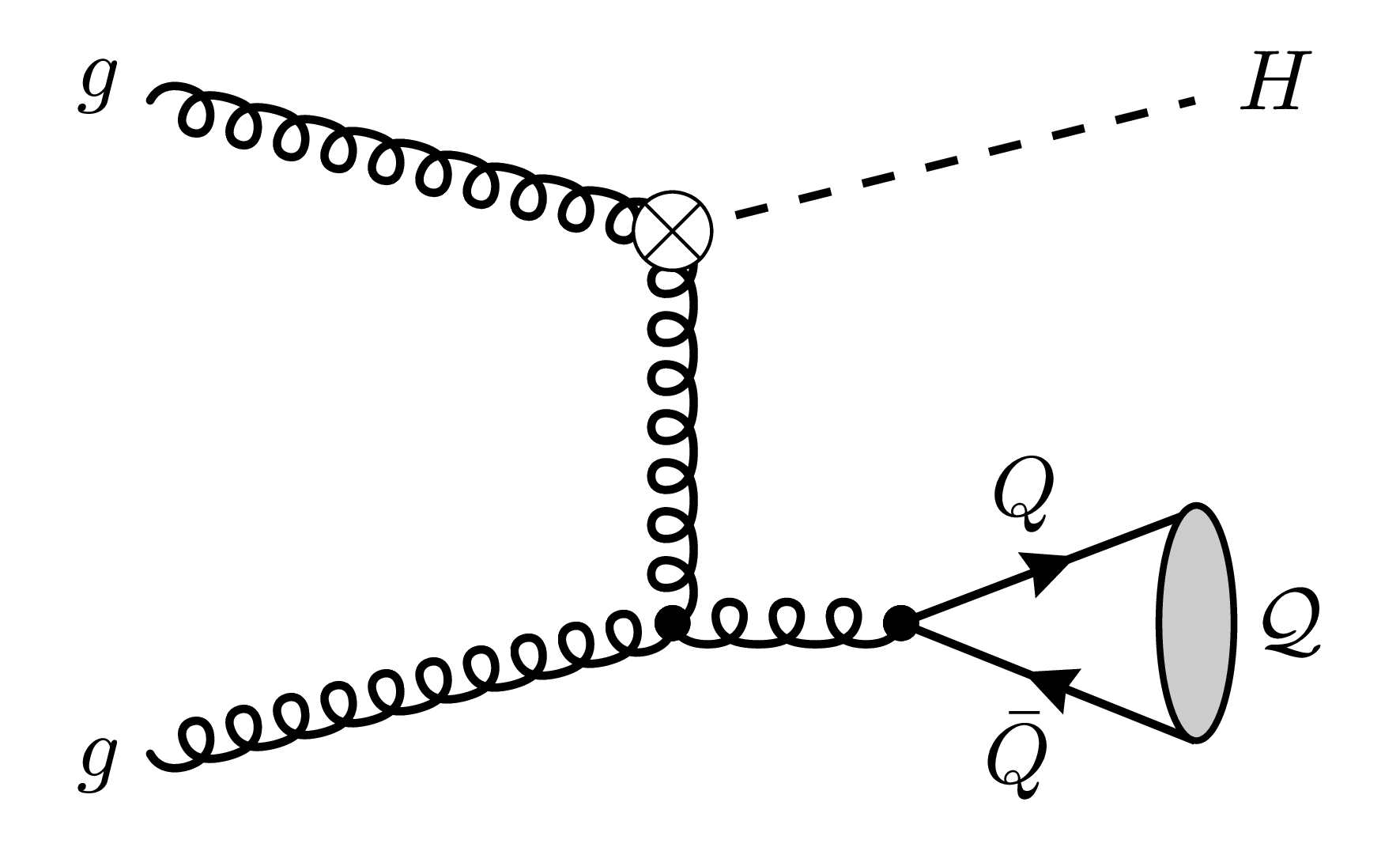}
        \caption{}
        \label{fig:diag_heft_b}
    \end{subfigure}
    \caption{Example Feynman diagrams: (a) the final-state Higgs radiates off a heavy-quark line, and (b) it radiates off a gluon via the effective operator in HEFT. The crossed vertex represents the effective coupling between two gluons and a Higgs boson, as given in Eq.~\eqref{eq:lagrangian_heft}.}
	\label{fig:diag_heft}
\end{figure}\noindent

As a unique feature, \mgshort\ can calculate quarkonium production not only in the SM but also in other theories, provided an appropriate \ufo\ model is available. This extends its capabilities beyond those of other frameworks such as \helaconia. To demonstrate this flexibility, we employ HEFT to compute the process $pp \to gg \to \mathcal Q + H$, where a Higgs boson $H$ is produced on-shell together with a heavy vector quarkonium $\mathcal Q\in\{J/\psi,\Upsilon\}$. At LO in the full SM, the final-state Higgs boson can be radiated off the heavy-quark line of the bound-state constituents (figure~\ref{fig:diag_heft_a}), or off a top-quark via a loop-induced contribution. These processes have been considered in the literature for $\mathcal Q\in\{J/\psi, \psi(2S)\}$~\cite{Kniehl:2002wd,Pan:2021nij}. In HEFT, the
top quark is integrated out under the assumption $m_t\gg m_H$, yielding an effective interaction between gluons and the Higgs particle:
\begin{equation}\label{eq:lagrangian_heft}
    \mathcal{L}_{\rm HEFT} \supset \frac{\alpha_s}{12 \pi v} H G^{a\mu \nu} G^a_{\mu \nu}\,,
\end{equation}
where $v$ is the vacuum expectation value of the Higgs field and $G^{a\mu \nu}$ is the non-Abelian gluon field-strength tensor. This effective vertex reduces the full one-loop gluon-gluon-Higgs interaction to a tree-level $H\!gg$ coupling (see figure~\ref{fig:diag_heft_b}), which can be directly treated with our quarkonium implementation together with a modified version of the \texttt{heft} model, dubbed \texttt{heft\_onia}, adapted to ensure compatibility with the newly introduced \ufo\ format in \mgshort.

To generate the Feynman diagrams in HEFT, we use the following prompts:

\vskip 0.25truecm
\noindent
~~\prompt\ {\tt ~import~model~heft\_onia}

\noindent
~~\prompt\ {\tt ~generate g g > Upsilon H}

\vskip 0.25truecm

\noindent
For the case $\mathcal Q=J/\psi$, we replace the bottom-quark parameters with the corresponding charm-quark values, updating both the mass and Yukawa coupling in the \texttt{param\_card.dat}, as well as the LDME values in \texttt{onia\_card.dat}.~\footnote{By default, the \texttt{heft} model in \mgshort\ includes a massive $b$ quark but not a massive $c$ quark. Since initial-state quark contributions are neglected for this process, these minor modifications are sufficient to obtain $J/\psi$ production for our purposes.} 

Using the $pp$ collision setup introduced in section~\ref{sec:setup}, we obtain the total cross sections
\begin{equation}
    \begin{aligned}
        \sigma(pp\to gg \to J/\psi+H) &= 1.53^{+0.40}_{-0.29}\,\mathrm{fb}\,,\\
        \sigma(pp\to gg \to \Upsilon+H) &= 91^{+23}_{-17}\,\mathrm{ab}\,.
    \end{aligned}
\end{equation}
The uncertainties are estimated via a 7-point scale variation with $\mu_{R/F}\in\{\mu_0/2,\mu_0,2\mu_0\}$ around the central scale $\mu_0=H_T/2$, enforcing $1/2\le\mu_{R}/\mu_{F}\le2$, with $H_T/2$ defined in Eq.~\eqref{eq:HTovertwo}. Only the colour-octet $^3S_1^{[8]}$ and $^1S_0^{[8]}$ Fock states contribute at LO; the $^3S_1^{[1]}$ state is forbidden by C-parity conservation. Colour-singlet contributions first appear at NLO via the real-emission process $pp \to gg \to \mathcal Q + H g$. Although this channel is na\"ively suppressed by an additional power of the strong coupling $\alpha_s$, the larger colour-singlet LDME partially compensates for this suppression. Indeed, we find
\begin{equation}
\begin{aligned}
    \sigma(pp \to gg \to J/\psi\!\left[^3S_1^{[1]}\right]+Hg) &= 50^{+21}_{-14}\,\mathrm{ab}\,,\\
    \sigma(pp \to gg \to \Upsilon\!\left[^3S_1^{[1]}\right]+Hg) &= 16.9^{+6.4}_{-4.5}\,\mathrm{ab}\,.
\end{aligned}
\end{equation}
Here, we observe that $\sigma(pp \to gg \to J/\psi\!\left[^3S_1^{[1]}\right]+Hg)$ is significantly suppressed compared to $\sigma(pp\to gg \to J/\psi+H)$, while this suppression is less pronounced in the $\Upsilon$ case. This behaviour arises for the same reason as in the quarkonium–$Z$ boson associated production discussed in section~\ref{sec:quarkoniumplusboson}. Namely, the propagator enhancement in $gg\to J/\psi\!\left[^3S_1^{[8]}\right]+H$ is stronger than in $gg\to \Upsilon\!\left[^3S_1^{[8]}\right]+H$ due to the lighter heavy-quark mass. Nevertheless, our current framework does not yet allow for a complete NLO computation, and therefore these partial NLO contributions are neglected in the following discussion.

\begin{figure}[t]
    \begin{subfigure}[b]{0.49\textwidth}
        \includegraphics[width=\textwidth]{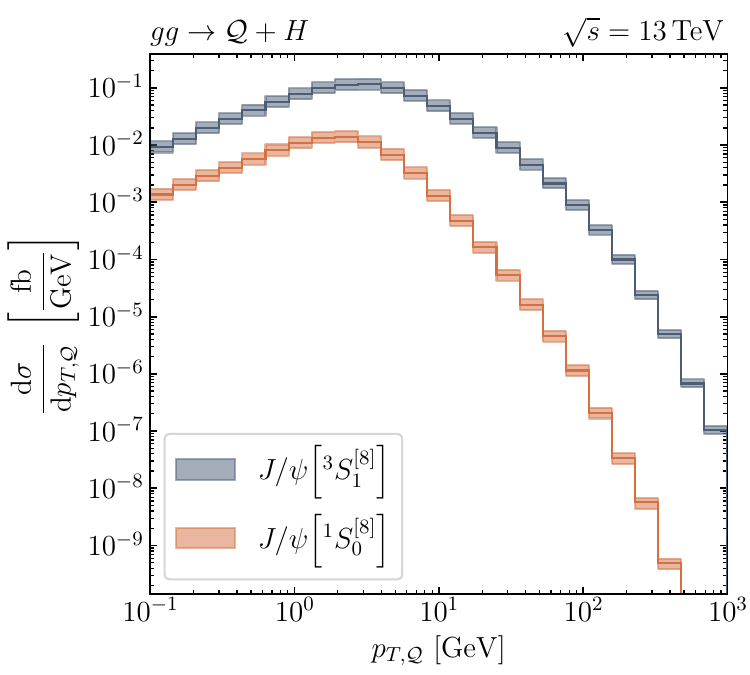}
        \caption{}
        \label{fig:heft_a}
    \end{subfigure}
    \begin{subfigure}[b]{0.49\textwidth}
        \includegraphics[width=\textwidth]{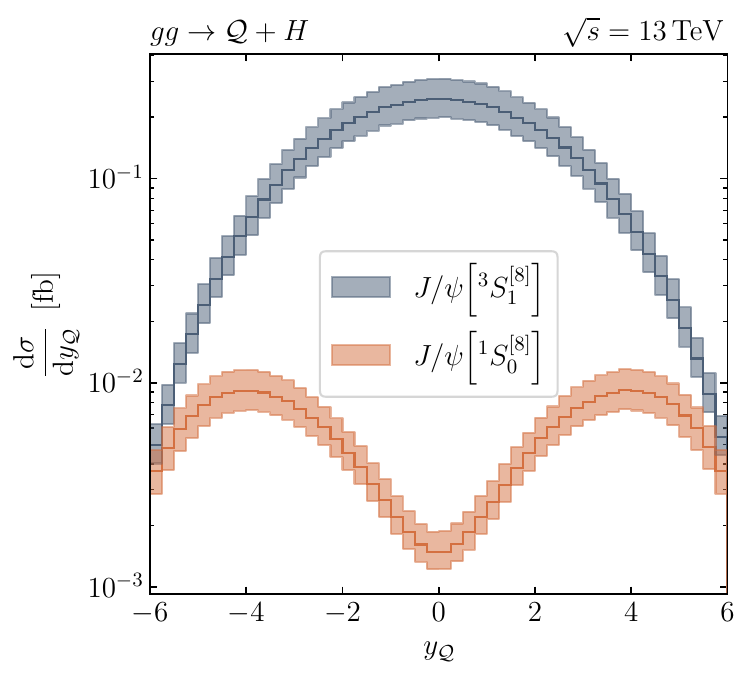}
        \caption{}
        \label{fig:heft_b}
    \end{subfigure}\\
    \begin{subfigure}[c]{0.49\textwidth}
        \includegraphics[width=\textwidth]{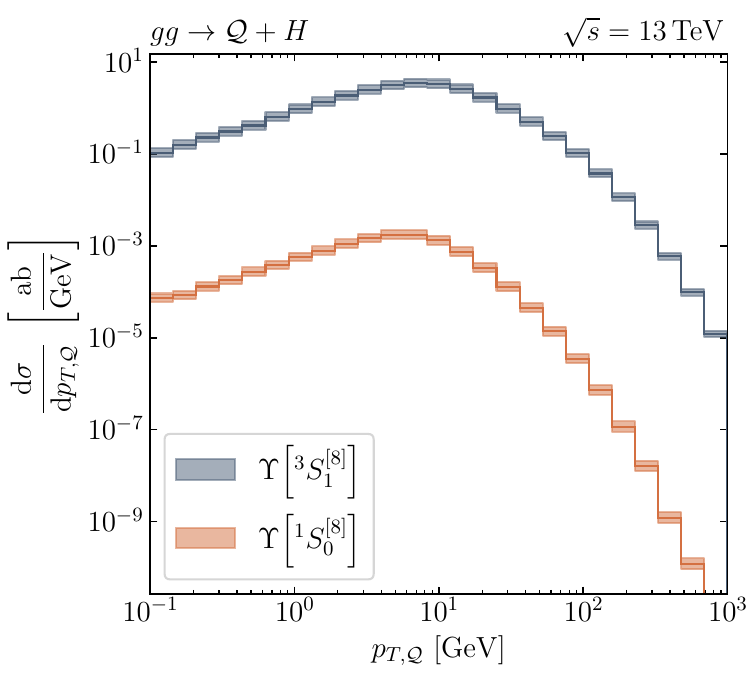}
        \caption{}
        \label{fig:heft_c}
    \end{subfigure}
    \begin{subfigure}[d]{0.49\textwidth}
        \includegraphics[width=\textwidth]{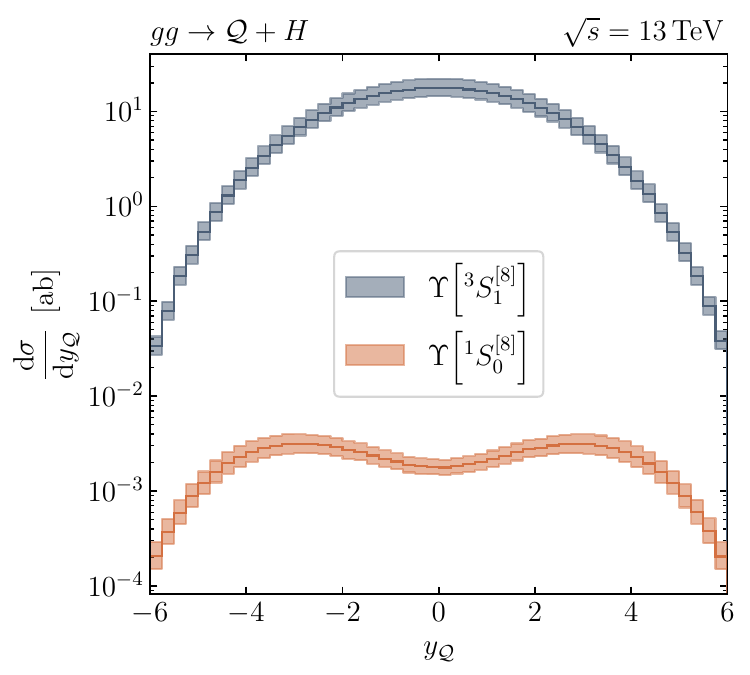}
        \caption{}
        \label{fig:heft_d}
    \end{subfigure}
    \caption{Transverse momentum ($p_{T,\mathcal Q}$) and rapidity ($y_{\mathcal Q}$) differential cross-section distributions for $J/\psi$ (top) and $\Upsilon$ (bottom) production in association with a Higgs boson $H$ in $pp$ collisions at $\sqrt{s}=13\,\mathrm{TeV}$.  The contributions from the $^1S_0^{[8]}$ and $^3S_1^{[8]}$ Fock states are shown in orange and blue, respectively, with the error bands representing the standard 7-point scale variation.}
	\label{fig:heft}
\end{figure}\noindent

In figure~\ref{fig:heft}, we show the cross sections differential in the transverse momentum $p_{T,\mathcal Q}$ and rapidity $y_{\mathcal Q}$  of the final-state quarkonium $\mathcal Q$. In both $J/\psi$ and $\Upsilon$ production, the $^3S_1^{[8]}$ contribution dominates across the entire kinematic range. The $p_{T,\mathcal{Q}}$ distributions of the $^1S_0^{[8]}$ and $^3S_1^{[8]}$ Fock states have similar shapes for both $J/\psi$ and $\Upsilon$ production (figures~\ref{fig:heft_a} and \ref{fig:heft_c}), whereas the $y_\mathcal{Q}$ spectra (figures~\ref{fig:heft_b} and \ref{fig:heft_d}) show pronounced differences. The $^3S_1^{[8]}$ contribution peaks at central rapidities, while the $^1S_0^{[8]}$ contribution exhibits a dip at mid-rapidity with symmetric peaks at forward and backward rapidities around $|y_\mathcal{Q}|\approx 4$ for $J/\psi$ and $|y_\mathcal{Q}|\approx 3$ for $\Upsilon$. The dip is much more prominent for $J/\psi$ than for $\Upsilon$. This behaviour can be traced back to the relative impact of different classes of Feynman diagrams. The rapidity dip in $^1S_0^{[8]}$ production arises from diagrams involving the effective $H\!gg$ vertex (left of figure~\ref{fig:diag_heft_b}), whereas other diagram classes (figure~\ref{fig:diag_heft}) peak at $y_\mathcal{Q}=0$. Diagrams with Higgs emission via a Yukawa coupling (left of figure~\ref{fig:diag_heft_a}) partially mitigate the dip. For $J/\psi$, the small charm-quark mass suppresses these contributions, leaving the dip largely intact. For $\Upsilon$, the larger bottom-quark Yukawa coupling enhances these diagrams, making both contributions comparable and significantly reducing the mid-rapidity dip.

\subsection{Quarkonium production in electron-proton collisions}
\label{sec:ep_collider}

At electron-proton colliders, inelastic quarkonium production can occur via either DIS or photoproduction processes. In the following two subsections, we consider both types of collisions and compute cross sections for charmonium production. We retain the same SM input parameters as discussed in section~\ref{sec:setup}, and here specify the relevant collider settings, which were not introduced previously.

For our simulations, we adopt the maximum c.m. energy configuration of the EIC~\cite{Accardi:2012qut,Boer:2024ylx},
with $\sqrt{s} = 140.7\,\mathrm{GeV}$, corresponding to beam energies of $E_\mathrm{beam}^{e^-} = 18\,\mathrm{GeV}$ and $E_\mathrm{beam}^p = 275\,\mathrm{GeV}$. For the proton beam, we use the \texttt{PDF4LHC21\_40} PDF set. The central renormalisation and factorisation scales are set to $\mu_R = \mu_F = H_T/2$, with $H_T/2$ defined in Eq.~\eqref{eq:HTovertwo}.

For recent phenomenological studies, the reader is guided to refs.~\cite{Artoisenet:2009xh,Butenschoen:2009zy,Sun:2017nly,Sun:2017wxk,Zhang:2019wxo,Qiu:2020xum,Flore:2020jau,ColpaniSerri:2021bla,Lansberg:2023kzf,Boer:2024ylx,Maxia:2024cjh}. The same processes have also received considerable attention in recent years in the study of TMD dynamics}~\cite{Godbole:2013bca,Mukherjee:2016qxa,Rajesh:2018qks,Bacchetta:2018ivt,DAlesio:2019qpk,Boer:2020bbd,Boer:2021ehu,Boer:2023zit,Kishore:2024bdd,Echevarria:2024idp,Copeland:2025vop,Echevarria:2025oab}.

\subsubsection{Quarkonium production in deep-inelastic scattering}

At HERA, $J/\psi$ production in $ep$ interactions has been extensively studied in the elastic and inelastic regime~\cite{H1:1996kyo,  ZEUS:1997wrc,  ZEUS:2002src, H1:2002voc, ZEUS:2009qug, H1:2010udv, ZEUS:2012qog,H1:2013okq} to gain a better understanding of  the quarkonium-production mechanism and to probe the gluon structure of the proton. To illustrate the capabilities of \mgshort\ with a simple example, we consider the DIS of an electron and a proton to produce a charmonium final state together with the measured scattered electron and a jet originating from the proton dissociation. The processes can be generated and run with the commands:
\vskip 0.25truecm
\noindent
~~\prompt\ {\tt ~import~model~sm\_onia-c\_mass}

\noindent
~~\prompt\ {\tt ~generate~e- p > etac e- j}

\noindent
~~\prompt\ {\tt ~output;~launch}
\vskip 0.25truecm
\noindent
for producing an $\eta_c$ meson. Replacing \texttt{etac} with \texttt{Jpsi} in the \texttt{generate} command yields $J/\psi$ production instead. Thanks to the use of the \textit{boundstates} syntax, all relevant S-wave Fock state contributions considered in this work are automatically included. To ensure infrared-safe results, we apply the following kinematic cuts: $p_{T,j}>5\,\mathrm{GeV}$ and $|\eta_j|<5$ for the jet, and $p_{T,e}>1\,\mathrm{GeV}$ and $|\eta_{e}|<3$ for the scattered electron, thereby avoiding contributions from on-shell internal photon exchange in the hard scattering. This results in the LO fiducial cross sections at $\mathcal{O}(\alpha_s^2\alpha^2)$
\begin{align}
    \sigma(e^-p\to \eta_c+e^-j+X) &= 394.8(1)\,\mathrm{fb}\,,\label{eq:DISetacjet}\\
    \sigma(e^-p\to J/\psi+e^-j+X) &= 2.5651(8)\,\mathrm{pb}\,,\label{eq:DISpsijet}
\end{align}
for $\eta_c$ and $J/\psi$ production, respectively. A breakdown into the individual partonic channels~\footnote{Channels with a light quark $q$ include the sum over all light quark and antiquark flavours.} gives the following contributions. For the pseudoscalar $\eta_c$:
{\allowdisplaybreaks
\begin{align}
    \sigma(e^-p\to e^-g\to \eta_c\!\left[{}^1S_0^{[1]}\right]+e^-g) &= 17.9(5)\,\mathrm{ab}\,,\\
    \sigma(e^-p\to e^-q\to \eta_c\!\left[{}^1S_0^{[1]}\right]+e^-q) &= 0\,,\\
    \sigma(e^-p\to e^-g\to \eta_c\!\left[{}^3S_1^{[8]}\right]+e^-g) &= 33.47(4)\,\mathrm{fb}\,,\\
    \sigma(e^-p\to e^-q\to \eta_c\!\left[{}^3S_1^{[8]}\right]+e^-q) &= 147.81(9)\,\mathrm{fb}\,,\\
    \sigma(e^-p\to e^-g\to \eta_c\!\left[{}^1S_0^{[8]}\right]+e^-g) &= 155.42(8)\,\mathrm{fb}\,,\\
    \sigma(e^-p\to e^-q\to \eta_c\!\left[{}^1S_0^{[8]}\right]+e^-q) &= 58.06(5)\,\mathrm{fb}\,,
\end{align}
}
and for the vector $J/\psi$:
{\allowdisplaybreaks
\begin{align}
    \sigma(e^-p\to e^-g\to J/\psi\!\left[{}^3S_1^{[1]}\right]+e^-g) &= 1.4171(6)\,\mathrm{pb}\,,\\
    \sigma(e^-p\to e^-q\to J/\psi\!\left[{}^3S_1^{[1]}\right]+e^-q) &= 0\,,\\
    \sigma(e^-p\to e^-g\to J/\psi\!\left[{}^1S_0^{[8]}\right]+e^-g) &= 754.3(4)\,\mathrm{fb}\,,\\
    \sigma(e^-p\to e^-q\to J/\psi\!\left[{}^1S_0^{[8]}\right]+e^-q) &= 281.5(3)\,\mathrm{fb}\,,\\
    \sigma(e^-p\to e^-g\to J/\psi\!\left[{}^3S_1^{[8]}\right]+e^-g) &= 20.65(6)\,\mathrm{fb}\,,\\
    \sigma(e^-p\to e^-q\to J/\psi\!\left[{}^3S_1^{[8]}\right]+e^-q) &= 91.5(2)\,\mathrm{fb}\,.
\end{align}
}
We note that the tiny cross section for the colour-singlet contribution in the gluon channel, $e^-g\to\eta_c\!\left[^1S_0^{[1]}\right]+e^-g$, arises from $Z$-boson exchange, since photon exchange is again forbidden by C-parity. In contrast, the cross section for $J/\psi$ production is dominated by the colour-singlet channel via $\gamma^*g\to J/\psi\left[^3S_1^{[1]}\right]+g$. For this reason, the $J/\psi$ cross section is much larger than that of the $\eta_c$, as shown in Eqs.~\eqref{eq:DISetacjet} and \eqref{eq:DISpsijet}.

\subsubsection{Photoproduction}

A second mechanism for quarkonium production in $ep$ collisions proceeds via photoproduction, where a photon and a parton annihilate into a quarkonium state. We model the initial-state photon flux, originating from electron splitting, using the improved Weizsäcker-Williams (iWW) approximation~\cite{Frixione:1993yw}. Adopting the iWW electron PDF requires specifying a maximum allowed photon virtuality, $Q^2_\mathrm{max}$, for the quasi-real incoming photon.~\footnote{This is implemented in \mgshort\ by setting the factorisation scale for the electron beam PDF to $Q_\mathrm{max}$ in the \texttt{run\_card.dat} file.} In our study, we set $Q_\mathrm{max}=1\,\mathrm{GeV}$.

We focus here on the production of an $\eta_c$ meson via photoproduction.~\footnote{We do not consider here the contributions from resolved photons which are discussed in ref.~\cite{Zhang:2019wxo} and which would require resorting to the asymmetric version of \mgshort~\cite{Flore:2025ync}. Along the same lines, we exclude processes involving light (anti)quarks since their total cross sections are infrared divergent and would require additional fiducial cuts.\label{fnt:photoproduction}} At LO, the three different S-wave Fock states that can be produced in $\gamma g$ collisions require a different number of final-state jets. The ${}^1S_0^{[8]}$ Fock state can be produced through the $2\to 1$ process 
\begin{equation}
    e^-p \to \gamma g \to \eta_c\!\left[{}^1S_0^{[8]}\right]\,,\label{eq:etac1S08photoproduction}
\end{equation}
while the production of a ${}^3S_1^{[8]}$ state requires one additional jet,
\begin{equation}
    e^-p \to \gamma g \to \eta_c\!\left[{}^3S_1^{[8]}\right]+g\,,
\end{equation}
and the colour-singlet state ${}^1S_0^{[1]}$ appears for the first time in association with two jets,
\begin{equation}
    e^-p \to \gamma g \to \eta_c\!\left[{}^1S_0^{[1]}\right]+gg\,.
\end{equation}
This can again be easily understood: the tree-level process $\gamma g\to \eta_c\!\left[{}^3S_1^{[8]}\right]$ is forbidden by the Landau-Yang theorem, while both $\gamma g \to \eta_c\!\left[{}^1S_0^{[1]}\right]$ and $\gamma g \to \eta_c\!\left[{}^1S_0^{[1]}\right]+g$ vanish because they violate colour and C-parity conservation, respectively.~\footnote{The situation is similar for other, more complex processes, such as $J/\psi\!\left[^3S_1^{[1]}\right]+J/\psi\!\left[^3S_1^{[1]}\right]$ production in $\gamma g$ collisions. The lowest-order contribution occurs at $\mathcal{O}(\alpha_s^5 \alpha)$ via $\gamma g \to J/\psi\!\left[^3S_1^{[1]}\right] + J/\psi\!\left[^3S_1^{[1]}\right] + gg$. In contrast, a non-zero contribution arises from the quark-initiated process $\gamma q \to J/\psi\!\left[^3S_1^{[1]}\right] + J/\psi\!\left[^3S_1^{[1]}\right] + q$ at $\mathcal{O}(\alpha_s^4 \alpha)$. However, without proper regularisation, the quark-initiated channel suffers from an initial-state QED collinear singularity, $\gamma \to q\bar{q}$, which generally mixes with the resolved-photon contribution at NLO.}

The first process in Eq.~\eqref{eq:etac1S08photoproduction}, for instance, can be generated with the prompts
\vskip 0.25truecm
\noindent
~~\prompt\ {\tt ~import~model~sm\_onia-c\_mass}

\noindent
~~\prompt\ {\tt ~generate~a g > etac(1|1S08)}

\noindent
~~\prompt\ {\tt ~output;~launch}
\vskip 0.25truecm
\noindent
We set \texttt{lpp1=3}, \texttt{pdlabel1=iww}, \texttt{fixed\_fac\_scale1=True}, and \texttt{dsqrt\_q2fact1=1.0} in the run card to indicate that the photon originates from the electron beam, parametrised using the iWW PDF with $Q_\mathrm{max}=1\,\mathrm{GeV}$. The total LO cross sections for these tree-level processes are
\begin{align}
    \sigma(e^-p \to \gamma g\to \eta_c\!\left[^1S_0^{[8]}\right]) &= 590.02(3)\,\mathrm{pb}\,,\\
    \sigma(e^-p \to \gamma g\to \eta_c\!\left[^3S_1^{[8]}\right]+g) &= 58.27(1)\,\mathrm{pb}\,,\\
    \sigma(e^-p \to \gamma g\to \eta_c\!\left[^1S_0^{[1]}\right]+gg) &= 12.342(5)\,\mathrm{pb}\,.
\end{align}

However, since individual Fock states are not physical, only the inclusive production of physical $\eta_c$ mesons can be compared with experimental measurements. A meaningful observable should therefore combine all relevant Fock states and account for processes involving light (anti)quarks.~\footnote{Of course, contributions from P-wave states and possible feeddown effects are still missing.} For instance, $\eta_c$ production in association with two jets includes the three subprocesses
\begin{align}
    e^-p \to \gamma p\to \eta_c\!\left[^1S_0^{[1]}\right]+jj+X\,,\\
    e^-p \to \gamma p\to \eta_c\!\left[^3S_1^{[8]}\right]+jj+X\,,\\
    e^-p \to \gamma p\to \eta_c\!\left[^1S_0^{[8]}\right]+jj+X\,.
\end{align}
To avoid infrared divergences from these processes, we apply fiducial cuts of $p_{T,j}>1\,\mathrm{GeV}$ and $|\eta_j|<5$. In addition, to ensure the two final-state jets are well separated, we impose a jet isolation criterion $\Delta R_{jj}>0.4$, where the angular distance is defined as
\begin{equation}\label{eq:Rjj_hadronic}
    \Delta R_{jj} = \sqrt{\Delta \eta_{jj}^2 + \Delta \phi_{jj}^2}\,,
\end{equation}
with $\Delta \eta_{jj}$ and $\Delta \phi_{jj}$ denoting the pseudorapidity and azimuthal-angle differences between the two jets, respectively. Under these conditions, we obtain the cross section
\begin{equation}
    \sigma(e^-p \to \gamma p\to \eta_c+jj+X)=163.28(6)\,\mathrm{pb}\,,
\end{equation}
which can be decomposed into the three contributions
\begin{align}
    \sigma(e^-p \to \gamma p\to \eta_c\!\left[^1S_0^{[1]}\right]+jj+X)&=26.24(3)\,\mathrm{pb}\,,\\
    \sigma(e^-p \to \gamma p\to \eta_c\!\left[^3S_1^{[8]}\right]+jj+X)&=17.25(2)\,\mathrm{pb}\,,\\
    \sigma(e^-p \to \gamma p\to \eta_c\!\left[^1S_0^{[8]}\right]+jj+X)&=119.80(5)\,\mathrm{pb}\,.
\end{align}

\subsection{Quarkonium production at electron-positron colliders}\label{sec:lepton_collider}

To complete the picture of quarkonium production across different collider types, we present in this section numbers for cross sections in $e^+e^-$ collisions. Measurements from the Belle and BaBar $B$ factories provide one of the most precise data on quarkonium production at $e^+e^-$ colliders, including both inclusive processes such as $e^+e^-\to J/\psi+X$~\cite{BaBar:2001lfi,Belle:2001lqi,Belle:2009bxr} and $e^+e^-\to J/\psi+c\bar{c}+X$~\cite{Belle:2002tfa,Belle:2009bxr}, and exclusive processes such as $e^+e^-\to J/\psi + \eta_c$~\cite{Belle:2002tfa,Belle:2004abn,BaBar:2005nic,Belle:2023gln} and $e^+e^-\to J/\psi + \gamma$~\cite{BaBar:2003waz}. In the following, we present examples of both inclusive and exclusive quarkonium production processes in sections~\ref{sec:eeincl} and~\ref{sec:eeexcl}, respectively.

\subsubsection{Inclusive quarkonium production\label{sec:eeincl}}

As a demonstration, we present results for inclusive quarkonium production processes in association with either a single jet or a jet pair,
\begin{align}
        e^+e^- &\to \mathcal{Q}+j+X\,,\label{eq:ee2Qj}\\
        e^+e^- &\to \mathcal{Q}+jj+X\,,\label{eq:ee2Qjj}
\end{align}
where $\mathcal{Q}$ denotes either a charmonium state ($\eta_c$ or $J/\psi$) or a bottomonium state ($\eta_b$ or $\Upsilon$). Here, $X$ represents any possible soft parton radiation arising during the hadronisation of the colour-singlet quarkonium $\mathcal{Q}$ from intermediate colour-octet states.

The baseline setup follows the configuration described in section~\ref{sec:setup}. Among these processes, the single-jet final state in Eq.~\eqref{eq:ee2Qj} is infrared safe and does not require any cuts. The di-jet processes, however, suffer from infrared divergences that must be regulated by suitable kinematic cuts. Such singularities arise in the di-jet process of Eq.~\eqref{eq:ee2Qjj} when one of the final-state massless partons becomes soft or when the two partons are emitted collinearly. These can be controlled by applying jet isolation cuts. Specifically, we require a minimal invariant mass for the jet pair, $m_{jj}>1\,\mathrm{GeV}$, and a minimum angular separation of $\Delta R_{jj}>0.4$, where the angular distance is defined in Eq.~\eqref{eq:Rjj_hadronic}.

The resulting cross sections at $\sqrt{s}=10.58\,\mathrm{GeV}$ in $e^+e^-$ collisions are summarised in table~\ref{tab:ee2Qj}. They correspond to $\mathcal{O}(\alpha_s\alpha^2)$ and $\mathcal{O}(\alpha_s^2\alpha^2)$ for the single-jet and di-jet cases, respectively. For single-jet processes, only the colour-octet ${}^1S_0^{[8]}$ state contributes, since the colour-singlet states and the ${}^3S_1^{[8]}$ state are forbidden by colour and C-parity conservation.~\footnote{The $(Q\bar{Q})\!\left[{}^3S_1^{[8]}\right]$ channels receive negligible contributions from $s$-channel $Z$-boson exchange diagrams, which violate C-parity.} Consequently, the spin-singlet to spin-triplet quarkonium cross-section ratios reflect the corresponding LDME ratios (cf.\@\xspace table~\ref{tab:ldme}). In contrast, for di-jet processes, the colour-singlet channel $e^+e^-\to \gamma^* \to J/\psi\!\left[^3S_1^{[1]}\right]+gg$ is allowed, while the processes $e^+e^-\to \gamma^* \to \eta_c\!\left[^1S_0^{[1]}\right]+gg$ and $e^+e^-\to \gamma^* \to \eta_c\!\left[^1S_0^{[1]}\right]+q\bar{q}$ remain forbidden by C-parity and colour conservation, respectively. In addition, a new colour-octet channel, $e^+e^-\to \gamma^* \to (c\bar{c})\!\left[^3S_1^{[8]}\right]+gg$, becomes accessible compared with the single-jet case. This makes the di-jet cross sections larger than the single-jet ones for both the charmonium and $\Upsilon$ cases. For bottomonium production, however, the cross sections are more strongly suppressed by phase-space effects in the di-jet case than in the single-jet case, since $\sqrt{s}\sim 2m_b$. This suppression causes the $e^+e^-\to \eta_b+jj+X$ cross section to remain smaller than that of $e^+e^-\to \eta_b+j+X$. In table~\ref{tab:ee2Qj}, effects from initial-state photon radiation~\cite{Shao:2014rwa,Gong:2019rpd}, QCD radiative corrections~\cite{Ma:2008gq,Gong:2009kp,Zhang:2009ym}, and relativistic corrections~\cite{He:2009uf,Jia:2009np} are not included, although they could be significant. Note that the bottomonium plus (di-)jet cross sections in table~\ref{tab:ee2Qj} are probably not reliable, as the c.m. energy is very close to the production threshold, where non-perturbative effects are not well controlled.

\begin{table}[t!]
\centering
\renewcommand*{\arraystretch}{1.4}
\begin{tabular}[t]{lc|lc}
\toprule
\textbf{process}& $\sigma$ & \textbf{process}& $\sigma$ \\
\midrule
$e^+e^- \to \eta_c+j+X$ & $33.652(2)\,\mathrm{fb}$ & $e^+e^- \to \eta_b+j+X$ & $1.9367(1)\,\mathrm{fb}$ \\
$e^+e^- \to J/\psi+j+X$ & $163.23(1)\,\mathrm{fb}$ & $e^+e^- \to \Upsilon+j+X$ & $34.002(2)\,\mathrm{ab}$ \\
$e^+e^- \to \eta_c+jj+X$ & $99.31(4)\,\mathrm{fb}$ & $e^+e^- \to \eta_b+jj+X$ & $15.029(3)\,\mathrm{ab}$ \\
$e^+e^- \to J/\psi+jj+X$ & $606.9(3)\,\mathrm{fb}$ & $e^+e^- \to \Upsilon+jj+X$ & $1.1239(2)\,\mathrm{fb}$ \\
\bottomrule
\end{tabular}
\caption{Total (fiducial) cross sections for charmonium and bottomonium production in association with a single jet (or a jet pair) at $\sqrt{s}=10.58$ GeV, based on the setup described in section~\ref{sec:setup}. For di-jet processes, the fiducial cuts applied are $m_{jj}>1\,\mathrm{GeV}$ and $\Delta R_{jj}>0.4$. The numbers in parentheses indicate the numerical uncertainties from the MC phase-space integration. These results are provided for illustration only: they do not include feeddown effects and are specific to the chosen LDME values (cf.\@\xspace table \ref{tab:ldme}).}
\label{tab:ee2Qj}
\end{table}

\subsubsection{Exclusive quarkonium production \label{sec:eeexcl}}

As a showcase, we now consider the following exclusive quarkonium production processes:
\begin{align}
        e^+e^- &\to \mathcal{Q}+\gamma\,,\label{eq:ee2Qa}\\
        e^+e^- &\to \mathcal{Q}+\gamma\gamma\,,\label{eq:ee2Qaa}\\
        e^+e^- &\to \eta_c+J/\psi\,,\label{eq:ee2Jpsietac}\\
        e^+e^- &\to J/\psi+J/\psi\,,\label{eq:ee2JpsiJpsi}
\end{align}
where, as before, $\mathcal{Q}$ denotes either a charmonium state ($\eta_c$ or $J/\psi$) or a bottomonium state ($\eta_b$ or $\Upsilon$). Being exclusive processes, only colour-singlet channels contribute; otherwise, hadronisation of colour-octet states would produce additional light hadrons in the final state.

We adopt the baseline setup described in section~\ref{sec:setup}. For the processes with final-state photons in Eqs.~\eqref{eq:ee2Qa} and \eqref{eq:ee2Qaa}, infrared divergences may arise when photons are emitted from the incoming electron or positron. These singularities are regulated by imposing photon cuts of $p_{T,\gamma}>p_{T,\gamma}^{\mathrm{min}}=0.5\,\mathrm{GeV}$ and $|\eta_\gamma|<5$. To better mimic realistic detector conditions, where photons must be spatially separated, we further apply an isolation requirement of $\Delta R_{\gamma\gamma}>0.4$ for di-photon final states. The angular distance is defined analogously to Eq.~\eqref{eq:Rjj_hadronic} as
\begin{equation}\label{eq:Rgammagamma_leptonic}
\Delta R_{\gamma\gamma} = \sqrt{\Delta \eta_{\gamma\gamma}^2 + \Delta \phi_{\gamma\gamma}^2}\,,
\end{equation}
where $\Delta \eta_{\gamma\gamma}$ and $\Delta \phi_{\gamma\gamma}$ denote the differences in pseudorapidities and azimuthal angles between the two photons, respectively.

\begin{table}[t!]
\centering
\renewcommand*{\arraystretch}{1.4}
\begin{tabular}[t]{lc|lc}
\toprule
\textbf{process}& $\sigma$ & \textbf{process}& $\sigma$ \\
\midrule
$e^+e^- \to \eta_c+\gamma$ & $62.728(4)\,\mathrm{fb}$ & $e^+e^- \to \eta_b+\gamma$ & $2.0417(2)\,\mathrm{fb}$ \\
$e^+e^- \to J/\psi+\gamma$ & $15.272(6)\,\mathrm{pb}$ & $e^+e^- \to \Upsilon+\gamma$ & $4.2893(8)\,\mathrm{pb}$ \\
$e^+e^- \to \eta_c+\gamma\gamma$ & $5.480(3)\,\mathrm{fb}$ & $e^+e^- \to \eta_b+\gamma\gamma$ & $496.4(5)\,\mathrm{zb}$ \\
$e^+e^- \to J/\psi+\gamma\gamma$ & $413.1(2)\,\mathrm{fb}$ & $e^+e^- \to \Upsilon+\gamma\gamma$ & $1.711(2)\,\mathrm{fb}$ \\ \midrule
$e^+e^- \to J/\psi+J/\psi$ & $4.942(1)\,\mathrm{fb}$ & $e^+e^- \to \eta_c+J/\psi$ & $4.0722(3)\,\mathrm{fb}$ \\
\bottomrule
\end{tabular}
\caption{Fiducial cross sections for charmonium and bottomonium production in association with a single photon or a photon pair, and total cross sections for charmonium-pair production, are presented at $\sqrt{s}=10.58$ GeV, based on the setup described in section~\ref{sec:setup}. All photons are required to satisfy $p_{T,\gamma}>0.5\,\mathrm{GeV}$ and $|\eta_\gamma|<5$. For di-photon processes, an additional isolation cut of $\Delta R_{\gamma\gamma}>0.4$ is applied. The numbers in parentheses indicate the numerical uncertainties from the MC phase-space integration.}
\label{tab:ee2Qa}
\end{table}

The resulting cross sections are summarised in table~\ref{tab:ee2Qa}. For single-quarkonium processes, they are of orders $\mathcal{O}(\alpha^3)$ and $\mathcal{O}(\alpha^4)$ for single- and double-photon final states, respectively. The cross section for $e^+e^-\to \eta_c+\gamma$ is much smaller than that for $e^+e^-\to J/\psi+\gamma$, as can be understood from the analytic expressions given in Eqs.~(5) and (13) of ref.~\cite{dEnterria:2023yao}. The former is suppressed by a factor of $8m_c^2/3s\approx 0.057$, while the latter is enhanced by $\log{\left(s/(p_{T,\gamma}^{\mathrm{min}})^2\right)}\approx 6.1$. We expect similar suppression and enhancement mechanisms to apply to the double-photon processes $e^+e^-\to \eta_c+\gamma\gamma$ and $e^+e^-\to J/\psi+\gamma\gamma$ shown in table~\ref{tab:ee2Qa}. 

For bottomonium production, since $\sqrt{s}\sim 2m_b$, the cross section for $e^+e^-\to \eta_b+\gamma$ is suppressed by a factor of $\tfrac{2}{3}\left(1-4m_b^2/s\right)\approx 0.14$, whereas that for $e^+e^-\to \Upsilon+\gamma$ is enhanced by a factor of $\left(s/(p_{T,\gamma}^{\mathrm{min}})^2\right)^{5/4}\!\left(1-4m_b^2/s\right)^{3/2}\approx 200$. The bottomonium cross sections are therefore highly sensitive to the bottom-quark mass $m_b$, and the strong phase-space suppression leads to extremely small cross sections for bottomonium plus di-photon production.

For double-charmonium exclusive production, the process $e^+e^-\to \gamma^*\to J/\psi+J/\psi$ vanishes in both QCD and QED because of C-parity conservation.~\footnote{The $Z$-boson contribution is also included. It appears at $\mathcal{O}(\alpha^2\alpha_s^2)$ but is negligible.} Therefore, the leading contribution arises from the double-fragmentation mechanism, $e^+e^-\to \gamma^*(\to J/\psi)\,\gamma^*(\to J/\psi)$, which occurs at $\mathcal{O}(\alpha^4)$. The total LO cross section presented in table~\ref{tab:ee2Qa} at $\mathcal{O}(\alpha^4)$ is qualitatively consistent with that in table 4 of ref.~\cite{Shao:2012iz}, despite the different setup. In contrast, the process $e^+e^-\to \gamma^*\to \eta_c+J/\psi$ is non-zero at $\mathcal{O}(\alpha_s^2\alpha^2)$. Its cross section in table~\ref{tab:ee2Qa} also agrees well with table 3 of ref.~\cite{Shao:2012iz} when adopting the same setup. Large QCD corrections, positive for $e^+e^-\to \eta_c+J/\psi$~\cite{Zhang:2005cha,Gong:2007db,Feng:2019zmt,Huang:2022dfw,Chen:2025qgy} and negative for $e^+e^-\to J/\psi+J/\psi$~\cite{Gong:2008ce,Sang:2023liy,Huang:2023pmn}, are known in the literature but are not included here.

%% file: 07_leptonium.tex
\section{Leptonium production}\label{sec:leptonium}

Beyond quarkonium production, \mgshort\ now also supports NRQED bound states of leptons. These leptonic ``atoms'' can be computed within the factorisation framework as outlined in section~\ref{sec:theory}. In the following, we investigate the production of three purely leptonic bound states -- positronium ($e^+e^-$), true muonium ($\mu^+\mu^-$), and ditauonium ($\tau^+\tau^-$) -- at both $e^+e^-$ and $pp$ colliders. Although we restrict our discussion here to leptonia composed of same-flavour lepton–antilepton pairs, we emphasise that the framework can equally accommodate mixed-flavour bound states, such as muonium ($\mu^\pm e^\mp$), tauonium ($\tau^\pm e^\mp$), and mu–tauonium ($\tau^\pm\mu^\mp$).

By default, in the \texttt{sm\_onia} \ufo\ model, the electroweak coupling constant $\alpha$ is computed in the $G_\mu$ scheme, such that the effective electroweak coupling constant is fixed to
\begin{equation}
    \alpha_{G_\mu} = \dfrac{1}{132.507}\,,
\end{equation}
as introduced in section~\ref{sec:setup}. This scheme is well suited for processes involving $W$ or $Z$ bosons, or for describing interactions relevant to higher-order electroweak corrections~\cite{Denner:2019vbn,Shao:2025fwd}. However, in case of leptonium production, where the dynamics are typically governed by quasi-real photons with small virtuality, the $G_\mu$ scheme leads to artificially enhanced higher-order electroweak corrections. This arises from a ``misplaced'' running of $\alpha(\mu_R)$ from the electroweak scale down to the bound-state scale~\cite{Dittmaier:2025ikh}. To avoid such artefacts, it is more appropriate to employ a hybrid renormalisation scheme, using the fine-structure constant in the Thomson limit,
\begin{equation}
    \alpha(0) = \dfrac{1}{137.036}\,,
\end{equation}
for all couplings to quasi-real photons, while retaining the $G_\mu$ scheme for the remaining electroweak interactions. A similar hybrid renormalisation approach has been adopted within the \mgshort\ framework in studies involving tagged final-state photons~\cite{Pagani:2021iwa} and coherent initial-state photons in ultraperipheral collisions~\cite{Shao:2025bma} at NLO. To correct the mismatch between the default $G_\mu$ scheme used in the model setup and the physically more appropriate $\alpha(0)$ scheme for leptonium, we apply a global rescaling factor $(\alpha(0)/\alpha_{G_\mu})^{n_\gamma}$ to the computed cross sections, where the exponent $n_\gamma$ corresponds to the total number of vertices connected to quasi-real photons in the given process.~\footnote{This global rescaling is valid only at LO, where all relevant diagrams share the same power $n_\gamma$.}

\subsection{Positronium production}

Positronium ($\mathit{Ps}$), the simplest purely leptonic bound state composed of an electron and a positron, was first predicted theoretically by Mohorovičić in 1934~\cite{Mohorovicic1934} and later observed experimentally by Deutsch in 1951~\cite{PhysRev.82.455}. Remarkably, even modern experimental setups used to investigate positronium properties remain conceptually similar to those employed in its original discovery. However, direct observations of positronium produced in high-energy particle collisions are still lacking. Nonetheless, several theoretical studies have explored positronium production mechanisms at electron–ion and hadron colliders~\cite{Holvik:1986ty,Gevorkian:1998xj,Francener:2024eep,Meledin:1971fe,Kotkin:1998hu,Yu:2022hdt,refId0,dEnterria:2025ecx}. These studies suggest that a large number of positronia could be produced at facilities such as RHIC or the LHC, though their detection remains extremely challenging. The positronium large Bohr radius, approximately $1\,\text{\AA}$, makes it particularly susceptible to ionisation or dissociation in strong magnetic fields of detectors or through interactions with detector material effects that would likely destroy the bound state before it can be observed.

Nonetheless, to demonstrate the theoretical capabilities of our framework, we present in this section several cross sections for positronium production in $e^+e^-$ and $pp$ collisions. Here, we restrict ourselves to ground-state positronium with the principal number $N=1$, although \mgshort\ also supports calculations for excited leptonia with $N>1$. Depending on the relative spin orientation of the electron and positron, two distinct ground-state configurations can form. Para-positronium ($\mathit{Ps}_0$) with total angular momentum $J=0$ (corresponding to the Fock state $n={}^1S_0$) arises from antialigned spins, where the spin vectors point in opposite directions. Conversely, if the spins are aligned parallel, ortho-positronium ($\mathit{Ps}_1$) with total angular momentum $J=1$ (corresponding to the Fock state $n={}^3S_1$) is formed.

\begin{table}[t!]
\centering
\renewcommand*{\arraystretch}{1.4}
\begin{tabular}[t]{lc|lc}
\toprule
\textbf{process}& $\sigma$ & \textbf{process}& $\sigma$ \\
\midrule
& & $pp \to \mathit{Ps}_{1}+j+X$ & $9.483(2)\,\mathrm{fb}$ \\ \midrule
$e^+e^-\to\mathit{Ps}_0+\gamma$ & $3.300(1)\,\mathrm{ab}$ & $e^+e^- \to \mathit{Ps}_{1}+\gamma$ & $8.420(2)\,\mathrm{ab}$ \\
\bottomrule
\end{tabular}
\caption{Cross-section results for para- ($\mathit{Ps}_0$) and ortho-positronium ($\mathit{Ps}_1$) production at $\sqrt{s}=13$ TeV in $pp$ collisions and $\sqrt{s}=10.58\,\mathrm{GeV}$ in $e^+e^-$ collisions, based on the setup described in section~\ref{sec:setup}. Additional fiducial cuts of $p_{T,j}>2\,\mathrm{GeV}$ and $|\eta_j|<5$ ($p_{T,\gamma}>0.5\,\mathrm{GeV}$ and $|\eta_\gamma|<5$) are applied to final-state jets (photons) in the computations for ortho-positronium production. All cross sections are rescaled according to the $\alpha(0)$ scheme. The numbers in parentheses indicate the numerical uncertainties from the MC phase-space integration.}
\label{tab:positronium}
\end{table}

\begin{figure}[t]
    \begin{subfigure}[b]{0.5\textwidth}
        \includegraphics[width=\textwidth]{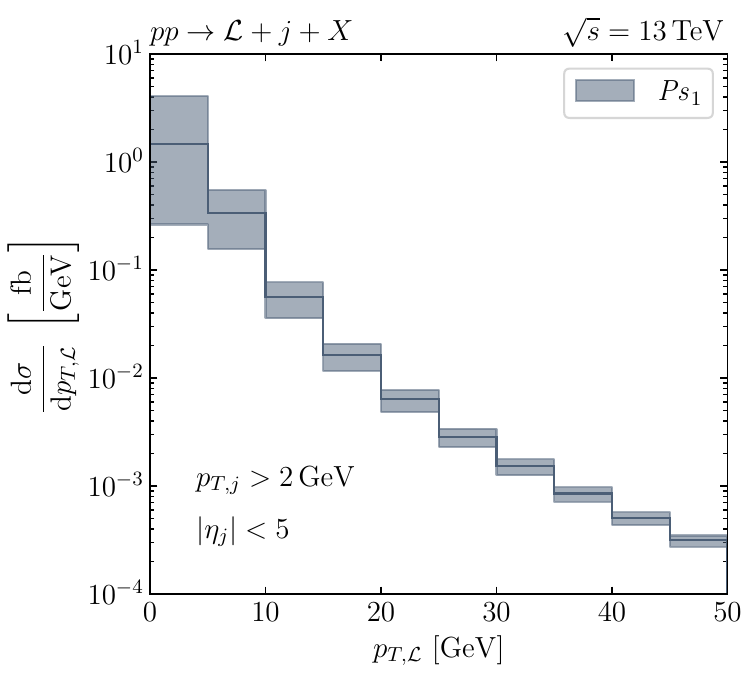}
        \caption{}
        \label{fig:positronium}
    \end{subfigure}
    \begin{subfigure}[b]{0.5\textwidth}
        \includegraphics[width=\textwidth]{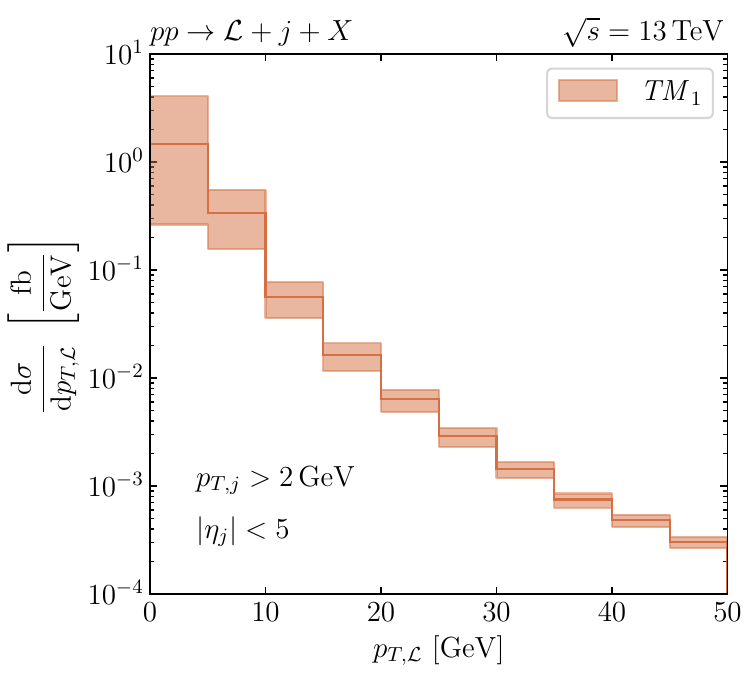}
        \caption{}
        \label{fig:truemuonium}
    \end{subfigure}\\
    \begin{subfigure}[c]{0.5\textwidth}
        \includegraphics[width=\textwidth]{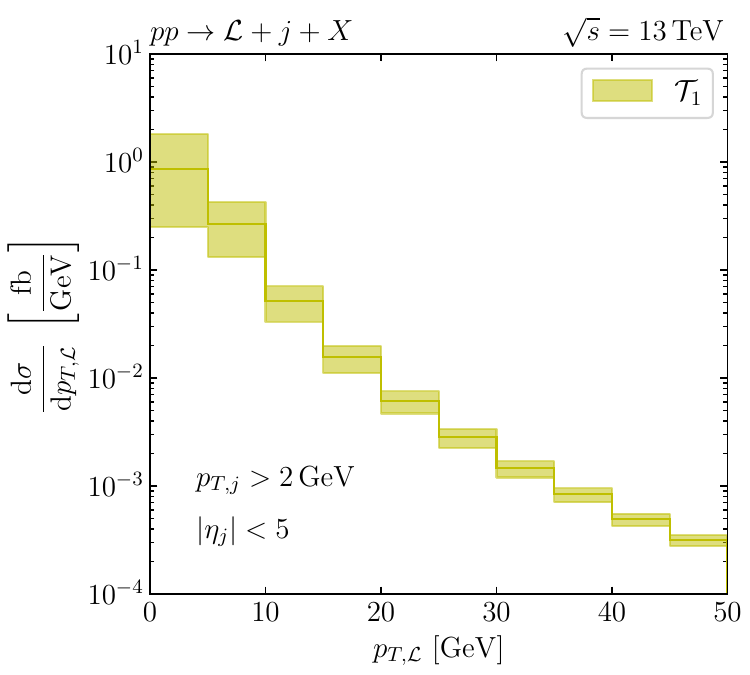}
        \caption{}
        \label{fig:ditauonium}
    \end{subfigure}
    \begin{subfigure}[c]{0.5\textwidth}
        \includegraphics[width=\textwidth]{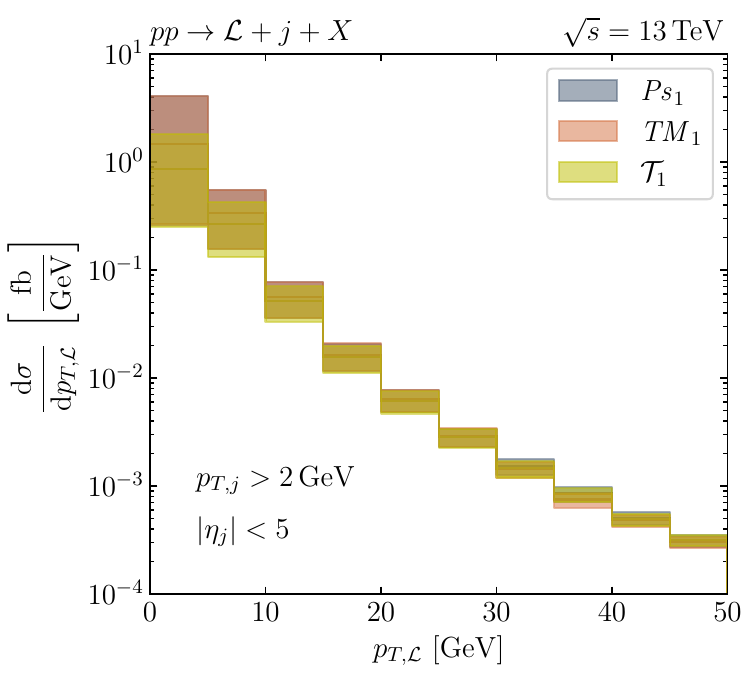}
        \caption{}
        \label{fig:leptonium_all}
    \end{subfigure}
    \caption{Differential cross sections with respect to the leptonium transverse momentum $p_{T,\mathcal{L}}$ at $\sqrt{s}=13$ TeV, shown for (a)~$\mathcal{L}=\mathit{Ps}_1$, (b)~$\mathcal{L}=\mathit{TM}_1$, (c)~$\mathcal{L}=\mathcal{T}_1$, and (d)~all three states combined. Fiducial jet cuts of $p_{T,j}>2\,\mathrm{GeV}$ and $|\eta_j|<5$ are applied. The error bands represent the standard 7-point renormalisation and factorisation scale variations.}
    \label{fig:leptionium}
\end{figure}

Ortho-positronium can, in principle, be produced in the $2\to1$ process~\footnote{For the process $pp\to\mathit{Ps}_1+X$, the factorisation scale is set to $\mu_F=1.4001\,\mathrm{GeV}$, which corresponds to the lowest scale permitted by the chosen PDF set. We emphasise that these studies are carried out purely for technical interest. At such low energy scales (around $2m_e\sim 1$ MeV), the assumption of massless initial-state partons becomes unreliable, and the applicability of the factorisation theorem is therefore questionable in this regime.}
\begin{equation}
    pp\to\mathit{Ps}_1+X\,,\label{eq:ops}
\end{equation}
but owing to the small invariant mass of the final state, this process cannot be reliably described within the standard factorisation framework. Nonetheless, investigating this process within our setup provides a valuable benchmark for testing the numerical stability of the computation at very low energy scales. We have verified our results against an independent implementation based on the analytic expression of the squared amplitude and found excellent agreement within the MC uncertainties. The physically more relevant production mechanism for ortho-positronium at hadron colliders is its associated production with a jet,
\begin{equation}
    pp\to \mathit{Ps}_1+j+X\,,\label{eq:ops_jet}
\end{equation}
where we impose fiducial jet cuts of $p_{T,j}>2\,\mathrm{GeV}$ and $|\eta_j|<5$ in addition to the common setup to ensure infrared safety. We obtain a cross section of $9.483(2)\,\mathrm{fb}$, which might make experimental detection conceivable. The corresponding $p_{T,\mathit{Ps}_1}$ distribution is shown in figure~\ref{fig:positronium}, where the uncertainty band represents a 7-point scale variation with $\mu_{R/F}\in\{\mu_0/2,\mu_0,2\mu_0\}$ around the central scale $\mu_0=H_T/2$, obeying $1/2\le\mu_{R}/\mu_{F}\le2$.

\noindent

Lepton colliders, in contrast, provide a cleaner environment and are better suited for studying positronium production in relativistic collisions. Using the $e^+e^-$ setup described in section~\ref{sec:setup}, we investigate positronium production at $\sqrt{s}=10.58\,\mathrm{GeV}$. Both para- and ortho-positronium can be produced in association with a photon. The total inclusive cross section for the para-positronium process
\begin{equation}\label{eq:ee2Ps0a}
    e^+e^-\to\mathit{Ps}_0+\gamma    
\end{equation}
is found to be $3.300(1)\,\mathrm{ab}$. For ortho-positronium production,
\begin{equation}\label{eq:ee2Ps1a}
    e^+e^-\to\mathit{Ps}_1+\gamma\,,
\end{equation}
we apply additional photon cuts of $p_{T,\gamma}>0.5\,\mathrm{GeV}$ and $|\eta_\gamma|<5$ to maintain consistency with the setup used in subsequent sections. The cross section obtained with these settings is $8.420(2)\,\mathrm{ab}$. Given the high luminosities of modern $e^+e^-$ colliders, these rates for para- and ortho-positronium may offer promising prospects for experimental observation.

Before turning to other leptonia, we analyse the numerical stability and accuracy of the values obtained with \mgshort\ by comparing them to analytic computations in the decoupling limit of the $Z$ and Higgs bosons. In this limit, the para-positronium cross section for the process in Eq.~\eqref{eq:ee2Ps0a} reduces to the expression
\begin{equation}
    \begin{aligned}
        &\quad\lim_{m_Z,m_H\to\infty}\sigma(e^+e^-\to\mathit{Ps}_0+\gamma) \\&= \dfrac{2}{3}\dfrac{\pi\alpha^6(0)}{N^3}\dfrac{s^3}{(s-4m_e^2)^4}\left\lbrace\sqrt{1-\dfrac{4m_e^2}{s}}\left[3+\dfrac{7}{2}\!\left(\dfrac{4m_e^2}{s}\right)-\dfrac{45}{2}\!\left(\dfrac{4m_e^2}{s}\right)^{\!2}\!+\dfrac{47}{2}\!\left(\dfrac{4m_e^2}{s}\right)^{\!3}\right.\right.\\
        &\quad\left.\left.-10\left(\dfrac{4m_e^2}{s}\right)^{\!4}\!-\left(\dfrac{4m_e^2}{s}\right)^{\!5}\!+\dfrac{1}{2}\!\left(\dfrac{4m_e^2}{s}\right)^{\!6}\right]+3\left(\dfrac{4m_e^2}{s}\right)\!\left[1-\dfrac{11}{4}\!\left(\dfrac{4m_e^2}{s}\right)+\left(\dfrac{4m_e^2}{s}\right)^{\!2}\right.\right.\\
        &\quad\left.\left.+\dfrac{3}{4}\!\left(\dfrac{4m_e^2}{s}\right)^{\!3}\!-\dfrac{1}{2}\!\left(\dfrac{4m_e^2}{s}\right)^{\!4}\right]\log\!\left(\dfrac{\sqrt{s}-\sqrt{s-4m_e^2}}{\sqrt{s}+\sqrt{s-4m_e^2}}\right)\right\rbrace \overset{s\gg m_e^2}{\longrightarrow} \frac{\pi \alpha^6(0)}{N^3}\frac{2}{s}\,.\label{eq:eexs4parapositronium}
    \end{aligned}
\end{equation}
Similarly, for ortho-positronium production as in Eq.~\eqref{eq:ee2Ps1a}, we have
\begin{equation}
    \begin{aligned}
        &\quad\lim_{m_Z,m_H\to\infty}\sigma(e^+e^-\to\mathit{Ps}_1+\gamma) \\&= \dfrac{\pi\alpha^6(0)}{N^3}\dfrac{s^3}{(s-4m_e^2)^4}\left\lbrace\sqrt{1-\dfrac{4m_e^2}{s}}\left[9-64\left(\dfrac{4m_e^2}{s}\right)+94\left(\dfrac{4m_e^2}{s}\right)^{\!2}-44\left(\dfrac{4m_e^2}{s}\right)^{\!3}\right.\right.\\
        &\quad\left.\left.-\left(\dfrac{4m_e^2}{s}\right)^{\!4}\right]-\left[1+17\left(\dfrac{4m_e^2}{s}\right)-48\left(\dfrac{4m_e^2}{s}\right)^{\!2}+54\left(\dfrac{4m_e^2}{s}\right)^{\!3}-21\left(\dfrac{4m_e^2}{s}\right)^{\!4}\right]\right.\\
        &\quad\left.\times\log\!\left(\dfrac{\sqrt{s}-\sqrt{s-4m_e^2}}{\sqrt{s}+\sqrt{s-4m_e^2}}\right)\right\rbrace \overset{s\gg m_e^2}{\longrightarrow} \frac{\pi \alpha^6(0)}{N^3}\frac{9+\log{\left(\frac{s}{m_e^2}\right)}}{s}\,.\label{eq:eexs4orthpositronium}
    \end{aligned}
\end{equation}
\begin{figure}[t]
    \begin{subfigure}[b]{0.5\textwidth}
        \includegraphics[width=\textwidth]{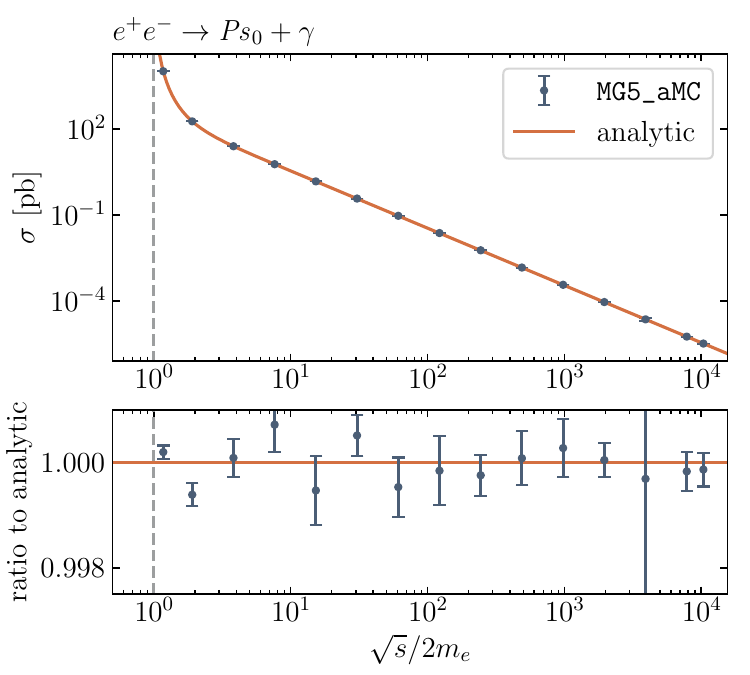}
        \caption{}
        \label{fig:ee_to_Ps0}
    \end{subfigure}
    \begin{subfigure}[b]{0.5\textwidth}
        \includegraphics[width=\textwidth]{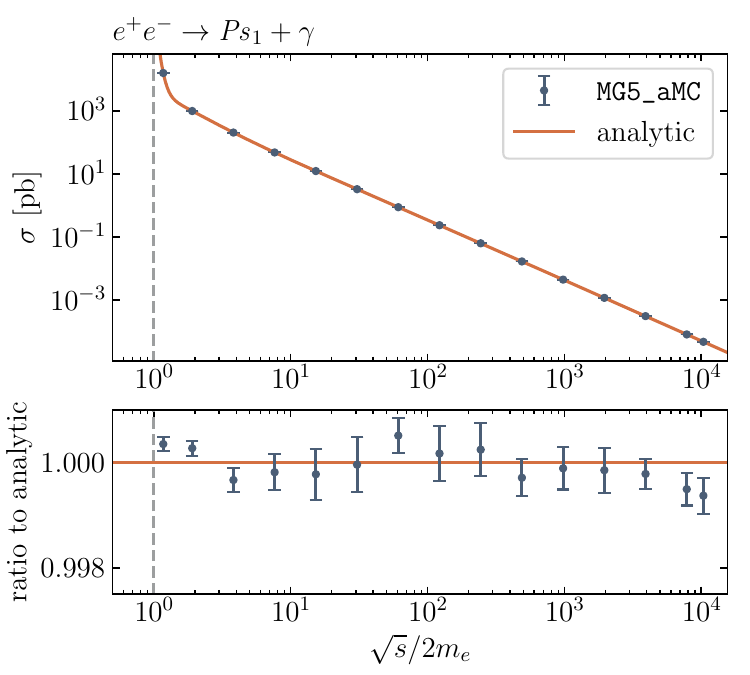}
        \caption{}
        \label{fig:ee_to_Ps1}
    \end{subfigure}
    \caption{Associated positronium-production cross sections for (a) para-positronium plus photon and (b) ortho-positronium plus photon in $e^+e^-$ collisions. The upper panels show the total cross sections as functions of the $e^+e^-$ c.m. energy, obtained with \mgshort\ (blue dots) and from analytic expressions (orange line). The lower panels display the ratio of the two results, normalised to the analytic reference values. Error bars on the \mgshort\ points indicate statistical MC uncertainties.}
    \label{fig:ee_to_Ps}
\end{figure}\noindent
Figure~\ref{fig:ee_to_Ps} compares the numerical outputs from \mgshort\ with these analytic results as a function of the ratio of the $e^+e^-$ c.m. energy to the positronium mass. To obtain the infinitely heavy $Z$ and $H$ boson limit in \mgshort, we use the prompts:
\vskip 0.25truecm
\noindent
~~\prompt\ {\tt ~import~model~sm\_onia-lepton\_masses}

\noindent
~~\prompt\ {\tt~generate~e+ e- > Ps(1|1S0) a / Z H}

\noindent
~~\prompt\ {\tt ~output;~launch}
\vskip 0.25truecm
\noindent
which explicitly exclude all diagrams containing virtual $Z$ or $H$ exchanges. Excellent agreement is found between the numerical and analytic calculations over a wide range of energies, from near threshold ($\sqrt{s}\sim 2m_e$) to the high-energy regime ($\sqrt{s}\gg 2m_e$), with \mgshort\ reproducing the analytic results to permille precision and demonstrating stable, reliable behaviour within the expected statistical uncertainties from MC integration.

\subsection{True muonium production}

The existence of a bound state of a muon and an antimuon, called true muonium ($\mathit{TM}$), was first proposed in the 1960s~\cite{Bilenky:1969zd}. Despite theoretical interest, its experimental discovery remains elusive. Various production mechanisms have been studied, including rare radiative decays of mesons such as $\eta$, $\eta^\prime$, $K_0$, and $B$~\cite{Nemenov:1972ph,Ji:2017lyh,Ji:2018dwx,Fael:2018ktm,CidVidal:2019qub}. Feasibility studies at fixed-target experiments, electron-positron, electron-ion, and hadron colliders~\cite{Banburski:2012tk,Gatto:2016rae,Gninenko:2025hsv,Moffat:1975uw,Brodsky:2009gx,Bogomyagkov:2017uul,Fox:2021mdn,Holvik:1986ty,Francener:2024eep,Ginzburg:1998df,Fael:2018ktm,CidVidal:2019qub,Azevedo:2019hqp,refId0,Yu:2022hdt,Bertulani:2023nch,Dai:2024imb,dEnterria:2025ecx,Liang:2025sol,Feng:2025ppc} suggest that its observation could be within reach of modern experiments. Due to the larger mass of the muon compared to the electron, true muonium has a Bohr radius roughly 200 times smaller than positronium, making its detection at collider experiments more feasible, as it is less susceptible to dissociation by external magnetic fields or material interactions.

\begin{table}[t!]
\centering
\renewcommand*{\arraystretch}{1.4}
\begin{tabular}[t]{lc|lc}
\toprule
\textbf{process}& $\sigma$ & \textbf{process}& $\sigma$  \\
\midrule
& & $pp \to \mathit{TM}_{1}+j+X$ & $9.460(2)\,\mathrm{fb}$ \\ \midrule
$e^+e^-\to\mathit{TM}_0+\gamma$ & $438.16(1)\,\mathrm{yb}$ & $e^+e^- \to \mathit{TM}_{1}+\gamma$ & $8.422(2)\,\mathrm{ab}$ \\
\bottomrule
\end{tabular}
\caption{Cross-section results for para- ($\mathit{TM}_0$) and ortho-true muonium ($\mathit{TM}_1$) production at $\sqrt{s}=13$ TeV in $pp$ collisions and at $\sqrt{s}=10.58\,\mathrm{GeV}$ in $e^+e^-$ collisions, based on the setup described in section~\ref{sec:setup}. For ortho-true muonium, additional fiducial cuts are applied: $p_{T,j}>2\,\mathrm{GeV}$ and $|\eta_j|<5$ for final-state jets, and $p_{T,\gamma}>0.5\,\mathrm{GeV}$ and $|\eta_\gamma|<5$ for photons. All cross sections are rescaled to the $\alpha(0)$ scheme. Numbers in parentheses indicate numerical uncertainties from the MC phase-space integration.}
\label{tab:truemuonium}
\end{table}

Following our analysis of positronium, we present cross sections for the ground-state configurations of $n={}^1S_0$ para-true muonium ($\mathit{TM}_0$) and $n={}^3S_1$ ortho-true muonium ($\mathit{TM}_1$). In table~\ref{tab:truemuonium}, we show the cross section for the hadronic process
\begin{equation}\label{eq:otm_jet}
    pp\to\mathit{TM}_1+j+X 
\end{equation}
based on the setup described in section~\ref{sec:setup}, with fiducial cuts $p_{T,j}>2\,\mathrm{GeV}$ and $|\eta_j|<5$. The production of ortho-true muonium in association with a jet provides a promising experimental signature. Our result of $9.460(2)\,\mathrm{fb}$ indicates that observing true muonium in $pp$ collisions remains challenging, but may still be within experimental reach. Moreover, the cross section for this channel is nearly identical to that of ortho-positronium plus jet production (Eq.~\eqref{eq:ops_jet}), suggesting that mass effects are negligible. This is expected when $p_{T,\mathit{TM}_1}\gg 2m_\mu$, which is satisfied here due to the $p_{T,j}$ cut. The $p_{T,\mathit{TM}_1}$ spectrum shown in figure~\ref{fig:truemuonium} confirms this, matching the positronium case across the entire range.

\begin{figure}[t]
    \center
    \begin{subfigure}[b]{\textwidth}\center
        \includegraphics[width=0.4\textwidth]{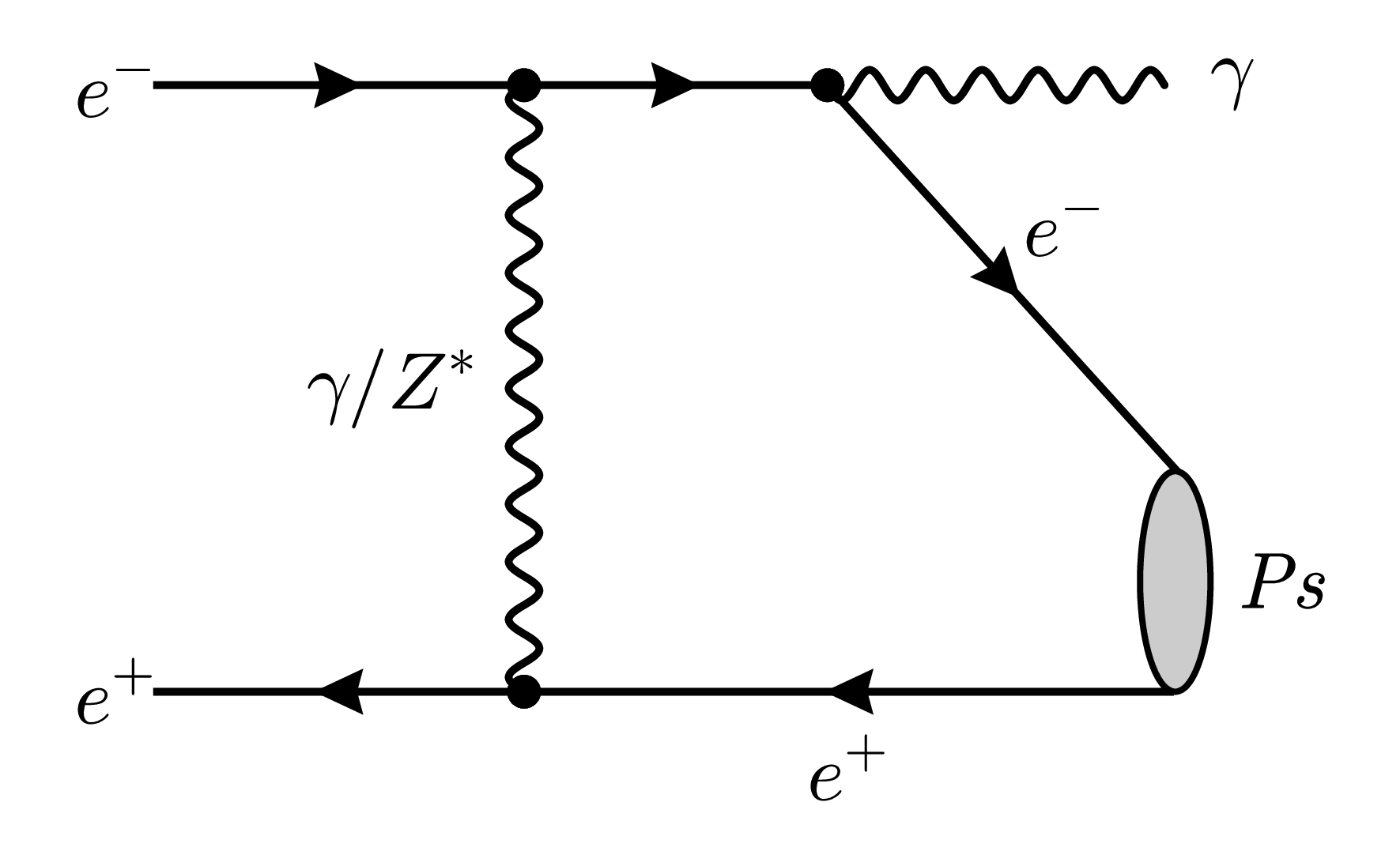}
        \includegraphics[width=0.4\textwidth]{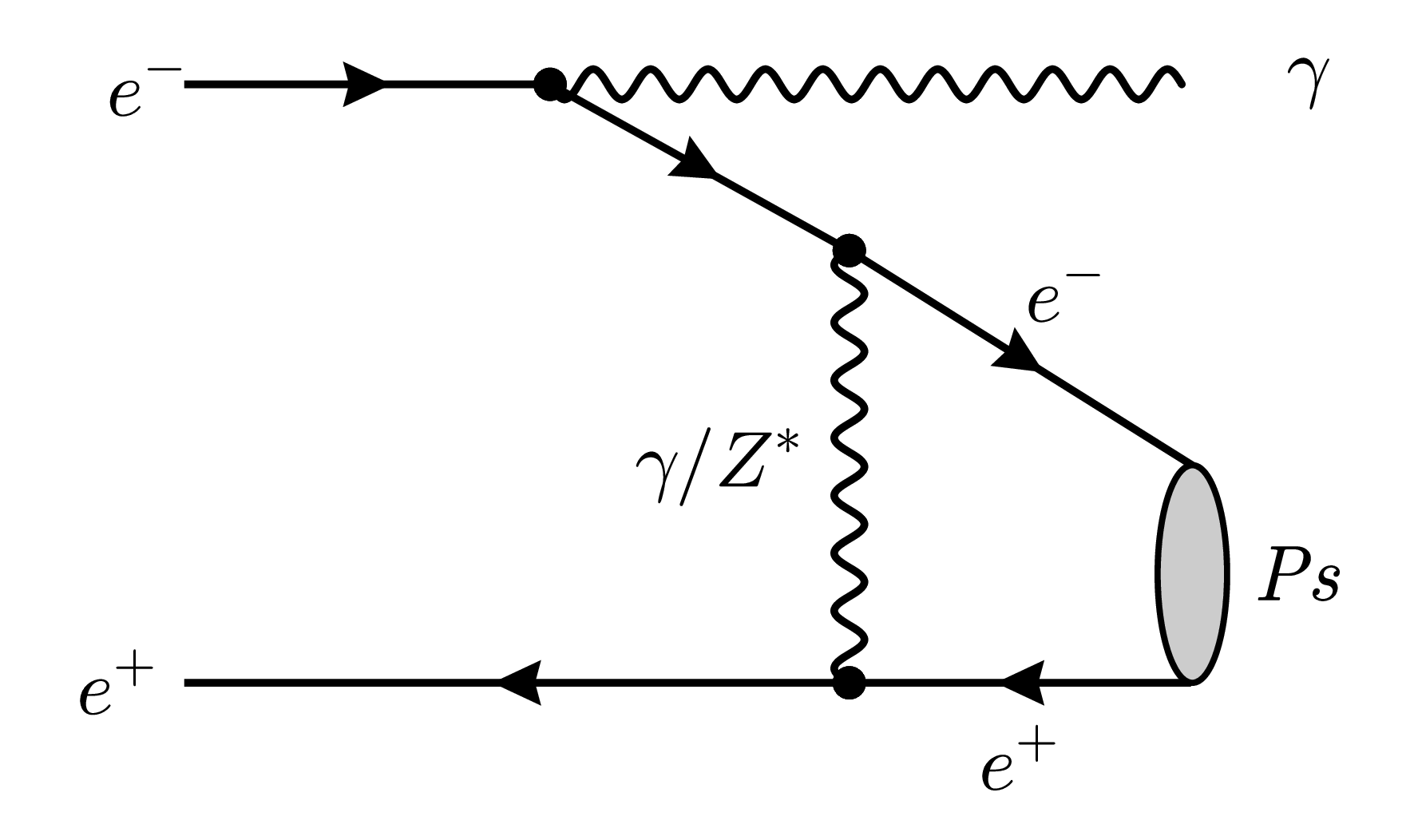}
        \caption{}\label{fig:diag_leptonium_a}
    \end{subfigure}
    \begin{subfigure}[b]{0.4\textwidth}
        \includegraphics[width=\textwidth]{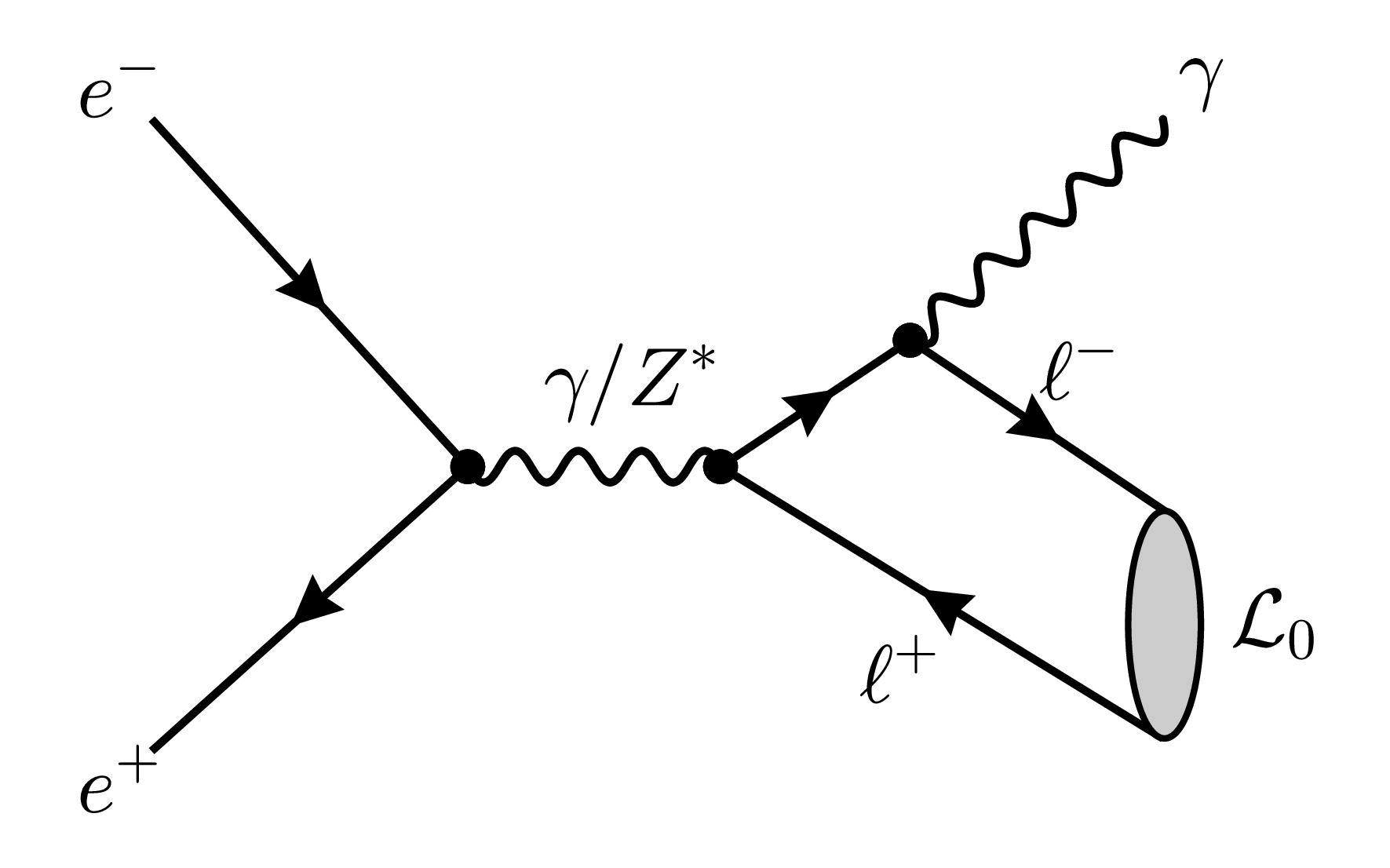}
        \caption{}\label{fig:diag_leptonium_b}
    \end{subfigure}
    \begin{subfigure}[b]{0.4\textwidth}
        \includegraphics[width=\textwidth]{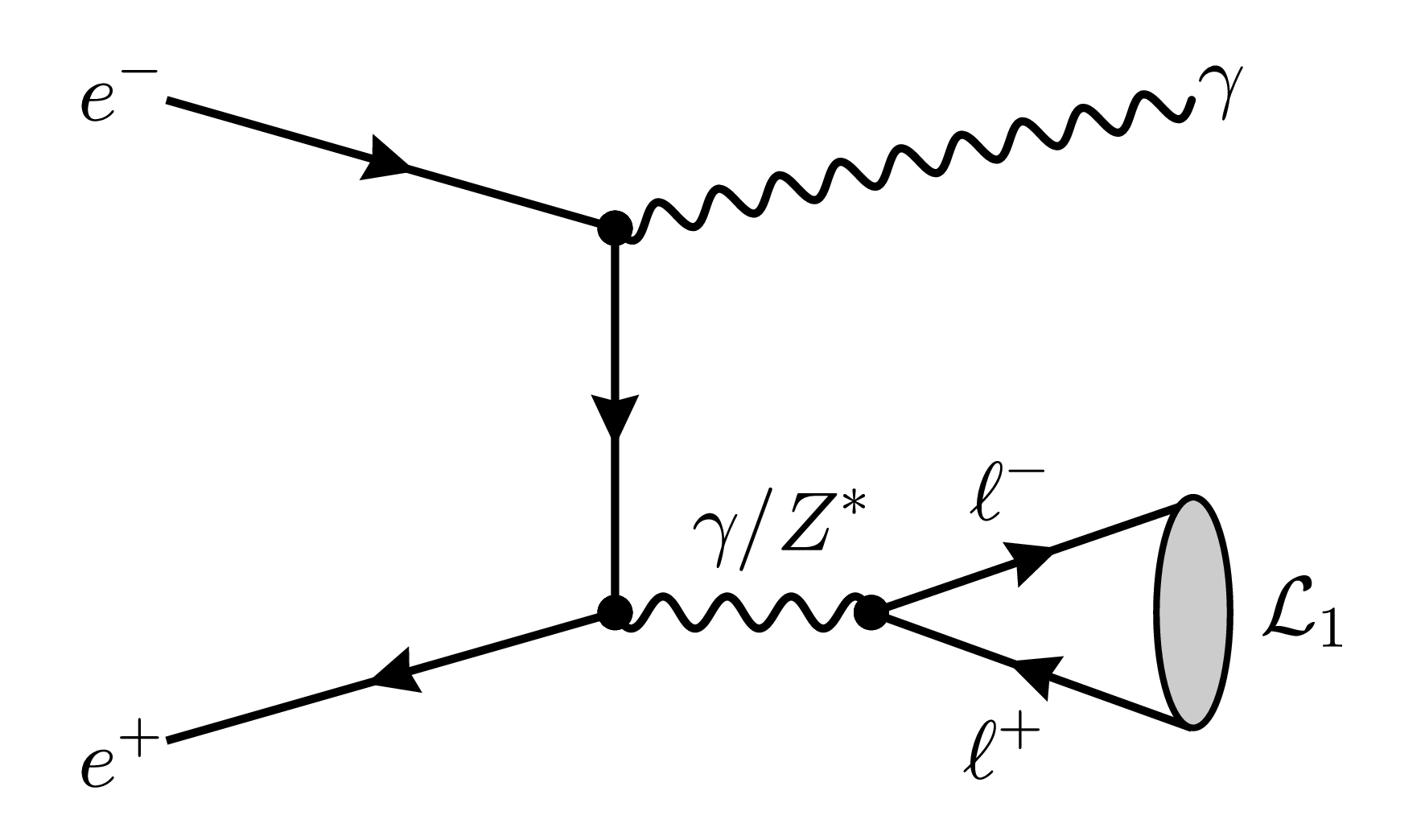}
        \caption{}\label{fig:diag_leptonium_c}
    \end{subfigure}
    \caption{Example Feynman diagrams: (a) para- or ortho-positronium production via $t$-channel $\gamma/Z^\ast$ exchange; (b) para-leptonium production in association with a photon from FSR; and (c) ortho-leptonium production in association with a photon from ISR at $e^+e^-$ colliders. The leptonium $\mathcal{L}$ can be positronium ($\mathit{Ps}$), true muonium ($\mathit{TM}$), or ditauonium ($\mathcal{T}$), corresponding to $\ell^\pm = e^\pm$, $\mu^\pm$, or $\tau^\pm$, respectively.}
	\label{fig:diag_leptonium}
\end{figure}\noindent

Searching for a $\mathit{TM}$ plus photon final state in $e^+e^-$ collisions has already been proposed in ref.~\cite{Brodsky:2009gx}, concluding that this process may be accessible at Belle II. Both para- and ortho-true muonium can be produced in association with a photon, so we consider the $e^+e^-$ annihilation processes
\begin{align}
    e^+e^-&\to\mathit{TM}_0+\gamma\,,\\
    e^+e^-&\to\mathit{TM}_1+\gamma\,.
\end{align}
Unlike the previously considered positronium production processes, no $t$-channel mediated diagrams contribute here (cf.\@\xspace figure~\ref{fig:diag_leptonium_a}); the final-state photon in para- and ortho-true muonium production originates from final-state radiation (FSR) and initial-state radiation (ISR), respectively. Exemplary Feynman diagrams are shown in figures~\ref{fig:diag_leptonium_b} and \ref{fig:diag_leptonium_c}, respectively.

For para-true muonium production in association with an FSR photon, we compute a total inclusive cross section of $438.16(1)\,\mathrm{yb}$, which is too small to be experimentally detectable. Although this process may not be phenomenologically relevant, it serves as a useful benchmark for validating the \mgshort\ implementation. Using the analytic expression for $e^+e^-\to\mathit{TM}_0+\gamma$ in the limit of infinitely heavy $Z$ and $H$ bosons~\cite{dEnterria:2023yao},
\begin{equation}\label{eq:leptonium_analytic}
    \lim_{m_Z,m_H\to\infty}\sigma(e^+e^-\to\mathit{TM}_0+\gamma) = \dfrac{2}{3}\dfrac{\pi\alpha^6(0)}{N^3}\dfrac{4m_\mu^2}{s^2}\left(1-\dfrac{4m_\mu^2}{s}\right)\,,
\end{equation}
where the electron mass is neglected, we obtain a cross section of $438.721\,\mathrm{yb}$, in agreement at the permille level. The extremely small cross section at $\sqrt{s}=10.58$ GeV can be understood from Eq.~\eqref{eq:leptonium_analytic}, which is suppressed by a factor of $4m_\mu^2/3s \approx 1.3 \cdot 10^{-4}$ compared to the para-positronium case (Eq.~\eqref{eq:eexs4parapositronium}).

For ortho-true muonium produced in association with an ISR photon, fiducial photon cuts of $p_{T,\gamma}>0.5\,\mathrm{GeV}$ and $|\eta_\gamma|<5$ are applied to regulate collinear initial-state singularities. This results in a significantly larger cross section of $8.422(2)\,\mathrm{ab}$ compared to para-true muonium production, confirming the potential for measuring true muonium at $e^+e^-$ colliders. Moreover, this cross section is quite similar to that of the corresponding ortho-positronium process.

\subsection{True tauonium production}

The heaviest possible atomic state of two leptons is a bound state of a $\tau^+\tau^-$ pair, commonly referred to as ditauonium or true tauonium. Its existence was proposed nearly five decades ago~\cite{Moffat:1975uw,Avilez:1977ai,Avilez:1978sa}, but experimental evidence remains elusive. In recent years, however, ditauonium has regained attention in theoretical studies. Its spectroscopic properties have been investigated in ref.~\cite{dEnterria:2022alo}, and feasibility studies of its production at electron-positron, electron-ion, and hadron colliders~\cite{Malik:2008pn,refId0,dEnterria:2022ysg,Yu:2022hdt,dEnterria:2023yao,Fu:2023uzr,Francener:2024eep} suggest that observing ditauonium may be challenging but achievable in the near future -- for instance, ortho-ditauonium at a future super tau-charm factory~\cite{dEnterria:2023yao} and para-ditauonium at FCC-ee or CEPC~\cite{dEnterria:2022ysg}.

As in the cases of positronium and true muonium discussed in the previous sections, we present cross-section calculations for ditauonium production in its ground-state configurations with principal quantum number $N=1$. Specifically, we consider the $n={}^1S_0$ singlet para-ditauonium, denoted $\mathcal{T}_0$, and the $n={}^3S_1$ triplet ortho-ditauonium, $\mathcal{T}_1$. Calculations are performed using the baseline setup described in section~\ref{sec:setup} for both $e^+e^-$ and $pp$ colliders. A summary of the results is provided in table~\ref{tab:ditauonium}.

\begin{table}[t!]
\centering
\renewcommand*{\arraystretch}{1.4}
\begin{tabular}[t]{lc|lc}
\toprule
\textbf{process}& $\sigma$ & \textbf{process}& $\sigma$ \\
\midrule
$pp\xrightarrow{\gamma\gamma} p~\mathcal{T}_{0}~p$ & $108.055(7)\,\mathrm{ab}$ & $pp \to \mathcal{T}_{1}+X$ & $9.1906(4)\,\mathrm{fb}$ \\
$pp\to \mathcal{T}_{0}+\gamma+X$ & $3.4522(3)\,\mathrm{ab}$ & $pp \to \mathcal{T}_{1}+j+X$ & $6.055(1)\,\mathrm{fb}$ \\ \midrule
$e^+e^-\to\mathcal{T}_0+\gamma$ & $110.001(3)\,\mathrm{zb}$ & $e^+e^- \to \mathcal{T}_{1}+\gamma$ & $9.576(2)\,\mathrm{ab}$ \\
\bottomrule
\end{tabular}
\caption{Cross-section results for para- ($\mathcal{T}_0$) and ortho-ditauonium ($\mathcal{T}_1$) production at $\sqrt{s}=13$ TeV in $pp$ collisions and at $\sqrt{s}=10.58$ GeV in $e^+e^-$ collisions, based on the setup described in section~\ref{sec:setup}. Additional fiducial cuts, $p_{T,j}>2\,\mathrm{GeV}$ and $|\eta_j|<5$ ($p_{T,\gamma}>0.5\,\mathrm{GeV}$ and $|\eta_\gamma|<5$), are applied to final-state jets (photons) in the computations for ortho-ditauonium production. All cross sections are rescaled to the $\alpha(0)$ scheme. Numbers in parentheses indicate numerical uncertainties from the MC phase-space integration.}
\label{tab:ditauonium}
\end{table}

The four hadronic processes we consider are
\begin{align}
    pp&\xrightarrow{\gamma\gamma} p~\mathcal{T}_{0}~p\,,\label{eq:pdt_upc}\\
    pp&\to\mathcal{T}_0+\gamma+X\,,\label{eq:pdt_photon}\\
    pp&\to\mathcal{T}_1+X\,,\label{eq:odt}\\
    pp&\to\mathcal{T}_1+j+X \label{eq:odt_jet}\,,
\end{align}
where the last process, Eq.~\eqref{eq:odt_jet}, is computed in analogy to Eqs.~\eqref{eq:ops_jet} and \eqref{eq:otm_jet}, applying the same fiducial jet cuts, $p_{T,j}>2\,\mathrm{GeV}$ and $|\eta_j|<5$.
In addition, we are now able to investigate three processes that could not previously be described consistently in perturbation theory for positronium and true muonium production. In the ultraperipheral collision (UPC) process of Eq.~\eqref{eq:pdt_upc}, para-ditauonium is produced via photon–photon fusion. This is simulated by combining \mgshort\ with \gammaupc~\cite{Shao:2022cly}, employing the charge form factor (ChFF) to model the photon flux. Owing to the larger hard scale involved in ditauonium production compared to positronium and true muonium, the equivalent photon approximation used here is well justified. Similar arguments apply to the photon-associated production of para-ditauonium, Eq.~\eqref{eq:pdt_photon}, and to the direct production of ortho-ditauonium, Eq.~\eqref{eq:odt}, both of which can now be consistently described within collinear factorisation.
We obtain cross sections of the order of $1$–$100\,\mathrm{ab}$ for para-ditauonium production and approximately $10\,\mathrm{fb}$ for ortho-ditauonium production at the LHC, consistent with previous findings~\cite{dEnterria:2023yao}. For the $\mathcal{T}_1+j$ process, the impact of the heavy $\tau$-lepton mass becomes clearly evident when compared to the corresponding positronium ($\mathit{Ps}_1+j$) and true muonium ($\mathit{TM}_1+j$) processes, cf.\@\xspace tables~\ref{tab:positronium} and \ref{tab:truemuonium}.
As shown by the $p_{T,\mathcal{T}_1}$ spectrum in figure~\ref{fig:ditauonium}, and even more clearly in figure~\ref{fig:leptonium_all}, these mass effects are prominent at low $p_T$, whereas at higher momenta, $p_T \gtrsim 15\,\mathrm{GeV}$, the distributions for positronium, true muonium, and ditauonium converge.

In electron–positron collisions, we consider the production of para- and ortho-ditauonium in association with a photon,
\begin{align}
    e^+e^-&\to\mathcal{T}_0+\gamma\,,\\
    e^+e^-&\to\mathcal{T}_1+\gamma\,,
\end{align}
originating from FSR and ISR, respectively (see figures~\ref{fig:diag_leptonium_b} and~\ref{fig:diag_leptonium_c}).

The para-ditauonium production process has a cross section of $110.001(3)\,\mathrm{zb}$ and, once again, can be validated against the analytic expression in the limit of decoupled $Z$ and $H$ bosons~\cite{dEnterria:2023yao},
\begin{equation}
\lim_{m_Z,m_H\to\infty}\sigma(e^+e^-\to\mathcal{T}_0+\gamma) = \dfrac{2}{3}\dfrac{\pi\alpha^6(0)}{N^3}\dfrac{4m_\tau^2}{s^2}\left(1-\dfrac{4m_\tau^2}{s}\right)\,,
\end{equation}
which yields a cross section of $110.133\,\mathrm{zb}$, in agreement with the \mgshort\ result at the permille level.

For ortho-ditauonium production, we apply fiducial cuts of $p_{T,\gamma}>0.5\,\mathrm{GeV}$ and $|\eta_\gamma|<5$ to photons from ISR, obtaining a cross section of $9.576(2)\,\mathrm{fb}$. The cross section is of the same order as those for ortho-positronium and ortho-true muonium. The same reasoning as for the exclusive charmonium-plus-photon production discussed in section~\ref{sec:eeexcl} can be invoked to understand why ortho-ditauonium exhibits a much larger yield than para-ditauonium.

In conclusion, with cross sections of the order of $0.1\,\mathrm{ab}$ and $10\,\mathrm{ab}$ for para- and ortho-ditauonium, respectively, our findings are consistent with previous dedicated feasibility studies. We do not perform an independent sensitivity analysis here but instead follow the conclusions of earlier works, such as ref.~\cite{dEnterria:2023yao}, which indicate that $B$-factory experiments like Belle~II are unlikely to observe ortho-ditauonium. As pointed out in ref.~\cite{dEnterria:2023yao}, the most promising approach for experimentally observing ortho-ditauonium is at a future super tau-charm factory~\cite{Achasov:2023gey} via a threshold scan around $2m_\tau$, while it might also be worthwhile to attempt its observation at the LHC by measuring the process $pp\to \mathcal{T}_1(\to \mu^+\mu^-)+j+X$.

%% file: 08_conclusions.tex
\section{Conclusions}\label{sec:conclusions}

We have developed a comprehensive implementation of S-wave quarkonium and leptonium production within the \mgshort\ framework, enabling automated event generation for these processes based on the LO NRQCD and NRQED formalisms, respectively. The implementation supports inclusive single, multiple, and associated quarkonium and leptonium production across a wide range of collider environments, from lepton–lepton collisions to hadron and photon–induced processes, as well as exclusive $\gamma\gamma$ fusion and $e^+e^-$ annihilation reactions. It further provides, for the first time, an automated treatment of leptonium production, all accessible through a user-friendly, \ufo-based interface. In particular, we have studied single- and associated-quarkonium production with a heavy-quark pair, electroweak boson(s), or jet(s), as well as di- and tri-quarkonium production. On the leptonium side, we have investigated associated production with an electroweak boson or a jet, and assessed its feasibility at current and future experiments. Through numerous concrete examples, one of the most intriguing conclusions we wish to highlight is that theoretical studies of various quarkonium processes usually require careful consideration. Owing to factors such as quantum-number conservation or kinematic/dynamical enhancement or suppression of certain channels, the impact of subleading contributions can easily be underestimated if one relies solely on simple counting arguments based on the hierarchy of couplings and velocity-scaling rules.

Benchmark comparisons with existing tools and analytic expressions confirm the accuracy and reliability of our implementation. Cross-section computations have been provided for a broad set of final-state bound states and observables, demonstrating the potential of our extension. In addition, our implementation is fully compatible with and seamlessly integrates into other \mgshort\ features, including custom \ufo\ models adapted for bound-state production, the parton-shower interface, and \gammaupc\ for studying photon–photon collisions in proton or nuclear UPCs.

Our work represents an important first step toward delivering state-of-the-art theoretical predictions for bound-state production studies within the collinear factorisation framework in the widely-used event generator \mgshort. Future studies include the automation of P-wave quarkonium production and, in the near term, extension to NLO accuracy. These developments will further expand the scope of bound-state studies in collider environments and facilitate future global NRQCD analyses and comparisons with experimental data. All implementations will be made publicly available via the standard \mgshort\ distribution on Launchpad~\footnote{\url{https://launchpad.net/mg5amcnlo}}, as well as via the \texttt{NLOAccess} EU Virtual Access~\footnote{\url{https://nloaccess.in2p3.fr/}}, which provides the high-energy, hadronic, and heavy-ion physics communities with versatile, high-precision tools for quarkonium and leptonium physics in a user-accessible format.

%% file: acknowledgements.tex
\begin{acknowledgments}
We would like to thank C.~Flore, R.~Frederix, K.~Lynch, F.~Maltoni, M.~Mangano, M.~Nefedov, and H.-F.~Zhang for useful discussions. L.S.\@\xspace thanks the Centre for Cosmology, Particle Physics and Phenomenology (CP3) at Universit\'e Catholique de Louvain for hospitality, where part of this work was carried out.

C.F.\@\xspace acknowledges support from the Marie Skłodowska-Curie Action (``AutomOnium''), funded by the European Union under grant agreement No.\@\xspace 101204057.

C.F., J.-P.L., and H.-S.S.\@\xspace are supported by the Agence Nationale de la Recherche (ANR) via the grant ANR-20-CE31-0015 (``PrecisOnium'').

The work of C.F.\@\xspace and J.-P.L.\@\xspace is supported by the IDEX Paris-Saclay ``Investissements d'Avenir'' (ANR-11-IDEX-0003-01) through the GLUODYNAMICS project funded by the ``P2IO LabEx (ANR-10-LABX-0038)'', the French CNRS via the IN2P3 projects ``GLUE@NLO” and ``QCDFactorisation@NLO'' as well as via the COPIN-IN2P3 project \#12-147 ``kT factorisation and quarkonium production in the LHC era''.

O.M.\@\xspace and the \madgraph\ project are supported by FRS-FNRS (Belgian National Scientific Research Fund) IISN projects 4.4503.16 (MaxLHC) and DR-Weave grant FNRS-DFG num\'ero T019324F (40020485). 

The work of H.-S.S.\@\xspace and L.S.\@\xspace is supported by the ERC grant 101041109 (``BOSON''). Views and opinions expressed are however those of the authors only and do not necessarily reflect those of the European Union or the European Research Council Executive Agency. Neither the European Union nor the granting authority can be held responsible for them.

\end{acknowledgments}